\documentclass[twocolumn, trackchanges]{aastex701}

\usepackage[T1]{fontenc}
\usepackage{ae,aecompl}

\usepackage{graphicx}	
\usepackage{amsmath}	
\usepackage{amssymb}	

\usepackage{natbib}

\bibpunct{(}{)}{;}{a}{}{,}

\usepackage{url}

\usepackage{longtable}
\usepackage{booktabs}

\tabletypesize{\scriptsize}

\usepackage[flushleft, referable]{threeparttablex}

\usepackage{subfloat}

\DeclareGraphicsExtensions{.jpg,.pdf,.png}

\usepackage{parskip}

\usepackage{lipsum}

\usepackage{dsfont}

\usepackage{comment}

\usepackage{tablefootnote}

\newcommand\lenstool{\textsc{lenstool}}

\newcommand\Neff{$\mathcal{N}_{\rm eff}$}

\graphicspath{{./}{figs/}}

\begin{document}

\title{Forecasting the Observable Rates of Gravitationally Lensed Supernovae for the PASSAGES Dusty Starbursts}

\author[0000-0001-9394-6732]{Patrick S. Kamieneski}
\affiliation{School of Earth and Space Exploration, Arizona State University, 
PO Box 876004,
Tempe, AZ 85287-6004, USA}
\email[show]{pkamiene@asu.edu}

\author[0000-0001-8156-6281]{Rogier A. Windhorst}
\affiliation{School of Earth and Space Exploration, Arizona State University, 
PO Box 876004,
Tempe, AZ 85287-6004, USA}
\email{rogier.windhorst@gmail.com}

\author[0000-0003-1625-8009]{Brenda L. Frye}
\affiliation{Department of Astronomy/Steward Observatory, University of Arizona, 933 N Cherry Ave, Tucson, AZ, 85721-0009}
\email{brendafrye@gmail.com}

\author[0000-0001-7095-7543]{Min S. Yun}
\affiliation{Department of Astronomy, University of Massachusetts, Amherst, MA 01003, USA}
\email{myun@astro.umass.edu}

\author[0000-0001-5429-5762]{Kevin C. Harrington}
\affiliation{European Southern Observatory, Alonso de C{\'o}rdova 3107, Vitacura, Casilla 19001, Santiago de Chile, Chile}
\affiliation{Joint ALMA Observatory, Alonso de C{\'o}rdova 3107, Vitacura, Casilla 19001, Santiago de Chile, Chile}
\affiliation{National Astronomical Observatory of Japan, Los Abedules 3085 Oficina 701, Vitacura 763 0414, Santiago, Chile}
\email{Kevin.Harrington@alma.cl}

\author[0000-0002-5573-9131]{Simon D. Mork}
\affiliation{School of Earth and Space Exploration, Arizona State University, 
PO Box 876004,
Tempe, AZ 85287-6004, USA}
\email{sdmork@asu.edu}

\author[0000-0002-7460-8460]{Nicholas Foo}
\affiliation{School of Earth and Space Exploration, Arizona State University, 
PO Box 876004,
Tempe, AZ 85287-6004, USA}
\email{nfoo1@asu.edu}

\author[0000-0003-3418-2482]{Nikhil Garuda}
\affiliation{Department of Astronomy/Steward Observatory, University of Arizona, 933 N Cherry Ave, Tucson, AZ, 85721-0009}
\email{nikhilgaruda@arizona.edu}

\author[0000-0002-2282-8795]{Massimo Pascale}
\affiliation{Department of Astronomy, University of California, 501 Campbell Hall \#3411, Berkeley, CA 94720, USA}
\email{massimopascale@berkeley.edu}

\author[0000-0002-4140-0428]{Bel\'{e}n Alcalde Pampliega}
\affiliation{European Southern Observatory, Alonso de C{\'o}rdova 3107, Vitacura, Casilla 19001, Santiago de Chile, Chile}
\affiliation{SKA Observatory, Jodrell Bank, SK11 9FT, UK}
\email{Belen.AlcaldePampliega@skao.int}

\author[0000-0001-6650-2853]{Timothy Carleton}
\affiliation{School of Earth and Space Exploration, Arizona State University, 
PO Box 876004,
Tempe, AZ 85287-6004, USA}
\email{tmcarlet@asu.edu}

\author[0000-0003-3329-1337]{Seth H. Cohen}
\affiliation{School of Earth and Space Exploration, Arizona State University, 
PO Box 876004,
Tempe, AZ 85287-6004, USA}
\email{seth.cohen@asu.edu}

\author[0000-0002-4223-2016]{Carlos Garcia Diaz}
\affiliation{Department of Astronomy, University of Massachusetts, Amherst, MA 01003, USA}
\email{cgarciadiaz@umass.edu}

\author[0000-0003-1268-5230]{Rolf A. Jansen}
\affiliation{School of Earth and Space Exploration, Arizona State University, 
PO Box 876004,
Tempe, AZ 85287-6004, USA}
\email{Rolf.Jansen@asu.edu}

\author[0000-0002-2640-5917]{Eric F. Jim\'{e}nez-Andrade}
\affiliation{Instituto de Radioastronom\'{i}a y Astrof\'{i}sica, Universidad Nacional Aut\'{o}noma de M\'{e}xico, Antigua Carretera a P\'{a}tzcuaro \# 8701, Ex-Hda. San Jos\'{e} de la Huerta, Morelia, Michoac\'{a}n, C.P. 58089, M\'{e}xico}
\email{e.jimenez@irya.unam.mx}

\author[0000-0002-6610-2048]{Anton M. Koekemoer}
\affiliation{Space Telescope Science Institute, 3700 San Martin Dr., Baltimore, MD 21218, USA}
\email{koekemoer@stsci.edu}

\author[0000-0001-9969-3115]{James D. Lowenthal}
\affiliation{Smith College, Northampton, MA 01063, USA}
\email{jlowenth@smith.edu}

\author[0000-0003-1832-4137]{Allison Noble}
\affiliation{School of Earth and Space Exploration, Arizona State University, 
PO Box 876004,
Tempe, AZ 85287-6004, USA}
\email{anoble5@asu.edu}

\author[0000-0002-2361-7201]{Justin D. R. Pierel}
\affiliation{Space Telescope Science Institute, 3700 San Martin Dr., Baltimore, MD 21218, USA}
\email{jpierel@stsci.edu}

\author[0000-0002-4444-8929]{Amit Vishwas}
\affiliation{Cornell Center for Astrophysics and Planetary Science, Cornell University, Space Sciences Building, Ithaca, NY 14853, USA}
\email{vishwas@cornell.edu}

\author[0000-0002-9279-4041]{Q. Daniel Wang}
\affiliation{Department of Astronomy, University of Massachusetts, Amherst, MA 01003, USA}
\email{wqd@astro.umass.edu}

\author[0000-0001-9163-0064]{Ilsang Yoon}
\affiliation{National Radio Astronomy Observatory, 520 Edgemont Road, Charlottesville, VA 22903, USA}
\email{iyoon@nrao.edu}

\begin{abstract}
More than 60 years have passed since the 
first formal suggestion
to use strongly-lensed supernovae to
measure the expansion rate of the Universe 
through 
time-delay cosmography.
Yet, fewer than 10 such objects have ever been discovered.
We 
consider
the merits of
a targeted strategy focused on lensed hyperluminous infrared galaxies\textemdash 
among the most rapidly star-forming galaxies known in the Universe.
With star formation rates (SFRs)
$\sim {200 - 6000}~{\rm M}_\odot~{\rm yr}^{-1}$, 
the $\sim 30$ objects in the Planck All-Sky Survey to Analyze Gravitationally-lensed Extreme Starbursts (PASSAGES) are excellent candidates for a case study, in particular, and have already led to the discovery of the multiply-imaged SN H0pe.
Considering their lens model-corrected SFRs,
we estimate their intrinsic supernova rates to be
an extraordinary 
${1.8 - 65}~{\rm yr}^{-1}$ (core-collapse) and ${0.2 - 6.4}~{\rm yr}^{-1}$ (Type Ia).
Moreover, 
these massive starbursts typically have star-forming companions 
which
are unaccounted for in 
this tally.
We demonstrate a strong correlation between Einstein radius and typical time delays, with cluster lenses often exceeding several months (and therefore most favorable for high-precision $H_0$ inferences).
A multi-visit monitoring campaign with 
a 
sensitive infrared telescope 
(namely, JWST)
is necessary to mitigate 
dust attenuation.
Still, 
a porous interstellar medium and clumpy star formation in these extreme galaxies might produce favorable conditions for detecting supernovae as transient point sources.
Targeted
campaigns
of known lensed galaxies
to discover
new lensed supernovae 
can greatly complement
wide-area cadenced surveys. 
Increasing the sample size 
helps to realize the potential of supernova time-delay cosmography to elucidate the Hubble tension 
through a single-step measurement,
independent of other $H_0$ techniques.
%
%
%
\end{abstract}





\section{Introduction} 
\label{sec:intro}

Gravitationally lensed supernovae (glSNe) are invaluable cosmological tools that are finally now being discovered and leveraged 
to make
unique, one-step measurements of the Hubble constant, $H_0$, describing the expansion rate of the Universe \citep{Suyu:2024aa}.
For multiply-imaged lensed objects, the arrival of the different images is staggered in time due to 
$i)$ the difference in path lengths along the different null geodesics;
and
$ii)$ the Shapiro effect \citep{Shapiro:1964aa}, by which the light rays experience disparate amounts of time dilation from the differing gravitational potential along their paths. 
The magnitude of this effect for galaxies at cosmological distances\textemdash typically of order hours to decades\textemdash is usually far too small to be detectable, given the timescales of Myr for galaxy-scale variations. 
However, with rapidly varying or transient signals, like from quasars or supernovae, it is possible to capture and measure these lensing-induced time delays through light curve monitoring. 
Through a technique now known as time-delay (TD) cosmography first introduced by \citealt{Refsdal:1964aa}
(see also reviews by
\citealt{Blandford:1992aa, Treu:2016aa, Treu:2022ac, Birrer:2024aa}),
these delays may be inverted to infer 
distances within the Hubble flow, and therefore $H_0$.
This owes to the dependence of the time delays on a ratio of 3 angular-diameter distances (from the observer to the lens, from the observer to the source, and from the lens to the source; \citealt{Schneider:1985aa}), which in turn is proportional to $H_0^{-1}$ (with minimal dependence on other cosmological parameters).

The ``Hubble tension" 
between calculations of the Hubble constant $H_0$ 
still endures, despite an extraordinary number 
of measurements
that have been put forth in the last two decades (see recent review by \citealt{Verde:2024aa}).
The primary conflict arises from the smaller value of $H_0 =  67.4 \pm 0.5~{\rm km~ s}^{-1} ~{\rm Mpc}^{-1}$ \citep{Planck-Collaboration:2020aa} inferred from the cosmic microwave background (CMB) within the standard $\Lambda$CDM model (also \citealt{Aiola:2020aa, Louis:2025aa}), relative to the larger values found from the local Universe (e.g., $H_0 = 73.0 \pm 1.0~{\rm km~ s}^{-1} ~{\rm Mpc}^{-1}$ from SH0ES; \citealt{Riess:2022aa}). 
The latter measurements are predominantly determined through calibration of a distance ladder on cosmological scales, making use of indicators such as Type Ia supernovae in combination with, e.g., 
$i)$ Cepheid variable stars (e.g., \citealt{Riess:2022aa}), $ii)$ the tip of the red giant branch (TRGB; e.g., \citealt{Freedman:2019aa, Anand:2024aa}), or $iii)$ the J-Region Asymptotic Giant Branch (JAGB; e.g., \citealt{Lee:2021aa, Freedman:2025aa}).
Given the manifold challenges in this calibration (\citealt{Mortsell:2022aa, Mortsell:2022ab}), 
there is a drastic need for independent and orthogonal measurements of $H_0$ that do not suffer from the same systematics. 

One such example is the Megamaser Cosmology Project (\citealt{Reid:2009aa, Reid:2013aa, Pesce:2020aa}), which enables a ``single-rung" geometric distance inference through high-precision mapping of circumnuclear H$_2$O masers originating from accretion disks around active galactic nuclei (AGN) in low-$z$ galaxies.
Another, of course, is TD cosmography, which is capable of yielding a single-rung measurement well into the Hubble flow 
without hinging on the possible pitfalls from the usual distance ladder methods\footnote{To be clear, this technique does not rely on the nature of SNe Ia as ``standardizable candles" like other late-Universe efforts (although this property may assist in more objectively assessing the accuracy of various lens models; \citealt{Pascale:2025aa}).
Supernovae are simply a useful transient source by which time delays may be measured, revealing the underlying cosmology.}.

TD cosmography has perhaps matured more quickly for lensed quasars, which are variable but offer the benefit of persistence that supernovae lack \citep{Wong:2020aa, Birrer:2020ab, Birrer:2024aa}. 
Over 300 lensed quasars have been identified, most with image separations of only a few arcseconds and with time delays typically less than one year, although some (e.g., \citealt{Fohlmeister:2007aa, Fohlmeister:2008aa, Dahle:2015aa,  Munoz:2022aa, Napier:2023aa}) have measured time delays using wide-separation cluster-lensed quasars (which have some advantages over galaxy-scale lenses and their systematic uncertainties; \citealt{Kochanek:2020ab}).
However, inferring time delays from quasars requires long-term monitoring (of order 10 years) for accurate measurements \citep{Courbin:2005aa, Bonvin:2017aa}, and can be plagued by microlensing \citep{Irwin:1989aa, Witt:1995ab, Schechter:2003aa, Vernardos:2024aa},
degeneracies and systematics in lens models (\citealt{Keeton:2000aa, Koopmans:2003aa}),
and the well-known mass-sheet and mass-slope degeneracies (\citealt{Falco:1985aa, Gorenstein:1988aa, Wambsganss:1994aa,  Kochanek:2002aa, Schneider:2013aa}). 
While some of these challenges also affect TD cosmography with supernovae, this approach has some distinct advantages. Their rapid and more predictable variation in brightness means that time delays can be measured with much shorter monitoring campaigns\textemdash even as little as a few hours of spectroscopy \citep{Chen:2024ab}\textemdash and with a level of precision and accuracy that is actually more affected by photometric uncertainties than by microlensing (e.g. \citealt{Pierel:2019aa, Huber:2022aa, Pierel:2024aa}).

Perhaps the most important systematic plaguing the time-delay cosmography approach for supernovae at present is scarcity.
Supernovae are inherently rare events\textemdash $\lesssim 1\%$ of all stars will end their lifetimes in core-collapse supernovas (CCSNe).
As their duration above a survey's sensitivity limits is short-lived, especially 
at $z > 1$, their discovery essentially 
requires
a not insignificant amount of luck.
Strong gravitational lensing in the regime of producing multiple images is similarly quite rare \citep{Treu:2010aa}.
While the lensing optical depth (probability of being lensed above a certain magnification threshold or lensed into more than one image) increases monotonically with source-plane redshift, this largely asymptotes above $z\gtrsim 2$ (e.g., \citealt{Hilbert:2007aa, Hilbert:2008aa, Robertson:2020ac}).
As a result, barely more than a handful of other multiply-imaged lensed SNe have been discovered 
in the last decade since SN Refsdal was first identified behind the MACS J1149.6+2223 galaxy cluster (\citealt{Kelly:2015aa, Kelly:2016aa, Grillo:2018aa}; $H_0$ measurements in \citealt{Kelly:2023ab,Grillo:2024aa}).
Since then, three other glSNe suitable for TD cosmography have been discovered: 
SN H0pe behind the PLCK G165.7+67.0 cluster \citep{Frye:2023ab, Frye:2024aa}, and 
the pair of
SN Requiem \citep{Rodney:2021ab} and SN Encore \citep{Pierel:2024ab}, both within the same massive host galaxy lensed by MACS J0138.0-2155 and separated by 2.5 years in the galaxy's rest frame.
The analysis of SN H0pe, the first confirmed glSN Ia used for TD cosmography, recently produced a new $H_0$ measurement at $\sim 10\%$ precision \citep{Pascale:2025aa}.
This work also introduced a new weighting scheme for the seven independently-constructed lens models according to their ability to accurately predict observables like magnifications, made possible by H0pe's nature as a standardizable candle. 
The first $H_0$ result from SN Encore is imminent (\citealt{Ertl:2025aa}; S. Suyu et al., in prep.; J. D. R. Pierel et al., in prep.), and now there is also hope for a precise measurement from SN Requiem's return a decade later in 2026-2027 (HST program GO-18069).

Upcoming wide-area surveys like the Vera C. Rubin Observatory's Legacy Survey of Space and Time (LSST; \citealt{LSST-Science-Collaboration:2009aa})
promise to reveal $\sim 50 - 100$ glSNe yr$^{-1}$
\citep{Arendse:2024ab, Bag:2024aa, Melo:2025aa, Sainz-de-Murieta:2024aa},
but not all of these will be discovered sufficiently early to follow up with light curve monitoring for the purposes of TD cosmography.
Also, many of the glSNe behind galaxy-scale deflectors will have time delays too short to yield meaningful precision on $H_0$,
as in the case of iPTF16geu ($z=0.409$; \citealt{Goobar:2017aa}).
Moreover, in the pessimistic scenario, currently active facilities like the Zwicky Transient Facility (ZTF; \citealt{Bellm:2014aa}) were expected to yield at least one glSN per year, 
but only one was identified in four years (SN Zwicky at $z=0.354$; \citealt{Goobar:2023aa, Pierel:2023aa}), with few other compelling candidates
(\citealt{Magee:2023aa, Sagues-Carracedo:2024aa}; see also \citealt{Craig:2024aa}).
Still, the next decade is likely to see TD cosmography with glSNe mature more fully and reach a competitive precision on $H_0$, closer to 1\% than 10\% \citep{Treu:2022ac}.

What, then, is the best strategy at present for discovering ``live" glSNe\footnote{Here, ``live" glSNe draws a distinction from those discovered after the supernova has faded in all images, only apparent through its disappearance in a second epoch of observation, and therefore not suitable for monitoring to infer $H_0$.}?
One answer might come from remarking that some of the recent discoveries have been identified behind well-studied lensing clusters.
These include two of the famous Hubble Frontier Fields: SN Refsdal behind MACS J1149.6+2223 \citep{Kelly:2015aa} and a $z\approx 3$ CCSN behind Abell 370 \citep{Chen:2022ag}.
There is undoubtedly an observational bias here, given the attention these fields receive, resulting in a more well-sampled cadence, and the presence of numerous lens models to identify multiple imaging \citep{Bronikowski:2025aa}.
Yet, these massive lensing clusters also provide sizable magnification for a large number of background sources (e.g., \citealt{Gal-Yam:2002aa, Stanishev:2009aa, Petrushevska:2016aa, Petrushevska:2018ab, Petrushevska:2018aa,  Petrushevska:2020aa}).
However, 
SN H0pe, SN Requiem, and SN Encore were all identified
within high-mass host galaxies lensed by less massive, 
previously lesser-known clusters. 
These discoveries were facilitated not necessarily by the extraordinary nature of the foreground, but more so of the background lensed host galaxies.

A targeted search for glSNe in known lensed galaxies that are very massive or have a high SFR (or both) is not an entirely novel idea.
\citet{Shu:2018ab} estimated that monitoring a sample of $>100$ galaxy-scale lensed star-forming galaxies would yield rates of $\sim 3$ combined CCSN and SN Ia events per year \citep{Shu:2021ab},
reaching a greater efficiency than a blind search.
A targeted strategy also provides greater confidence that any discovered SNe are in fact lensed (see also \citealt{Holwerda:2021aa}).
Here, we modify the approach to encompass
a much smaller sample, but with commensurately larger rates of star formation. 
There is some difficulty in identifying sufficient numbers of starburst galaxies (with high intrinsic rates of supernovae) that are lensed into multiple images. 
One possible approach is to monitor samples of known lensed dusty star-forming galaxies (DSFGs),
also referred to as submillimeter galaxies (SMGs)
which now number in the hundreds \citep{Giulietti:2024ab},
as these objects are inferred to form stars at prodigious rates of $300 - 1000 ~ M_\odot ~{\rm yr}^{-1}$ (or even greater).
The prevailing selection method for this population is unique, relying essentially on submillimeter flux alone 
\citep{Blain:1996aa, Perrotta:2002aa, Negrello:2010aa, Negrello:2017aa},
as opposed to methods that identify distorted lensing morphologies or peaks in foreground mass density or Sunyaev-Zel'dovich effect signal.

As an application of this tactic,
we examine in particular
the set of lensed DSFGs identified from {\it Planck},
which is perhaps most well suited for this experiment due to its all-sky coverage.
For now, we consider the $\sim 30$ members of the {\it Planck} All-Sky Survey to Analyze Gravitationally-lensed Extreme Starbursts
(PASSAGES; \citealt{Harrington:2016aa, Berman:2022aa, Kamieneski:2024aa}), some of which were also identified independently through {\it Herschel} follow-up \citep{Canameras:2015aa, Planck-Collaboration:2015aa}.
Of these, 21 objects (summarized in Table \ref{tab:computations}) have the information necessary to compute expected glSN rates (\S \ref{sec:f_unobs} and \ref{sec:RCC}). 
\citet{Kamieneski:2024aa} determined that the median de-lensed (i.e. intrinsic) infrared luminosity for PASSAGES exceeded $10^{13}~L_\odot$, qualifying them as hyper-luminous infrared galaxies (HyLIRGs).
Nonetheless, the dust geometry of such extreme HyLIRGs remains quite uncertain at high-$z$. Any supernovae are likely to be attenuated by dust in even the best-case scenario,
so near-IR sensitivity is paramount.

HyLIRGs such as these are often signposts of significant over-dense regions (e.g., \citealt{Foo:2025aa}). 
While we consider 
primarily the contribution to the supernova rate from the primary starburst galaxies, there is a high likelihood of other star-forming objects at the same redshift that might also fall into the strongly lensed portion of the source plane, further boosting the chances of identifying lensed transients in these fields.
This is also borne out in part with the discovery of SN H0pe behind the lensing cluster G165.7+67.0, which was itself identified through its amplification of one of the PASSAGES DSFGs.
Initial JWST/NIRCam imaging of the field through the Prime Extragalactic Areas for Reionization and Lensing Science program \citep{Windhorst:2023aa} revealed the Type Ia supernova not within the giant arc of lensed DSFGs, but in a highly-magnified arc nearby in projection. Both this host galaxy and the DSFGs were found to be members of respective photometric-redshift overdensities at $z=1.78$ and $z=2.24$ \citep{Frye:2024aa}.

This work is organized as follows: in Section \ref{sec:methods}, we detail our novel approach to deriving forecasts for the intrinsic and observer-frame rates of supernovae in HyLIRGs, using the PASSAGES fields as an example. 
In Section \ref{sec:results}, we discuss our inference of star formation rates (both obscured and unobscured modes), lensing multiplicities, and the core-collapse and degenerate white dwarf supernova rates that result from these inferences.
We also examine the expected typical time delays and draw a connection to the angular separation between images.
In Section \ref{sec:summary}, we summarize our findings and reflect on the feasibility of our proposed approach in making a meaningful contribution to the rather saturated industry of deriving $H_0$ constraints.
Throughout, we adopt a fiducial $\Lambda$CDM concordance cosmology of $\Omega_m = 0.3$, $\Omega_\Lambda = 0.7$, and $H_0 = 70~{\rm km~s}^{-1}~{\rm Mpc}^{-1}$.

\section{Observations \& Methods} 
\label{sec:methods}

\subsection{Case study sample selection: PASSAGES}

In this work, we aim to forecast the expected rates of supernovae from
strongly-lensed DSFGs at $z>1$.
Lensed DSFGs are efficiently selected through their submillimeter flux owing to a steep drop-off in number counts \citep{Blain:1996aa}, such that an extragalactic source brighter than 100 mJy at 500 $\mu$m is $\gtrsim 30$ times more likely to be lensed than unlensed
\citep{Negrello:2017aa}.
However, they are still rare objects, with a modeled all-sky number density of $0.09 \pm 0.05~{\rm deg}^{-2}$ for $S_{500\mu{\rm m}} > 80$ mJy, while observations suggest up to $\sim 0.3~{\rm deg}^{-2}$
(\citealt{Sedgwick:2025aa}, and references therein including \citealt{Negrello:2010aa, Bussmann:2013aa, Wardlow:2013aa, Negrello:2014aa, Negrello:2017aa, Bakx:2018aa, Urquhart:2022ab, Borsato:2024aa}).
This model predicts $3600 \pm 1800$ such sources to exist in the entire sky (increasing by a factor of $\sim 40$ if the threshold is lowered by a factor of 10; \citealt{Bakx:2024aa,Sedgwick:2025aa}), but so far 
only 
a few hundred such objects have been uncovered (e.g. \citealt{Giulietti:2024ab}).
{\it Euclid}'s Wide Survey \citep{Euclid-Collaboration:2022ab} is expected to be sensitive to several thousand.

For now, while {\it Planck}'s surveys reach shallower depths than {\it Herschel}, its 
wide-area coverage has been an asset for identifying the very brightest lensed DSFGs in the sky \citep{Trombetti:2021ab}.
As discussed by \citet{Kamieneski:2024aa}, there is a complex balance between their intrinsic luminosities and their magnifications: HyLIRGs tend to have larger sizes to match their higher luminosities (in line with the Eddington limit; \citealt{Kamieneski:2024ab}), while size bias leads to more strongly-magnified objects correlating with more compact sizes \citep{Hezaveh:2012aa}. Together, this means that candidates with apparent luminosities $\mu_{\rm IR}{\rm L}_{\rm IR} \gtrsim 10^{14}~ L_\odot$ tend to still have intrinsic ${\rm L}_{\rm IR} \gtrsim 10^{12.5}~ L_\odot$.
For this reason, the greater submillimeter fluxes of {\it Planck}-identified lensed DSFGs are not simply the result of higher magnifications alone.
They are thus among the best candidates for high rates of glSNe given their high SFRs.
Moreover, their lower median redshift than 
lensed DSFGs identified in the millimeter regime (e.g., with the South Pole Telescope; \citealt{Weiss:2013aa}) lends well to greater ease in detecting glSNe in the observed-frame near-IR.

\subsection{Determining the expected value of image multiplicity}
\label{sec:multiplicity}

A galaxy that crosses the caustic network will have different parts of the source plane lensed into varying numbers of multiple images.
In the canonical lensing example of a singular isothermal ellipsoid, an astroid tangential caustic encloses a region with an image multiplicity of 4 (excluding demagnified images), whereas the region between the tangential caustic and the elliptical radial caustic is lensed into 2 images, and structures exterior to the radial caustic are weakly lensed into only a single image. 
For cluster-scale lenses, which have larger source-plane areas inside the caustics, it is more likely for most of a galaxy's light to be contained within a region with constant image multiplicity.
On the other hand, galaxy-scale lenses have much tighter caustic loci, so galaxies can easily cross caustics and have two (or even more) different image multiplicities of the source.

\begin{figure}[th]
\centering
\includegraphics[width=\columnwidth]{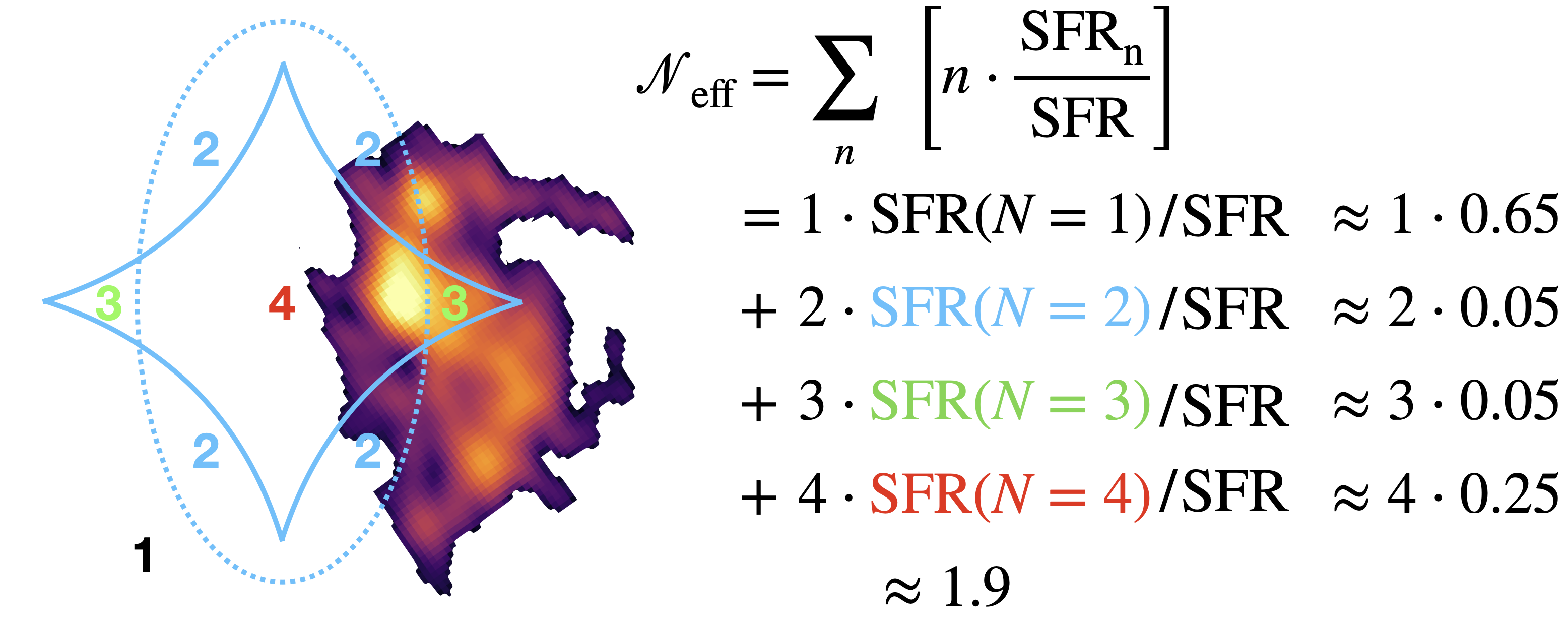}
\caption{
Cartoon schematic demonstrating the calculation of effective multiplicity \Neff\ per Equation \ref{eqn:Neff}, using an ALMA 870 $\mu$m image of the unlensed DSFG ALESS 112.1 as an example \citep{Hodge:2019aa}.
Mock caustic curves from an imaginary lensing galaxy are overlaid (radial=dashed, tangential=solid).
Put simply, $\mathcal{N}_{\rm eff}$ is the weighted average of multiplicities ($n=1,2,3,...$), where the weights are the respective portions of the total SFR.
}
\label{fig:Neff}
\end{figure}

The lensing multiplicity of a galaxy has a direct bearing on the average rate of SNe that can be observed on Earth. 
To account for this, we introduce the concept of {\it effective multiplicity}, \Neff, which
is the average multiplicity weighted by the distribution of star formation.
We define it as 
\begin{equation}
\mathcal{N}_{\rm eff} \equiv \sum_n \bigg[ n \cdot \frac{{\rm SFR}_{n}}{{\rm SFR}} \bigg]
\label{eqn:Neff}
\end{equation}
for the different multiplicities $n$, where SFR$_n$ is the total star formation rate within the region of multiplicity $n$.
This calculation is illustrated for a singular isothermal ellipsoid lens (with a large axis ratio) in a cartoon in Fig. \ref{fig:Neff}.
\Neff\ can also be functionally expressed as a pixel-based summation:
\begin{equation}
\mathcal{N}_{\rm eff} \equiv \frac{\sum_{i,j} N_{i,j} {\rm SFR}_{i,j}}{\sum_{i,j} {\rm SFR}_{i,j}}
\label{eqn:Neff2}
\end{equation}
where SFR$_{i,j}$ and $N_{i,j}$ are the star formation rate and multiplicity, respectively, at pixel $(i,j)$.

\subsection{ALMA 1mm continuum and VLA 6 GHz}

To infer a map of SFR to compute \Neff\ with the information available for the majority of PASSAGES, we rely on the 1mm dust continuum observed with the Atacama Large Millimeter/submillimeter Array (ALMA), by which we apportion the total IR SED-derived SFR (primarily from \citealt{Harrington:2016aa, Berman:2022aa}).
For northern-declination sources not covered in these observations, we instead use 6 GHz (5 cm) continuum imaging from the {\it Karl G. Jansky} Very Large Array in A-configuration, with angular resolutions relatively close to the ALMA images. The 6 GHz images are largely correlated with the dust continuum through the radio-FIR correlation (e.g., \citealt{Condon:1992aa}).
These observations are described in detail in 
Sections 2.4 and 2.5
of
\citet{Kamieneski:2024aa}, with images shown in their Figure 2.

We use these observations to map the source-plane spatial distribution of ${\rm SFR}_{\rm IR}$ (as inferred from unresolved far-IR photometry, the source of which is specified for each target in Table~\ref{tab:computations}).
Since the star formation is likely dominated by the obscured mode probed by SFR$_{\rm IR}$, the distribution $\Sigma_{\rm SFR_{\rm IR}}$ is our best estimate of the 2d probability density function of core-collapse supernovae (e.g., \citealt{Miluzio:2013aa}). While slightly less direct, this is also likely to closely match the spatial distribution of Type Ia supernovae, as well (especially if one assumes that $\Sigma_{\rm SFR}$ correlates with the distribution of stellar mass, $\Sigma_\star$; e.g., \citealt{Baker:2022aa}).
Of course, the {\it observed} distribution of supernovae is more complex, and depends on dust geometry, but for the purpose of computing \Neff, $\Sigma_{\rm SFR_{\rm IR}}$ should suffice.
Nonetheless, given the highly stochastic nature of supernovae, even for galaxies with very high SFRs, and imperfections in the lens models, we adopt a conservative fractional uncertainty\footnote{This uncertainty can be very roughly approximated as the fraction of a galaxy's star-forming area that is within the 1$\sigma$ uncertainty on the caustic's location, multiplied by the relative change in multiplicity on either side of the caustic.
In a canonical case of a galaxy straddling a caustic curve separating regions of double- vs. quad-imaging, 
this would be $\Delta N = 2$ from the $N=2$ to $N=4$ region, with average $N\sim 3$.
For a source-plane positional uncertainty of $\sim 0.1\arcsec$ and a typical source-plane size of $\sim 0.4\arcsec$ for PASSAGES \citep{Kamieneski:2024aa}, the typical uncertainty thus becomes $\sigma_{\mathcal{N}_{\rm eff}} / \mathcal{N}_{\rm eff} \sim (0.1/0.4) \cdot (4-2) / 3 \sim 20\%$.
} of 20\% on \Neff.
The achieved resolutions for the ALMA and VLA observations ranged from 
$\theta \approx 0.4 - 0.8\arcsec$
and
$\theta \approx 0.3 - 0.7\arcsec$, respectively.
While higher-resolution observations (e.g., spatially-resolved SED fitting with JWST) would improve the fidelity of \Neff, the location of the caustic network from lens models and systematics of SED-fitting dominate the uncertainty at present.

There is some concern that heavily dust-obscured active galactic nuclei (AGN) might contribute in the radio \citep{Donley:2005aa, Del-Moro:2013aa, Delvecchio:2017aa}, thus biasing the calculation of \Neff.
However, \citet{Kamieneski:2024aa} found that the PASSAGES DSFGs did not have obvious evidence of radio-loud AGN manifesting as an excess in spatially-integrated radio continuum relative to FIR, suggesting that this bias is not likely to have a sizable effect.

\subsection{Estimating the obscured \& unobscured SFR}
\label{sec:f_unobs}

\subsubsection{The situation at low-$z$ vs. $z>2$}

CCSNe have been discovered successfully in low-$z$ luminous infrared galaxies (LIRGs) in the near-IR within projected radii $< 1$ kpc (\citealt{Mannucci:2003aa, Mattila:2007aa, Kankare:2008aa, Kankare:2012aa, Miluzio:2013aa,  Kool:2018aa, Jencson:2019aa, Fox:2021ab, Kankare:2021aa, Perez-Torres:2021aa}). These include some radio supernovae and supernova remnants found through very long baseline interferometry (e.g., \citealt{Marti-Vidal:2007aa, Perez-Torres:2009aa, Bondi:2012aa, Varenius:2019aa}).
On the other hand, some campaigns have yielded zero, or at least
drastically fewer than expected, detected events
for starbursts and ultra-luminous infrared galaxies (ULIRGs), including \citet{Cresci:2007aa} with {\it HST}/NICMOS, or 
\citet{Vaisanen:2008aa} with adaptive optics-assisted VLT imaging.
\citet{Mantynen:2025aa} recently found that $\approx$90\% of CCSNe in local LIRGs might be missed if attenuations only up to $A_V < 3$ mag are detectable.
Even if hypothetically sensitive to extremely high values of $A_V \gtrsim 40$, the authors obtain an undetectable fraction $\gtrsim$50\% owing to the concentrated distribution of dust and star formation in local LIRGs.
\citet{Fox:2021ab} emphasized the importance of a stable, symmetric, high-resolution PSF in the near-IR (as achieved with space-based facilities) in order to combat this high concentration.

This might paint a dire picture for executing successful campaigns at high-$z$, although clever workarounds are possible. For example, \citet{Yan:2018aa} demonstrated that the presence of SNe may be revealed through variability in the integrated near-IR light of a galaxy (at the time with {\it Spitzer}). 
However, there also appear to be key distinctions between the typical structures of high-$z$ HyLIRGs and the LIRGs/ULIRGs in the low-$z$ Universe.
The typical star-dust geometry of DSFGs and starburst galaxies remains a matter of ongoing research, 
which of course has implications for the detectability of supernovae.
In particular, the presence of unobscured UV-bright sightlines in DSFGs may be responsible for bluer UV slopes $\beta_{\rm UV}$, resulting in the observed deviation from the IRX-$\beta$ relation \citep{Meurer:1999aa} where IRX$\equiv L_{\rm IR} / L_{\rm UV}$ is the IR excess
(e.g., \citealt{Cortese:2006ab, Boquien:2009aa, Casey:2014ab, Faisst:2017aa, Safarzadeh:2017ab, Narayanan:2018ab, Liang:2021aa}).
In essence, the exceptional rate of 
supernovae occurring per year in each of the PASSAGES galaxies (Table~\ref{tab:computations}) is likely to result in a
very porous ISM, 
especially considering the groundwork laid by pre-supernova feedback \citep{Lucas:2020aa, Chevance:2022ab, Watkins:2023aa}.
Supernovae are likely to occur within an inhomogeneous medium, and further contribute to this through compression of swept-up shells and rarefied bubbles 
(e.g., \citealt{Clarke:2002aa, Kim:2015aa}.
This ISM structure, in turn,
favors the ability to observe supernovae, even in dusty environments.

Local ULIRGs often host extremely concentrated starburst regions (although as \citealt{McQuinn:2012aa} point out, this may be partially a selection bias).
This high concentration reaches an extreme with compact obscured nuclei, or CONs (which have sizes of $10-100$ pc and very high attenuations of $A_V \gtrsim 10$; \citealt{Sakamoto:2013aa, Aalto:2015aa,  Falstad:2021aa}).
At Cosmic Noon ($z\sim 1 - 3$), their more luminous counterparts sustain high rates of star formation over the scale of several kpc \citep{Rujopakarn:2011aa, Hodge:2016aa, Hodge:2019aa, Harrington:2021aa, Kamieneski:2024aa, Kamieneski:2024ac, Mitsuhashi:2024aa, Polletta:2024ab}, often in clumps resulting from gravitational instabilities in gas-rich disks \citep{Immeli:2004aa, Tadaki:2018aa, Faisst:2025aa, Fujimoto:2024aa, Liu:2024ae, Kalita:2025ac} that are continually replenished through accretion \citep{Bournaud:2007aa, Dekel:2009aa}.

Also, the high rates of star formation likely lend well to an increased rate of star cluster formation, and thus a better sampling of the high-mass end of the cluster mass function.
For example, super star clusters (SSCs) are preferentially found in intense starburst regions, as exemplified by M82 and the Antennae Galaxies (e.g., \citealt{Smith:2001aa, Mengel:2002aa, de-Grijs:2003aa, Whitmore:2010aa, Linden:2017aa, Emig:2020aa, Levy:2024aa}).
If much of the star formation of DSFGs is concentrated in SSCs, which are  
progenitors of massive globular clusters, 
then their strong stellar feedback may contribute to clearing sightlines that both allows for Lyman continuum (LyC) leakage (e.g., \citealt{Bik:2018aa, Pascale:2023ab, Ji:2025aa})
and the
escape of less-attenuated optical light from SNe (e.g., \citealt{Leitet:2013aa, Roy:2024ae}).

\subsubsection{Adopting a fraction of unobscured SFR}

H$\alpha$ SFRs are typically significantly less than those derived from the infrared.
This discrepancy is primarily resolved after applying a Balmer decrement extinction correction to the H$\alpha$-derived value \citep{Kewley:2002ab, Takata:2006aa, Dominguez-Sanchez:2012aa, Timmons:2015aa, Puglisi:2016aa, Liu:2024ae}, although not in all cases (\citealt{Chen:2020aa}).
For the \citet{Calzetti:2000aa} attenuation law, SFR$_{\rm H\alpha}$ is corrected as 
\begin{equation}
{\rm SFR}_{\rm H\alpha, corr} = {\rm SFR}_{\rm H\alpha} \cdot 10^{0.4 \cdot E(B-V)_{\rm neb} \cdot k'({\rm H}\alpha)}
\end{equation}
where 
$k'(0.656 \mu {\rm m}) \approx 3.325$
and $E(B-V)_{\rm neb}$ is the color excess inferred from emission lines,
\begin{equation}
    E(B-V)_{\rm neb} 
    \approx 1.97 \cdot {\rm log}_{10}\bigg[\frac{L_{\rm H\alpha} / L_{\rm H\beta}}{2.86}\bigg]
\end{equation}
for case B recombination
\citep{Osterbrock:2006aa}.
Inherently, the strong attenuation of DSFGs has made it difficult to recover multiple Balmer lines.
In a composite spectrum of 31 DSFGs in the COSMOS field, \citet{Casey:2017aa} found a median Balmer decrement of $21.0 \pm 2.4$, yielding $E(B-V)= 1.7 \pm 0.1$ or $A_V = 6.9 \pm 1.4$ (using $R_V = 4.05 \pm 0.80$). This is higher than observed even for local ULIRGs on global scales (e.g., \citealt{Wild:2011aa}).
More recently, in a sample of typical DSFGs at $z\sim 2-3$, \citet{Cooper:2025aa} found consistency with the shape of the \citet{Calzetti:2000aa} attenuation curve, and $A_V \sim 3 - 4$, through combination of Balmer and Paschen lines obtained with JWST/NIRSpec.
As some star formation is likely to be hosted in regions with dust obscuration strong enough such that no H$\alpha$ emission would be detectable, these Balmer decrements are biased low \citep{Puglisi:2017aa}.
The extinction-corrected ${\rm SFR}_{\rm H \alpha}$ estimates are likely to give a more realistic assessment of the observable rates of supernovae, however. Therefore, in our analysis, we introduce an unobscured fraction $f_{\rm unobsc} \equiv {\rm SFR}_{\rm unobsc} / {\rm SFR}_{\rm IR} (\sim {\rm SFR}_{\rm H\alpha,corr} / {\rm SFR}_{\rm IR})$.
For example, \citet{Olivares:2016aa} found this fraction to be $<30\%$ in $z\sim 2$ SMGs (an upper limit since their H$\alpha$ SFRs are not corrected for extinction).
In other words, the discrepancy between the extinction-corrected SFR$_{\rm H\alpha, corr}$ and SFR$_{\rm IR}$ is a quantification of how much star formation is hosted in regions of extreme attenuation (from which neither H$\beta$ nor H$\alpha$ may be detected).

For the PASSAGES sample itself, two members already have measurements of the Balmer decrement.
For the pair of DSFGs PJ1127+42 at $z=2.24$ (comprising Arcs 1abc, 3abc, 4abc, and 6abc lensed by the G165.7+67.0 cluster), \citet{Frye:2024aa} found a value of $E(B-V) = 0.85 \pm 0.25$ through NIRSpec observations (corresponding to a Balmer decrement of $7.7 \pm 2.3$ and $A_V = 3.4 \pm 1.2$).
This yields a correction of log$_{10}$(SFR$_{\rm H\alpha} / {\rm SFR}_{\rm H\alpha,corr}) = -1.1 \pm 0.3$,
such that the demagnified
${\rm SFR}_{\rm H\alpha,corr} \approx 100^{+200}_{-70} ~ M_\odot ~{\rm yr}^{-1}$ for Arc 1 only.
Given that ${\rm SFR}_{\rm IR} = 320 \pm 70 ~ M_\odot ~{\rm yr}^{-1}$ for the same object \citep{Kamieneski:2024ac}, this gives
$f_{\rm unobsc} \approx 31\%$.
We adopt this fraction for the entire arc complex (Arcs 1, 3, 4, and 6, per notation from \citealt{Frye:2019aa}), which all lies at $z\approx 2.24$ \citep{Canameras:2018aa, Kamieneski:2024ac}.
Similarly, for PJ0116-24 ($z=2.13$), \citet{Liu:2024ae} derived a color excess through Balmer decrement of $E(B-V) = 0.95 \pm 0.13$ with VLT/ERIS observations (Balmer decrement $8.7 \pm  1.3$; $A_V = 3.9 \pm 0.9$),
suggesting log$_{10}$(SFR$_{\rm H\alpha} / {\rm SFR}_{\rm H\alpha,corr}) = -1.3 \pm 0.2$ and  
${\rm SFR}_{\rm H\alpha,corr} = 470 \pm 60~M_\odot ~{\rm yr}^{-1}$
and 
$f_{\rm unobsc} \approx 32\%$.
In this work, for objects without these Balmer decrement measurements at present, we adopt a range of $f_{\rm unobsc} = 20\pm 10\%$ in Table~\ref{tab:computations}
(see also, e.g., \citealt{Swinbank:2004aa,Chen:2020aa,Cheng:2020ac}).

This might at first seem to run counter to the results of \citet{Whitaker:2017aa} that $f_{\rm obsc} \equiv {\rm SFR}_{\rm IR} / {\rm SFR}_{\rm UV+IR}$ increases continually for higher SFRs and stellar masses (exceeding 90\% for massive galaxies ${\rm log}(M/M_\odot)>10.5$)\footnote{Note the different definition from \citet{Whitaker:2017aa} for $f_{\rm obsc}$, which also differs in their computation of SFR$_{\rm IR}$ using a calibration from UV and 24$\mu$m photometry \citep{Whitaker:2014ab}.}.
Higher dust obscuration may be due to elevated surface densities of both star formation and gas (which are themselves correlated; \citealt{Schmidt:1959aa, Kennicutt:1998ac}).
However, $\Sigma_{\rm SFR}$ does not seem to increase strongly with $L_{\rm FIR}$. 
The size-luminosity relation found by \citet{Fujimoto:2017aa} for $11 \lesssim {\rm log}(L_{\rm FIR} / L_\odot) \lesssim 13$, $R_e \propto L_{\rm FIR}^{0.28}$, implies a weak dependence of $\Sigma_{\rm SFR} \propto \Sigma_{\rm FIR} \propto L_{\rm FIR}^{0.44}$.
Empirically, HyLIRGs appear to remain at or below $\Sigma_{\rm SFR} \approx 100~M_\odot~{\rm yr}^{-1}~{\rm kpc}^{-2}$, with a similar distribution as ULIRGs with luminosities a decade lower
(e.g., \citealt{Hodge:2019aa, Bakx:2024ad, Kamieneski:2024aa}), suggesting that $\Sigma_{\rm SFR}$ may even saturate.
If this property plays an important role in the \citet{Whitaker:2017aa} relation, then the obscured fraction may saturate, as well. 

Drawing again a parallel to the escape of Lyman continuum, this fraction of 20\% has some precedent. For star-forming galaxies (SFGs) in UVCANDELS ($2.5 < z < 3.0$), \citet{Wang:2025an} reported an upper limit in the absolute Lyman continuum escape fraction of $f_{\rm esc} \lesssim 26\%$. 
\citet{Smith:2018ac} found that the average absolute $f_{\rm esc}$ for massive SFGs and weak AGN at $z\sim 2.4$ was $\sim 22^{+44}_{-22}\%$ (see also \citealt{Smith:2020ab} and \citealt{Smith:2024aa}, which found $f_{\rm esc} \sim 18\%$ for weak AGN in UVCANDELS).
If stellar and AGN feedback in galaxies at similar redshifts with comparable (or even much lower) SFRs can clear the way for this level of escape of Lyman continuum\textemdash which is much more affected by dust than the light from SNe\textemdash then it is likely not unreasonable to expect an unobscured fraction of $\sim 20\%$.

While clearly only a fraction of the star formation is visible in the unobscured mode for these objects, we emphasize that their total SFRs are large enough that the absolute unobscured star formation can still be quite substantial.
Moreover, these large color excesses $E(B-V)$ primarily quantify the dust attenuation within the more obscured star-forming regions (e.g., \citealt{Wild:2011aa, Price:2014aa, Chen:2020aa}), 
which \citet{Calzetti:1997ab} found differed as $E(B-V)_{\rm stellar} = 0.44~E(B-V)_{\rm nebular}$. 
Deviation from this towards greater color excess for the nebular component appears to correlate with specific SFR (sSFR $\equiv {\rm SFR}/M_\star$), as a greater share of star formation occurs in the obscured mode \citep{Reddy:2015aa}, but it also is likely to vary spatially within individual galaxies (e.g., \citealt{Robertson:2024aa}).
It is probably not reasonable to expect that DSFGs will have uniformly large dust attenuation of $A_V \sim 4 - 7$, but this remains to be tested more rigorously in larger samples.
It is also worth noting that multi-band JWST (or even HST) imaging enables robust spatially-resolved SED fitting for a wide range of redshifts (e.g., \citealt{Abdurrouf:2023aa, Gimenez-Arteaga:2023aa, Kamieneski:2024ac, Liu:2024af, Polletta:2024ab, Akins:2025ab, Harvey:2025ac}),
such that the distribution of star formation and dust attenuation can be characterized at sub-kpc scales.
This exercise is crucial to gain an understanding of where supernovae might be attenuated beyond detection limits, and even to better model the light curve of any SNe that are detected.
This can also help account for any contribution to $L_{\rm IR}$ from dust-obscured AGN in PASSAGES galaxies;
\citet{Harrington:2016aa} and \citet{Berman:2022aa} use mid- and far-IR photometry to show that they are star-formation-dominated, but not incontrovertibly.

\section{Results}
\label{sec:results}

\subsection{Computing expected intrinsic and observer-frame rates of CCSNe}
\label{sec:RCC}

As the progenitors of core-collapse supernovae are massive stars, the core-collapse rate is linearly related to the rate of star formation,
\begin{equation}
\label{eqn:RCC}
R_{\rm CC} = k_{\rm CC} \cdot {\rm SFR}
\end{equation}
(e.g., \citealt{Madau:1998aa}).
The factor $k_{\rm CC}$ is determined from the initial mass function (IMF), $\phi(M)$, integrated over the mass range of progenitors, typically 8 $M_\odot$ \citep{Smartt:2009aa} to 50 $M_\odot$ \citep{Tsujimoto:1997aa}:
\begin{equation}
k_{\rm CC} = \frac{\int_{8~M_\odot}^{50~M_\odot} \phi(M) {\rm d}M}
{\int_{0.1~M_\odot}^{125~M_\odot} M \cdot \phi(M) {\rm d}M},
\end{equation}
where the denominator is limited to the total mass range of 0.1 to 125 $M_\odot$.
As the SFR estimates used in this work are based on a \citet{Kroupa:2001aa} IMF, we adopt a value of $k_{\rm CC}=0.0104~M_\odot^{-1}$.
This is consistent with the optically-thick regime ($\epsilon \approx 1$) of equation 5 of \citet{Mattila:2001aa}, $R_{\rm CC}/{\rm yr}^{-1} = 1.1\times 10^{-12} \cdot \epsilon^{-1} \cdot L_{\rm FIR} / L_\odot$,
where $\epsilon$ is the fraction of optical/UV energy absorbed and re-radiated in the FIR.
In comparison, for a \citet{Salpeter:1955aa} IMF\footnote{ 
For completeness, a \citet{Chabrier:2003aa} IMF integrated over the same progenitor mass ranges gives $k_{\rm CC} = 0.0110~M_\odot^{-1}$, closer to that for the Kroupa IMF. However, since converting from a Salpeter to Chabrier IMF also lowers the SFR by a factor of 0.63 \citep{Madau:2014aa}, the resulting $R_{\rm CC} = 0.0069 \cdot {\rm SFR}_{\rm Salpeter}$, thus certainly consistent within uncertainties.}, this yields $k_{\rm CC} = 0.0070~M_\odot^{-1}$
(e.g., \citealt{Young:2008aa, Melinder:2012aa}).

\begin{figure}[th]
\centering
\includegraphics[width=\columnwidth]{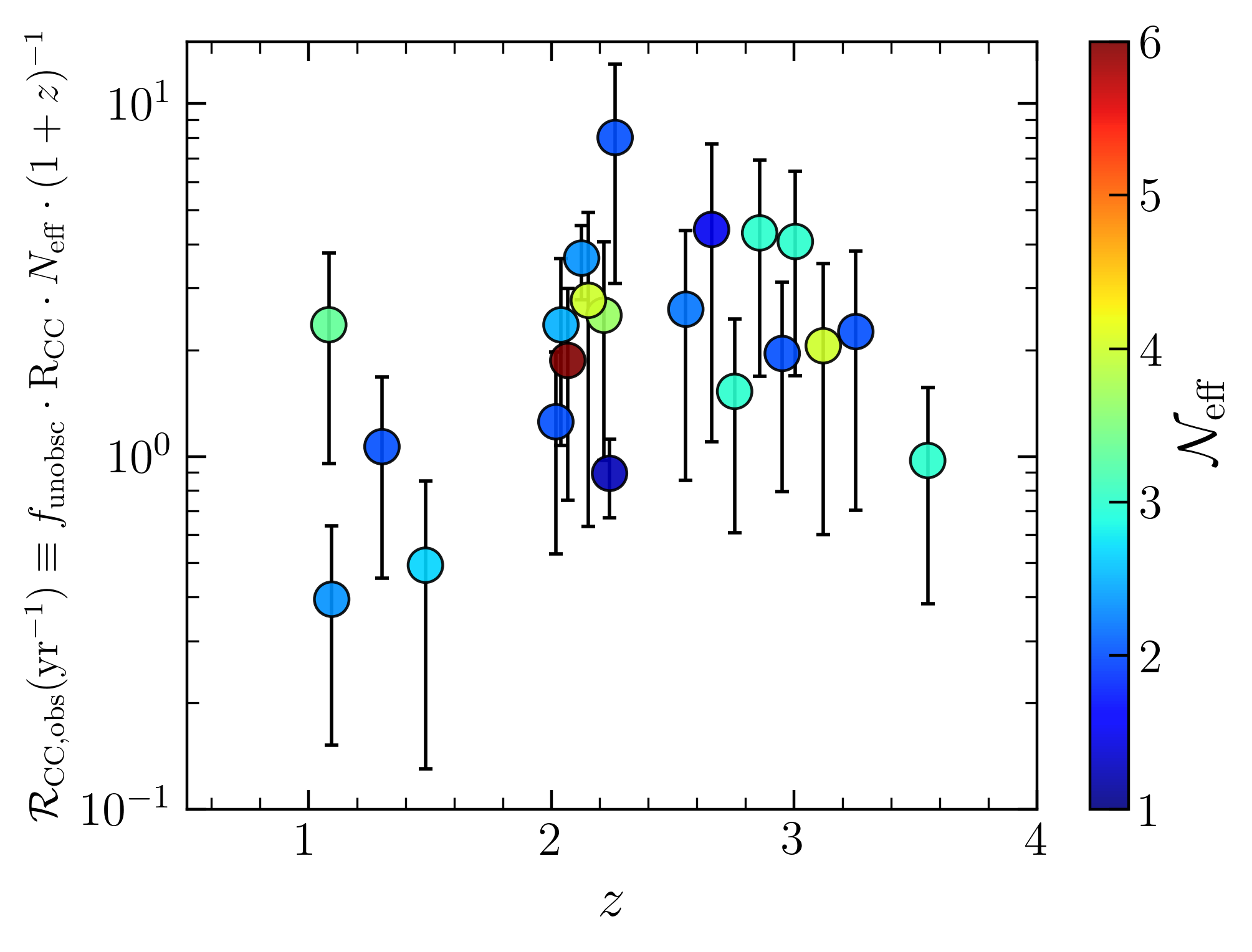}
\caption{
Observable rate of core-collapse supernovae $\mathcal{R}_{\rm CC,obs}$ vs. redshift for the 21 PASSAGES galaxies considered, using $\mathcal{R}_{\rm CC,obs}$ as defined in Equation~\ref{eqn:RCC_obs}.
The unobscured fractions adopted are provided in Table~\ref{tab:computations}.
Colors show the effective multiplicity, \Neff, described in Equations \ref{eqn:Neff} and \ref{eqn:Neff2}.
The increase with redshift is predominantly due to a general increase in SFR with $z$, which is also responsible for the trend observed in Fig.~\ref{fig:RIa_z} (see \S \ref{sec:independent}).
}
\label{fig:Rcc}
\end{figure} 

In Fig.~\ref{fig:Rcc}, we show the observable rate of CCSN images\footnote{To minimize confusion, we use $\mathcal{R}$ when referring to what we define as the observable rate, and simply $R$ when referring to an intrinsic, rest-frame rate.}
\begin{equation}
\label{eqn:RCC_obs}
\mathcal{R}_{\rm CC,obs} \equiv f_{\rm unobsc} \cdot R_{\rm CC} \cdot \mathcal{N}_{\rm eff} \cdot (1+z)^{-1}
\end{equation}
against redshift, with intrinsic rate $R_{\rm CC}$ determined from the intrinsic (i.e., magnification-corrected) SFR$_{\rm IR}$.
The apparent rates, $\mu$SFR, and magnifications $\mu$ are retrieved from the references given in Table~\ref{tab:computations}.
Given the dusty nature of these sources, we include $f_{\rm unobsc}$ in the equation {\it ab initio}, for a 
more realistic quantification of expected rates.
This does not quantify the fraction of supernovae that could be detectable in any observational campaign, but is more akin to a {\it maximum} recoverable fraction for what is reasonable with current telescope facilities.
For comparison, \citet{Mantynen:2025aa} recently found that the undetectable fraction of CCSNe in low-$z$ LIRGs for a survey sensitive to $A_V < 16$ mag is $66^{+8.6}_{-14.6}\%$. 
Using our formulation, this would suggest $f_{\rm unobsc} = 34^{+14.6}_{-8.6}\%$.
For a much shallower survey down to $A_V < 3$ mag, they found instead an undetectable fraction of $89.7^{+2.6}_{-4.4}\%$.

The large uncertainties in $\mathcal{R}_{\rm CC,obs}$ are the direct result of a combination of $i)$ the broad range chosen for $f_{\rm unobsc}$; $ii)$ inherently large uncertainty in ${\rm SFR}_{\rm IR}$; $iii)$ large uncertainty in magnification $\mu$ for the current lens models; and $iv)$ the conservative fractional uncertainties used for the effective multiplicity \Neff.
However, despite this uncertainty, the median $\mathcal{R}_{\rm CC,obs}$ is very high at {$\sim 2~{\rm yr}^{-1}$}, especially considering the average volumetric rates of $\sim 10^{-4}-10^{-3} ~ {\rm yr}^{-1}~{\rm Mpc}^{-3}$ at Cosmic Noon \citep{Dahlen:2012aa, Strolger:2015aa}.
As discussed in \S \ref{sec:f_unobs}, the IR-inferred SFRs are likely the closest to the true SFRs for these dust-obscured galaxies, but they are not necessarily representative of the rates of SNe that one could expect to observe. 
We adopt a typical value of $f_{\rm unobsc} = 0.2 \pm 0.1$,
following the findings of Balmer decrements for two members of the sample and for other DSFGs in the literature (\S \ref{sec:f_unobs})\footnote{As noted in Table \ref{tab:computations}, for two exceptions we use the direct measurements of $f_{\rm unobsc}$ that are available for PJ0116-24 \citep{Liu:2024ae} and PJ1127+42 \citep{Frye:2024aa}.}.

In \S \ref{sec:independent}, we compare the predicted rates $R_{\rm CC}$ (and the rate of SNe Ia) for PASSAGES with those from other works that also examine candidate samples for SN monitoring, alongside the expected rates for host galaxies of some of the already-discovered glSNe.
In Fig.~\ref{fig:Rcc}, we observe an increase in $\mathcal{R}_{\rm CC,obs}$ with redshift, which is likely to be driven primarily by the generally higher SFRs in the sample above $z\gtrsim 2$
(whereas \Neff\ is essentially constant with redshift).
In \S \ref{sec:independent}, we discuss how this compares with the redshift evolution of the cosmic star formation rate density.

\subsection{Estimating the rate of Type Ia SNe}
\label{sec:SNeIa}

As the progenitors of Type Ia supernovae are best understood to be binary systems that include a white dwarf star, the rate of SNe Ia inherently depends in part on the number of evolved lower-mass stars \citep{Ruiter:2025aa}.
In turn, since these make up the bulk of the stellar mass of a typical galaxy, it is reasonable to expect that the rate of Type Ia supernovae $R_{\rm Ia}$ would increase with $M_\star$ (either linearly or with some positive power law).
The donor star from which the progenitor white dwarf accretes may itself be another white dwarf 
or a main sequence or giant star, 
which implies that $R_{\rm Ia}$ should also have some dependence on the SFR, as with $R_{\rm CC}$.

\citet{Mannucci:2005aa} and \citet{Scannapieco:2005aa} found that the observed rate of SNe Ia could be described as a two-component linear combination of SFR and $M_\star$ (see also \citealt{Sullivan:2006aa}).
\citet{Smith:2012aa} included a power law for both terms in their bivariate parameterization, favoring 
\begin{eqnarray}
    \label{eqn:RIa}
    R_{\rm Ia} = & &(1.05 \pm 0.16) \times 10^{-10} ~ M_\star^{\ 0.68 \pm 0.01} \nonumber \\
    &+&
    (1.01 \pm 0.09) \times 10^{-3} ~ {\rm SFR}^{1.00 \pm 0.05} \nonumber \\
    \approx & &(6.6 \pm 1.0) \times 10^{-4} ~ \bigg(\frac{M_\star}{10^{10}~M_\odot} \bigg)^{0.68 \pm 0.01}\nonumber \\
    &+&
    (1.0 \pm 0.1) \times 10^{-1} ~ \bigg(\frac{\rm SFR}{100~M_\odot~{\rm yr}^{-1}} \bigg)^{1.00 \pm 0.05}
\end{eqnarray}
with a ``prompt" term and a ``delayed" term (although this is for a sample at $z<0.25$). 
\citet{Sullivan:2006aa} found that the rate of SNe Ia per unit mass increases with increasing sSFR, which may partly explain the sub-linear dependence on $M_\star$.
A more 
rigorous 
treatment would use the star formation history (SFH) convolved with the delay time distribution (DTD; \citealt{Madau:1998aa,Mannucci:2006aa}). Instead, the ``A+B" parameterization with $M_\star$ and SFR essentially samples the continuous SFH and DTD at two times \citep{Maoz:2014aa}.
However, as cautioned by \citet{Andersen:2018aa}, this calibration is unconstrained for starburst galaxies with log(sSFR) $ > -9$.
While this dependency may in principle also evolve with redshift, several combinations of cosmic star formation histories and delay time distributions tested by \citet{Palicio:2024aa} were able to adequately fit the observed rates of Type Ia supernovae (which are themselves still quite uncertain above $z > 2$). 
For our purposes, we consider it justifiable to use the \citet{Smith:2012aa} results\textemdash derived from a wide range of $8 < {\rm log}(M_\star / M_\odot) < 11$\textemdash to make predictions for our sample.

\begin{figure}[htb]
\centering
\includegraphics[width=\columnwidth]{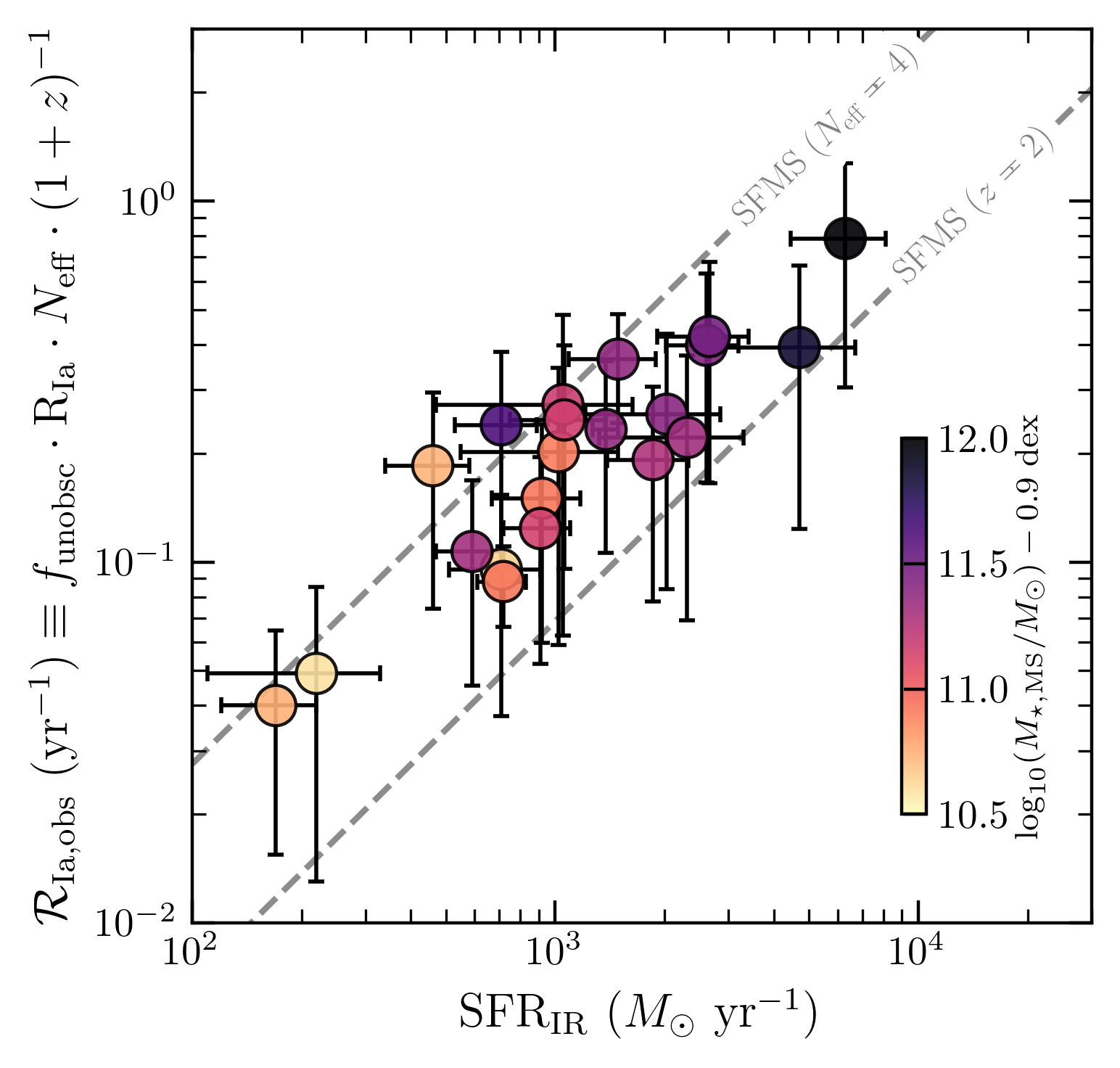}
\caption{
For the same galaxies as in Fig.~\ref{fig:Rcc}: predicted observer-frame rates $\mathcal{R}_{\rm Ia,obs}$ of Type Ia supernovae (the intrinsic rate $R_{\rm Ia}$ multiplied by the unobscured fraction $f_{\rm unobsc}$, the effective number of multiple images \Neff\ and the time dilation factor, $(1+z)^{-1}$).
In this regime of active star formation, $R_{\rm Ia}$ is primarily a function of SFR and weakly dependent on $M_\star$. Stellar masses are bounded by assuming that the sSFRs lie between 0.3 dex below and 0.9 dex above the star-forming main sequence.
Colors of data points show this effective lower limit on $M_\star$ for 0.9 dex 
above
the SFMS.
A dashed gray line shows the expected rate for a galaxy exactly on the SFMS at $z=2$ according to the \citet{Speagle:2014aa} treatment, and a second dashed line shows the observable rate for the same galaxy with an effective multiplicity of $\mathcal{N}_{\rm eff} = 4$.
}
\label{fig:RIa}
\end{figure} 

Unfortunately, the stellar masses for the majority of the PASSAGES DSFGs remain unconstrained, but they are at least very likely to be on the massive end of the stellar mass function (log($M_\star/M_\odot) > 10.5$) given their extreme luminosities and molecular gas masses \citep{Harrington:2021aa}.
While most are actually detected in $H$-band imaging with HST \citep{Kamieneski:2024aa}, 
insufficient observed-frame near-IR data have been collected so far to enable robust SED fitting for most of the background lensed sources\footnote{In principle, lensing-corrected rest-frame $K$-band luminosity may correlate with stellar mass (e.g., \citealt{Jarrett:2013aa}) and could be determined for $z\sim 2$ with WISE at 4.6$\mu$m, for example. However, there is likely significant differential magnification between the stellar continuum and dust continuum, and measuring the magnification factors for the rest-frame near-IR is beyond the scope of this work.}, apart from two exceptions.
For the pair of interacting DSFGs behind the G165 cluster, JWST NIRCam and NIRSpec observations yield stellar masses of ${\rm log}_{10} M_\star \approx 10.2$ and 10.3 \citep{Frye:2024aa, Kamieneski:2024ac}.
With their IR-inferred SFRs, their sSFRs are both 0.9 dex above the MS, but the likely tidal interaction between the two galaxies might be driving an elevated starburst phase above the rest of the sample.
Also, \citet{Liu:2024ae} found 
a large stellar mass of ${\rm log}(M_\star/M_\odot) \approx 11.2 \pm 0.4$ for PJ0116-24
with SFR$_{\rm IR} \approx 1480 ~ M_\odot {\rm yr}^{-1}$, which is 0.8 dex above the star-forming main sequence (SFMS).

\startlongtable
\begin{deluxetable*}{lccccccccccc}
\tablecaption{Estimated intrinsic and observable rates of SNe in a subset of PASSAGES \label{tab:computations}}
\tablehead{
\colhead{ID} &  \colhead{$z_s$} & \colhead{$\mu$} & \colhead{SFR$_{\rm IR, intr.}$} & \colhead{$f_{\rm unobsc}^\dagger$} & \colhead{$R_{\rm CC}({\rm SFR}_{\rm IR})$} & \colhead{$R_{\rm Ia}$} & \colhead{$\mathcal{N}_{\rm eff}$} & \colhead{$\mathcal{R}_{\rm CC,obs}$} & \colhead{$\mathcal{R}_{\rm Ia,obs}$} & \colhead{Refs.}  \\
\colhead{} & \colhead{} & \colhead{} & \colhead{($M_\odot~{\rm yr}^{-1}$)} & \colhead{(\%)} & \colhead{(${\rm yr}^{-1}$)} & \colhead{(${\rm yr}^{-1}$)} & \colhead{} & \colhead{(${\rm yr}^{-1}$)} & \colhead{(${\rm yr}^{-1}$)} & \colhead{}  
}
\startdata
PJ0116$-$24  & $2.125$ & $\approx 17$ 		& $1490 \pm 400$ &  $\approx 32\%^\ddagger$ & $15.5 \pm 4.2$& $1.5 \pm 0.4$ & $2.3 \pm 0.5$ & $3.6 \pm 0.9$ & $1.1 \pm 0.4$ & (1)	  	\\
PJ0143$-$01  & $1.096$ & $7.1 \pm 2.2$ 		& $170 \pm 50$   &  $20 \pm 10\%$ 	& $1.8 \pm 0.5$	& $0.2 \pm 0.1$	& $2.3 \pm 0.5$ &  $0.4 \pm 0.2$ & $0.2 \pm 0.1$ & (2) $\mu$ (3) SFR	  	\\
PJ0209+00  & $2.554$ & $10.5 \pm 3.7$ 		& $2030 \pm 820$    & $20 \pm 10\%$ 	& $21.1 \pm 8.5$	& $2.1 \pm 0.8$	& $2.2 \pm 0.4$ & $2.6 \pm 1.8$ & $1.3 \pm 0.6$ &  (2) $\mu$ (4) SFR	  	\\
PJ0226+23  & $3.120$ & $28.2 \pm 11.3$ 		& $1020 \pm 470$    &  $20 \pm 10\%$ 	& $10.6 \pm 4.9$	& $1.0 \pm 0.5$	& $4.0 \pm 0.8$ & $2.1 \pm 1.5$ & $1.0 \pm 0.5$ &  (2) $\mu$ (3) SFR	  	\\
PJ0305$-$30& $2.263$ & $2.2  \pm 0.6$  		& $6290 \pm 1850$    &  $20 \pm 10\%$ 	& $65 \pm 19$	& $6.4 \pm 1.9$	& $2.0 \pm 0.4$ &  $8.0 \pm 4.9$ & $3.9 \pm 1.4$ & (2) $\mu$ (3) SFR	  	\\
PJ0748+59  & $2.755$ & $\approx 15$ 			& $920 \pm 250$    & $20 \pm 10\%$ 	& $9.6 \pm 2.6$	& $0.9 \pm 0.3$	& $3.0 \pm 0.6$ &  $1.5 \pm 0.9$ & $0.8 \pm 0.3$ & (5) $\mu$ (3) SFR	  	\\
PJ0846+15  & $2.664$ & $7.7 \pm 1.5$ & ${5200 \pm 2600}$    &  $20 \pm 10\%$ 	& $54 \pm 27$	& $5.3 \pm 2.6$	& $1.5 \pm 0.3$ &  $4.4 \pm 3.3$ & $2.2 \pm 1.2$ & (6) $\mu$ (3) SFR	  	\\
PJ1053+60  & $3.549$ & $\approx 24$ 		& $710 \pm 200$    &  $20 \pm 10\%$ 	& $7.4 \pm 2.1$	& $0.7 \pm 0.2$	& $3.0 \pm 0.6$ &  $1.0 \pm 0.6$ & $0.5 \pm 0.2$ & (7) $\mu$ (3) SFR	  	\\
PJ1053+05  & $3.005$ & $7.6 \pm 0.7$ 		& $2610 \pm 590$    &  $20 \pm 10\%$ 	& $27.1 \pm 6.1$	& $2.7 \pm 0.6$	& $3.0 \pm 0.6$ & $4.1 \pm 2.4$ & $2.0 \pm 0.6$ & (2) $\mu$ (4) SFR	  	\\
PJ1127+46  & $1.303$ & $4.5 \pm 0.6$ 		& $590 \pm 120$    &  $20 \pm 10\%$ 	& $6.1 \pm 1.2$	& $0.6 \pm 0.1$	& $2.0 \pm 0.4$ & $1.1 \pm 0.6$ &  $0.5 \pm 0.2$ &  (2) $\mu$ (3) SFR	  	\\
PJ1127+42  & $2.236$ & $16.4 \pm 3.0$ 		& $720 \pm 110$    &  $\approx$ 31\%	& $7.5 \pm 1.1$	& $0.7 \pm 0.1$	& $1.3 \pm 0.3$ &  $0.9 \pm 0.2$ & $0.3 \pm 0.1$ & (8) $\mu$ (4) SFR	  	\\
PJ1138+32  & $2.019$ & $2.8  \pm 0.4$ 		& $910 \pm 190$    &  $20 \pm 10\%$ 	& $9.5 \pm 2.0$	& $0.9 \pm 0.2$	& $2.0 \pm 0.4$ &  $1.3 \pm 0.7$ & $0.6 \pm 0.2$ & (2) $\mu$ (3) SFR	  	\\
PJ1139+20  & $2.858$ & $4.8 \pm 1.1$ 		& $2660 \pm 750$    &  $20 \pm 10\%$ 	& $27.7 \pm 7.8$	& $2.7 \pm 0.8$	& $3.0 \pm 0.6$ &  $4.3 \pm 2.6$ & $2.1 \pm 0.7$ & (2) $\mu$ (3) SFR	  	\\
PJ1322+09  & $2.068$ & $\approx 20$ 	& $460 \pm 120$    &  $20 \pm 10\%$ 	& $4.8 \pm 1.2$	& $0.5 \pm 0.1$	& $6.0 \pm 1.2$ & $1.9 \pm 1.1$ & $0.9 \pm 0.3$ &  (5) $\mu$ (3) SFR	  	\\
PJ1326+33  & $2.951$ & $4.3 \pm 0.6$ 		& $1860 \pm 470$    &  $20 \pm 10\%$ 	& $19.3 \pm 4.9$	& $1.9 \pm 0.5$	& $2.0 \pm 0.4$ &  $2.0 \pm 1.2$ & $1.0 \pm 0.3$ & (2) $\mu$ (3) SFR	  	\\
PJ1329+22  & $2.040$ & $11 \pm 2$ 			& $1380 \pm 110$    &  $20 \pm 10\%$ 	& $14.4 \pm 1.1$	& $1.4 \pm 0.1$	& $2.5 \pm 0.5$ & $2.4 \pm 1.3$ &  $1.2 \pm 0.3$ & (6) $\mu$ (3) SFR	  	\\
PJ1336+49  & $3.254$ & $8.3 \pm 3.1$ 		& $2310 \pm 990$    &  $20 \pm 10\%$ 	& $24 \pm 10$	& $2.4 \pm 1.0$	& $2.0 \pm 0.4$ & $2.3 \pm 1.6$ & $1.1 \pm 0.5$ &  (2) $\mu$ (3) SFR	  	\\
PJ1446+17  & $1.084$ & $3.9 \pm 0.7$ 		& $710 \pm 180$    &  $20 \pm 10\%$ 	& $7.4 \pm 1.9$	& $0.7 \pm 0.2$	& $3.3 \pm 0.7$ &  $2.4 \pm 1.4$ & $1.2 \pm 0.4$ & (2) $\mu$ (3) SFR	  	\\
PJ1449+22  & $2.153$ & $10.1 \pm 5.2$ 		& $1050 \pm 580$    &  $20 \pm 10\%$ 	& $10.9 \pm 6.0$	& $1.1 \pm 0.6$	& $4.0 \pm 0.8$ & $2.8 \pm 2.1$ & $1.4 \pm 0.8$ &  (2) $\mu$ (3) SFR	  	\\
PJ1607+73  & $1.482$ & $6.8 \pm 1.7$ 		& $220 \pm 110$    &  $20 \pm 10\%$ 	& $2.3 \pm 1.1$	& $0.2 \pm 0.1$	& $2.7 \pm 0.5$ &  $0.5 \pm 0.4$ & $0.2 \pm 0.1$ & (2) $\mu$ (4) SFR	  	\\
PJ2313+01  & $2.217$ & $6.1 \pm 1.2$ 		& $1060 \pm 310$    &  $20 \pm 10\%$ 	& $11.0 \pm 3.2$	& $1.1 \pm 0.3$	& $3.7 \pm 0.7$ &  $2.5 \pm 1.5$ & $1.2 \pm 0.4$ & (2) $\mu$ (3) SFR	  	\\
\enddata
\tablenotetext{^\dagger}{Unobscured SFR estimated using the adopted $f_{\rm unobsc}$, except for PJ0116-24 \citep{Liu:2024ae} and G165, which are estimated directly from H$\alpha$ \citep{Frye:2024aa} following correction from the measured Balmer decrement. For G165, we assume that the $f_{\rm unobsc}$ for Arc 1 ($\approx 31\%$) applies also for its counterpart, Arc 3.
}
\tablenotetext{^\ddagger}{Derived using ${\rm SFR}_{\rm H\alpha,corr} = 470 \pm 60~M_\odot~{\rm yr}^{-1}$ \citep{Liu:2024ae}.}
\tablecomments{
{\bf References} for SFR and $\mu$ estimates: 
(1) \citealt{Liu:2024ae}; 
(2) \citealt{Kamieneski:2024aa} 
(3) \citealt{Berman:2022aa}; 
(4) \citealt{Harrington:2016aa}; 
(5) P. Kamieneski et al., in prep.;
(6) \citealt{Foo:2025aa};
(7) \citealt{Wang:2024ab};
(8) \citealt{Kamieneski:2024ac}; 
(9) \citealt{Diaz-Sanchez:2017aa}.
}
\end{deluxetable*}

\begin{figure}[htb]
\centering
\includegraphics[width=\columnwidth]{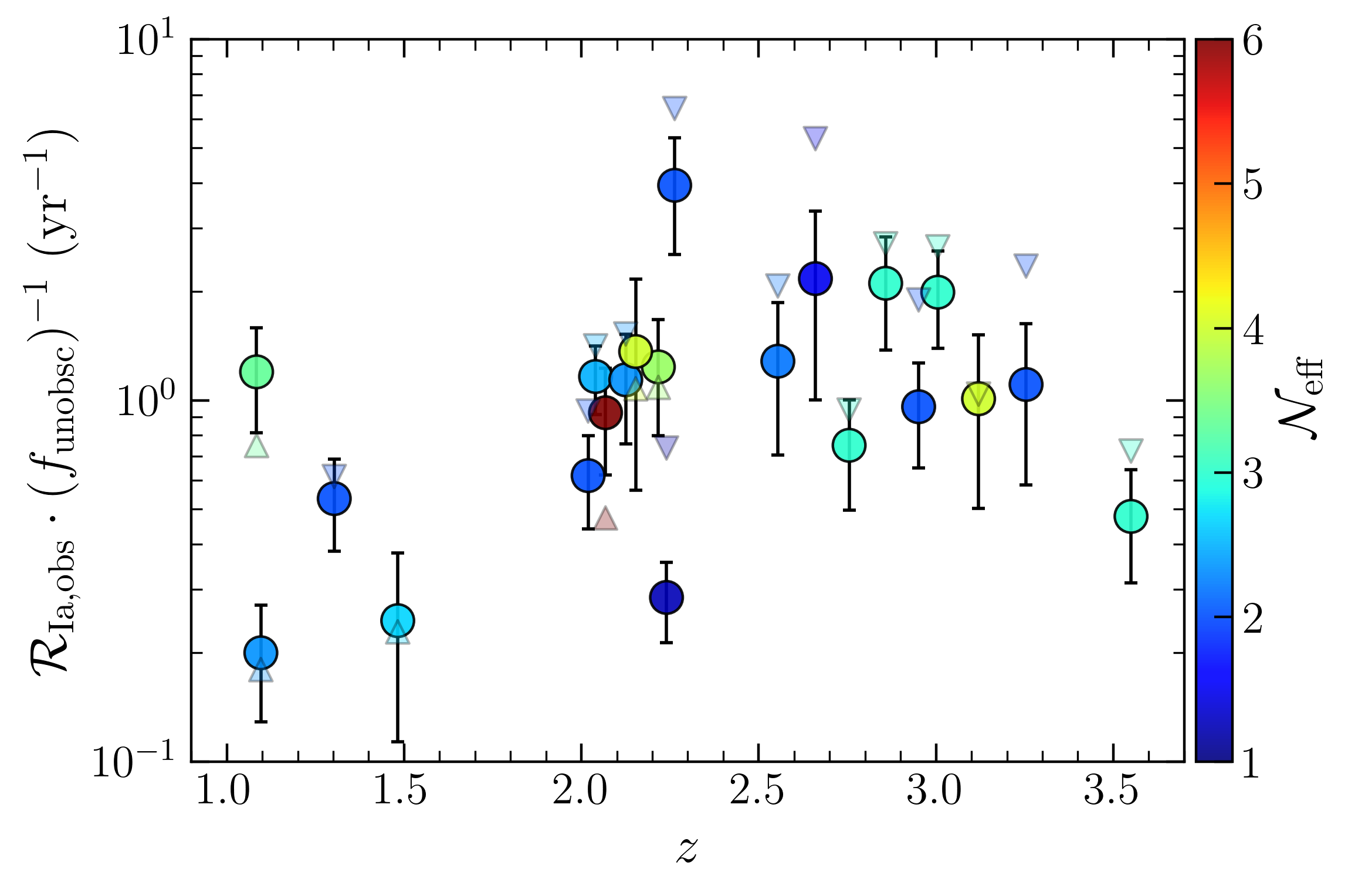}
\caption{
Redshift distribution of the observable rate of SNe Ia from PASSAGES estimated in Fig.~\ref{fig:RIa}, $\mathcal{R}_{\rm Ia}$, but now removing the factor of $f_{\rm unobsc}$ (thus reducing uncertainties and improving legibility).
Colors indicate the effective multiplicity \Neff. 
Transparent, smaller triangles indicate the intrinsic rates ($R_{\rm Ia}$) for each target (i.e., without factoring in the boost from multiple-imaging or the $1+z$ decrement from cosmological time dilation).
The triangles for intrinsic rates are oriented to point towards their corresponding observable rates.
This illustrates that since $\mathcal{N}_{\rm eff} \cdot (1+z)^{-1}$ is typically of order unity, these effects roughly cancel out.
}
\label{fig:RIa_z}
\end{figure}

In Appendix \ref{sec:appendix_RIa}, we use the expected offset of DSFGs from the SFMS to gain an imprecise estimate of reasonable stellar masses. However, ultimately the dependence of Equation \ref{eqn:RIa} on $M_\star$ is highly subdominant to the contribution from SFR in the star-forming regime (${\rm SFR}\gtrsim 10~M_\odot~{\rm yr}^{-1}$). 
Fortunately, this allows us to still predict $R_{\rm Ia}$ for the sample within reason, which we include in Table~\ref{tab:computations}.
In Fig.~\ref{fig:RIa}, we show the predicted observable rates of Type Ia supernovae for PASSAGES, $\mathcal{R}_{\rm Ia,obs} \equiv f_{\rm unobsc} \cdot R_{\rm Ia} \cdot \mathcal{N}_{\rm eff} \cdot (1+z)^{-1}$, as a function of their SFR$_{\rm IR}$.
In this regime of star formation, the SN Ia rates are dominated by the contribution from SFR, but the data points' colors correspond to the lower limits in stellar masses that we adopt, at 0.9 dex above the star-forming main sequence.
In Fig.~\ref{fig:RIa_z}, we show the observable rates $\mathcal{R}_{\rm Ia,obs} \cdot (f_{\rm unobsc})^{-1}$ vs. redshift\footnote{Dividing $\mathcal{R}_{\rm Ia,obs}$ by the obscured fraction in this version reduces the relative uncertainties and thus makes it easier to interpret. Given that $R_{\rm Ia}$ is dominated by the SFR, we expect the SNe Ia to be subject to roughly the same average attenuation as the CCSNe, but few works have yet examined this at these redshifts.}, with transparent points indicating the intrinsic rates $R_{\rm Ia}$.
This particularly emphasizes our finding that the decrement in observable rate due to redshift time dilation is approximately counteracted by the boost provided by multiple imaging.

\begin{figure*}[htb]
\centering
\includegraphics[width=0.495\textwidth]{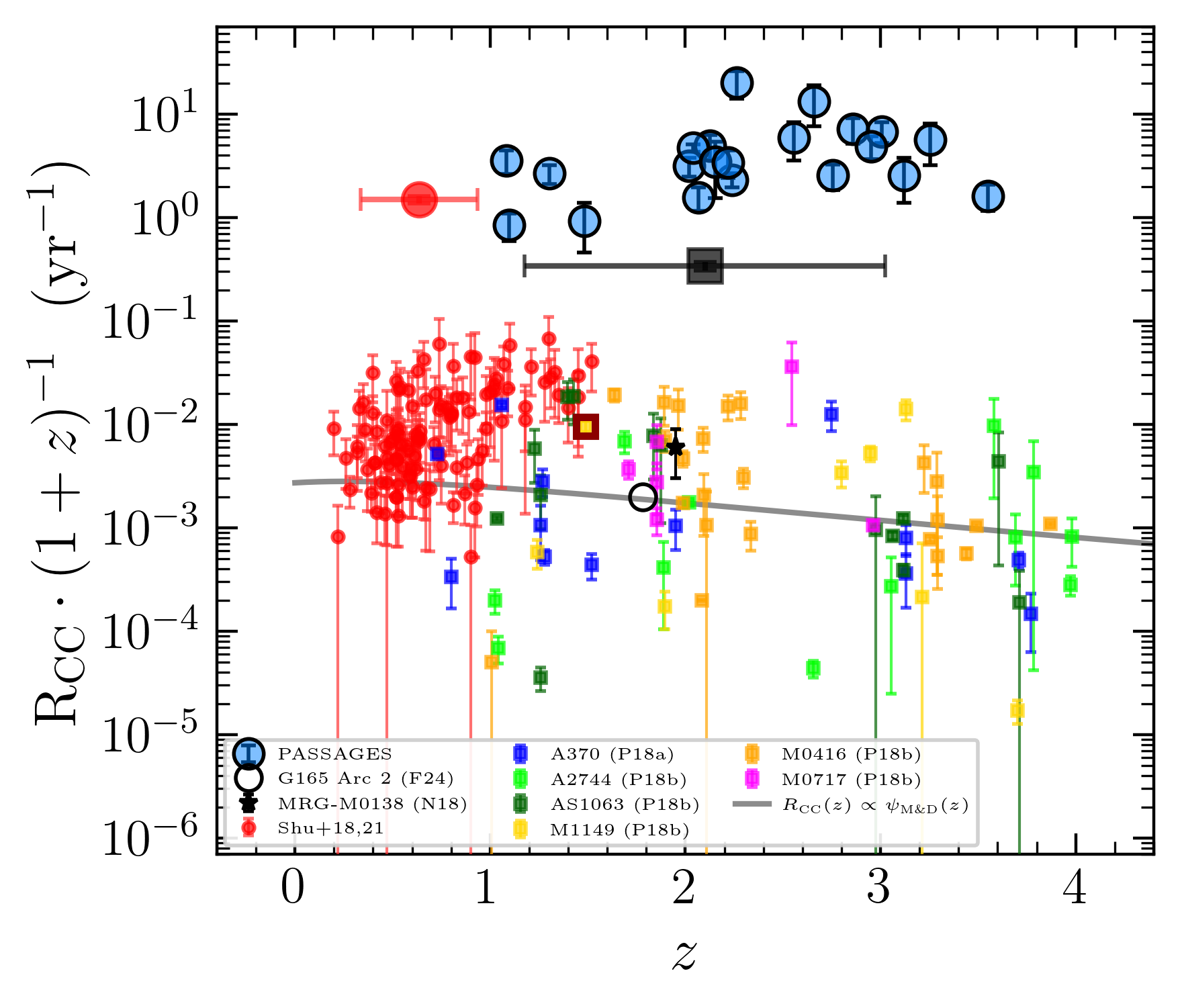}
\includegraphics[width=0.495\textwidth]{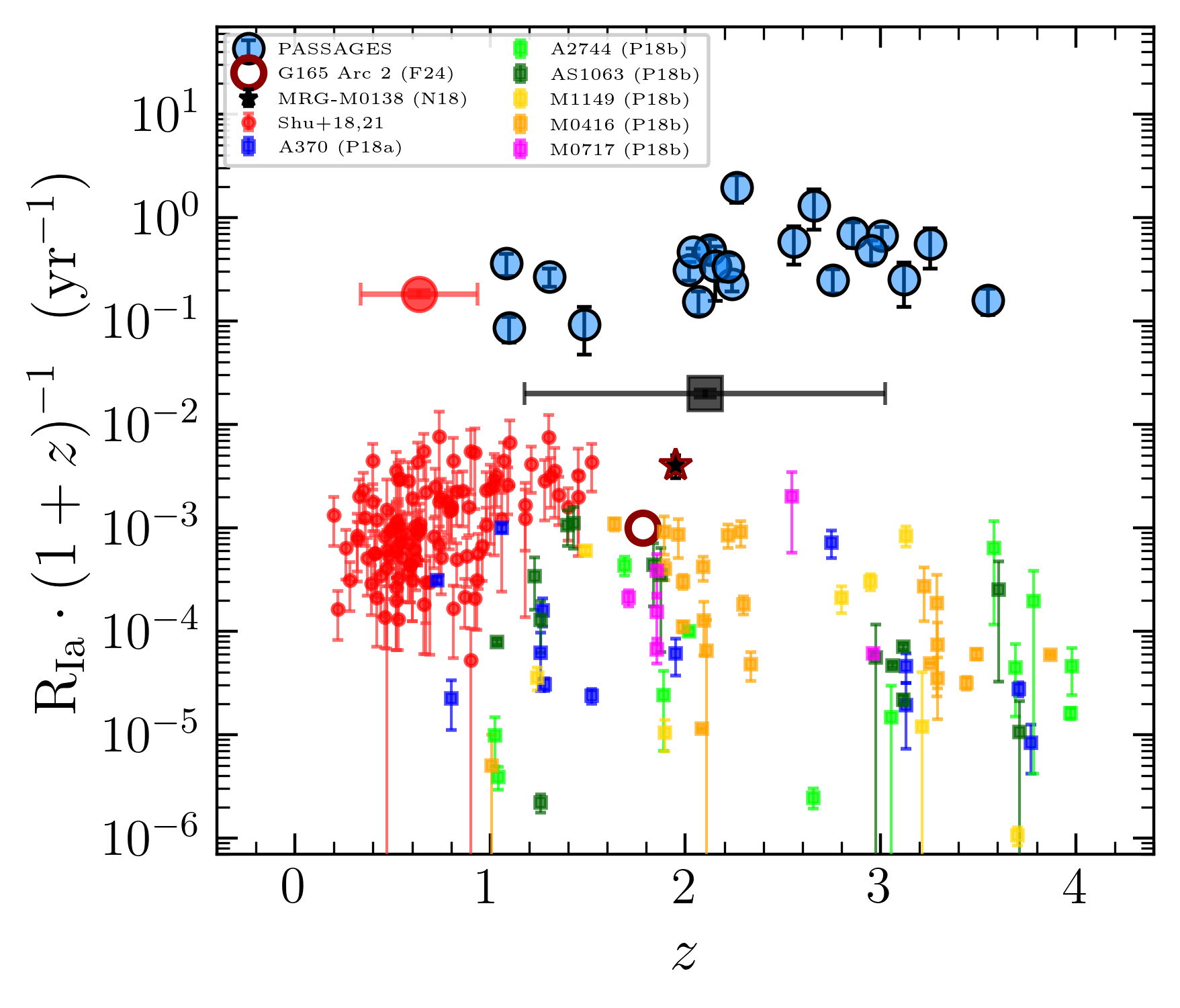}
\caption{
Intrinsic supernova rates (core-collapse on {\it left}, Type Ia on {\it right}), still rescaled by time dilation but excluding multiple imaging (and $f_{\rm unobsc.}$ correction), vs. redshift. 
Whereas $\mathcal{R}_{\rm CC,obs}$ and $\mathcal{R}_{\rm Ia,obs}$ represented the average rate at which supernova images appear, $R_{\rm CC}\cdot (1+z)^{-1}$ and $R_{\rm Ia}\cdot (1+z)^{-1}$ are the actual rate of individual supernova events, as if the galaxies were not lensed.
We compare with the predictions made for a large sample of galaxy-galaxy lenses by 
\citet{Shu:2018ab}, incorporating the corrections from \citet{Shu:2021ab}, and the Frontier Fields considered by \citet{Petrushevska:2018ab,Petrushevska:2018aa}.
The sums of the \citeauthor{Shu:2018ab} and \citeauthor{Petrushevska:2018ab} samples are shown as large red circles and large black squares, respectively.
Predicted rates for G165 Arc 2, the host galaxy of the Type Ia SN H0pe, are shown with open circles (based on SFR/$M_\star$ from \citealt{Frye:2024aa}).
Similarly, the predicted rates for MRG-M0138 (based on \citealt{Newman:2018aa}) are indicated as black stars (with a dark red border on the right panel, highlighting SN Requiem/Encore).
In the left panel, the MACS J1149 point with a dark red border is the host galaxy of the discovered core-collapse glSN Refsdal. 
The PASSAGES objects considered in this work have CCSN rates about 100 times greater than the \citeauthor{Shu:2018ab} sample (and similarly for SNe Ia).
In the left panel, the gray curve shows the expectation of $R_{\rm CC}(z) \propto \psi(z)$ for the \citet{Madau:2014aa} parameterization of the cosmic star formation rate density $\psi(z)$, which has units of $M_\odot~{\rm yr}^{-1}~{\rm Mpc}^{-3}$, rescaled arbitrarily to approximately match the Frontier Fields data points.
}
\label{fig:comparison}
\end{figure*} 

\subsection{Predicting the rate of independent supernova events}
\label{sec:independent}

Up to this point, we have considered the rate of observable supernovae from Earth's perspective, taking into account both cosmological time dilation by $(1+z)^{-1}$ and amplification by multiple imaging \Neff. 
This rate is useful if one's goal is to catch any image of a glSN, but if the goal is to perform TD cosmography, then the rate of {\it independent} supernovae events is of greater interest\footnote{Here, this rate still accounts for $R_{\rm CC}$ and $R_{\rm Ia}$ being modulated by $(1+z)^{-1}$ time dilation, but not by $f_{\rm unobsc}$, to allow for a more direct comparison with literature values.}.
We show these rates for PASSAGES in Fig.~\ref{fig:comparison}, along with the predicted rates from similar works forecasting glSN rates in targeted monitoring searches, including the Frontier Fields \citep{Petrushevska:2018aa, Petrushevska:2018ab} and $128$ galaxy-scale lenses from \citet{Shu:2018ab, Shu:2021ab}.
We also indicate rates from the host galaxies of a number of the already-discovered glSNe\textemdash e.g., the core-collapse SN Refsdal \citep{Kelly:2015aa} in a galaxy lensed by MACS J1149, and the set of Type Ia supernovae that includes SN Requiem/SN Encore \citep{Rodney:2021ab, Pierel:2024ab} in MRG-M0138 \citep{Newman:2018aa} and SN H0pe in G165 Arc 2abc \citep{Frye:2024aa}.

While $R_{\rm CC}$ and $R_{\rm Ia}$ might appear at first to be equivalent to within a constant factor, this is not quite the case. 
By equations \ref{eqn:RCC} and \ref{eqn:RIa}, the ratio $R_{\rm Ia} / R_{\rm CC} \sim {\rm sSFR}^{-1} \cdot M_\star^{-0.32}$ (plus a constant term).
By the star-forming main sequence, many of the star-forming galaxies included in Fig.~\ref{fig:comparison} will have comparable values of sSFR, while the ratio is more weakly dependent on $M_\star$.
For a $10^{10}~M_\odot$ galaxy on the main sequence at $z\sim 2$, for example, $R_{\rm Ia} / R_{\rm CC} \approx 0.1$ (adopting a Kroupa IMF).

Every PASSAGES object included in this work exceeds the rate of all other targets considered in Fig.~\ref{fig:comparison}, demonstrating the dramatic potential of this sample.
Most are even $>2$ dex above typical arcs in the Frontier Fields; while these clusters are remarkable for their lensing power, the background objects are largely unremarkable in terms of their SFR. 
Perhaps more surprising is the observation that our predictions for $R_{\rm CC}$ are all $\gtrsim 1.5$ dex above the host galaxy of SN Refsdal, and even our predicted $R_{\rm Ia}$ are $\gtrsim 1$ dex above MRG-M0138, the host of two recent SNe Ia, Requiem and Encore.
We also observe that the redshift-dilated $R_{\rm CC}$ and $R_{\rm Ia}$ rates for PASSAGES roughly increase with redshift (likely owing to their identification by flux; see Fig. 2 of \citealt{Wang:2024ab}). This trend is contrary to the slow decline of the (dilated) cosmic star formation rate density,
$\psi(z) \cdot (1+z)^{-1}$ \citep{Madau:2014aa}, shown as a gray curve in Fig.~\ref{fig:comparison}.

In absolute terms, the median (Earth-frame) occurrence rate of individual CCSNe is 
$3.5~{\rm yr}^{-1}$
and $\sim 0.3~{\rm yr}^{-1}$ for SNe Ia.
This does not imply, we caution, that any 1-year monitoring program would yield multiple glSNe per field, as the actual yield is the occurrence rate multiplied by the {\it control time} \citep{Zwicky:1938aa}. 
This is effectively the cumulative duration for which a given supernovae would fall above the detection threshold, which itself depends on a combination of observing sensitivity, dust attenuation, lensing magnification, and the precise supernova type.
We do not simulate a mock observing campaign in this work, but simply remark that this 
sample has $\gtrsim 18$ members with core-collapse occurrence rates $R_{\rm CC}\cdot (1+z)^{-1}$ in excess of 1 yr$^{-1}$ (with a median closer to 4 per year).
As a conservative example, an observer-frame control time of 1 month per epoch, with 3 epochs over the course of a year, would result in a yield of $N_{\rm CC} \sim 0.5$ detections for each field.
If these 18 fields were all monitored at this rate, one could expect to detect $\sim 9$ CCSNe with this campaign alone (not to mention any SNe Ia that are uncovered).
With the depths of $28 - 29$ mag routinely reachable by JWST at $\sim 1.5~\mu$m in $\sim 1$ hr, it is highly feasible to extend the per-epoch control time well beyond 1 month (e.g., Fig. 12 of \citealt{Petrushevska:2016aa}), even for only modestly magnified arcs, such that even a subsample would be expected to result in at least one detection.

As another example, we consider SNe Ia, which of course are more uniform in their light curve behavior, and which are themselves predicted to be quite frequent in HyLIRGs. With SN H0pe as a recent example of a $z\sim 2$ glSN Ia in a relatively dusty environment, we note that its de-magnified peak was observed at $\sim 24$ mag in F150W (rest-frame $V$-band), but it also remained above 25 mag for a duration of $\approx 90$ days in the observed frame \citep{Pierel:2024aa}. 
The light curve model suggested $A_V = 1.21 \pm 0.11$ (also \citealt{Chen:2024ab}); if the supernova were subjected to substantially higher attenuation as might be expected for HyLIRGs, such as $A_V = 5$, then the peak would be $\sim 4$ mag fainter (i.e. $\sim 28$ mag) and the SN would remain above 29 mag for an observer for $\sim 3$ months. With only modest magnification $\mu = 5$ (or 1.7 mag), a 5$\sigma$ point-source limiting depth of 27.3 mag would be required for a detection, achieved in less than 10 minutes of integration with JWST/NIRCam\footnote{Sensitivity is estimated with the JWST Exposure Time Calculator \citep{Pontoppidan:2016aa}.} or $<2$ orbits with HST WFC3/IR in F160W.


As a caveat, however, the occurrence rates we estimate do not account for the possibility that some detected SN events might be caught only when the last image is still visible. 
In this hypothetical scenario, there is not likely to be any hope for performing TD cosmography, since no delays can actually be measured through light curves.
It is rather non-trivial to quantify the likelihood of this actually occurring, as it depends on both the time delays between images and the control time (such that the penultimate image to arrive has faded below the detection limit, but not the final image). 

To place these forecasts in context, we compare with the yields that are currently expected from other search strategies.
\citet{Sainz-de-Murieta:2024aa} predicted that LSST will uncover $\sim$7 Type Ia glSNe per year within known galaxy-lensed objects that are suitable for TD cosmography.
In this case, the authors note that an $H_0$ measurement is not feasible if the host galaxy is not itself lensed as well (thereby enabling precise lens modeling).
Similarly, \citet{Arendse:2024ab} predict a ``gold sample" of $\sim 10$ glSNe Ia yr$^{-1}$ uncovered with LSST that are sufficiently bright and with time delays that are long enough to enable follow-up observations (which are likely essential; \citealt{Huber:2019aa}). 
Complementary to this will be an approach of monitoring a ``watchlist" of lensing galaxy clusters \citep{Ryczanowski:2023aa}, which are beneficial for reasons discussed in \S \ref{sec:REin}.
\citet{Bronikowski:2025aa} predict $1-2$ such SNe will be discovered by LSST per year from a sample of 46 well-studied lensing clusters.
The SNe Ia cosmology program of the upcoming
{\it Nancy Grace Roman Space Telescope} \citep{Hounsell:2018aa}
will also reveal some fraction that are gravitationally lensed \citep{Oguri:2010ac}, which may number as high as $\sim 11$ glSNe Ia and $\sim 20$ cumulative lensed CCSNe \citep{Pierel:2021ab}.

\subsection{Identifying candidates with the longest median time delay}
\label{sec:REin}

Cluster-scale lenses are typically the best candidates for high rates of lensed SNe in the observed frame due to their large lensing cross sections $\sigma(z_s)$ in the source-plane at $z_s$ \citep{Turner:1984aa}. 
This cross-section is closely linked to the Einstein radius 
\begin{equation}
\theta_E = \bigg[\frac{4GM(<\theta_E)}{c^2} \cdot \frac{D_{LS}}{D_L D_S}\bigg]^{1/2}
\end{equation}
(e.g., \citealt{Narayan:1996aa})
for enclosed mass $M(<\theta_E)$ and angular-diameter distances to the lens ($D_L$), to the source plane ($D_S$), and between these two ($D_{LS}$).
To first order, the area that produces 4 lensed images from a singular isothermal ellipsoid (again excluding demagnified images) is 
$\sigma_4 = \pi \epsilon^2 \theta_E^2 / 6$ \citep{Kochanek:2006aa},
where $\epsilon \equiv 1- b/a$ is the ellipticity (from major and minor axes, $a$ and $b$).
The cross section then scales linearly with mass, $\sigma_4 \propto M(<\theta_E)$,
such that a cluster can have orders of magnitude larger areas.
As discussed in \S \ref{sec:multiplicity}, this larger area of magnification means that a galaxy is more likely to be lensed uniformly into a constant multiplicity of images,
but it also means that objects nearby in projected distance will also be multiply-imaged (\S \ref{sec:environments}). This further amplifies the expected rate of observable SNe.

Cluster lenses with large Einstein radii are also beneficial for following up lensed SNe for the purpose of TD cosmography (e.g., \citealt{Bronikowski:2025aa}).
It is true that time delays between any pair of cluster-lensed images can still be very short (e.g. a few days)\textemdash especially in the case where the deflection is dominated by a galaxy- or group-scale halo embedded within the cluster. For example, this was the case with the first four images of SN Refsdal, which formed an Einstein cross configuration ($\theta_{\rm Ein} \approx 1.3$; delays $\approx 10-20$ days; \citealt{Kelly:2023ab}) around an elliptical galaxy in the cluster at $z=0.54$ \citep{Kelly:2015aa}.
However, 
one should expect that
a larger Einstein radius (or angular separation between images) 
would correlate 
with a larger median delay, by virtue of the longer path lengths followed by the highly-deflected light rays.

In practice, varying approaches are used to quantify the Einstein radius of a system (see for example \S 3.5 of \citealt{Kamieneski:2024aa}).
For the PASSAGES objects in this work, the values of $\theta_{\rm Ein}$ shown in Table~\ref{tab:td_Ein} are derived according to the {\it equivalent} Einstein radius definition, which is the radius of the smallest circle centered on the peak of lensing convergence $\kappa$ that contains an average $\langle \kappa \rangle < 1$ \citep{Schneider:1992aa}.
For lensed systems with multi-modal mass distributions, different Einstein radii may be measured for the different mass components.
This notably includes G165, the field containing both SN H0pe and the PJ1127+42 DSFG.
With the model from \citet{Kamieneski:2024ac}, we measure $\theta_{\rm Ein} = 10\farcs7 \pm 1\farcs1$ for the $z=2.24$ DSFG arc on the southwest side of the cluster, and $21\farcs1 \pm 0\farcs5$ for the $z=1.78$ arc hosting SN H0pe on the northeast side.
\citet{Frye:2019aa} measured a lens size of $13\arcsec$, but with a different definition (based on the total area inside the critical curve for the full cluster).
\citet{Pascale:2022aa} measured $15\farcs3$ for $z=9$ using the same approach, or $\approx 14\farcs4$ rescaled to $z_s = 2.24$, although this and the \citeauthor{Frye:2019aa} measurement were made before new constraints from NIRCam were available \citep{Frye:2024aa}.

For other previously discovered glSNe summarized at the bottom of Table~\ref{tab:td_Ein}, the Einstein radii are estimated 
using 
the
publicly-available Frontier Fields lens models \citep{Lotz:2017aa}.
For these, we also employ a model-independent approach to determine
the radius of a circle that best captures the distribution of features/clumps within the lensed arcs (e.g., \citealt{Remolina-Gonzalez:2020aa, Mork:2025aa}).

\begin{figure*}[th]
\centering
\includegraphics[width=0.9\textwidth]{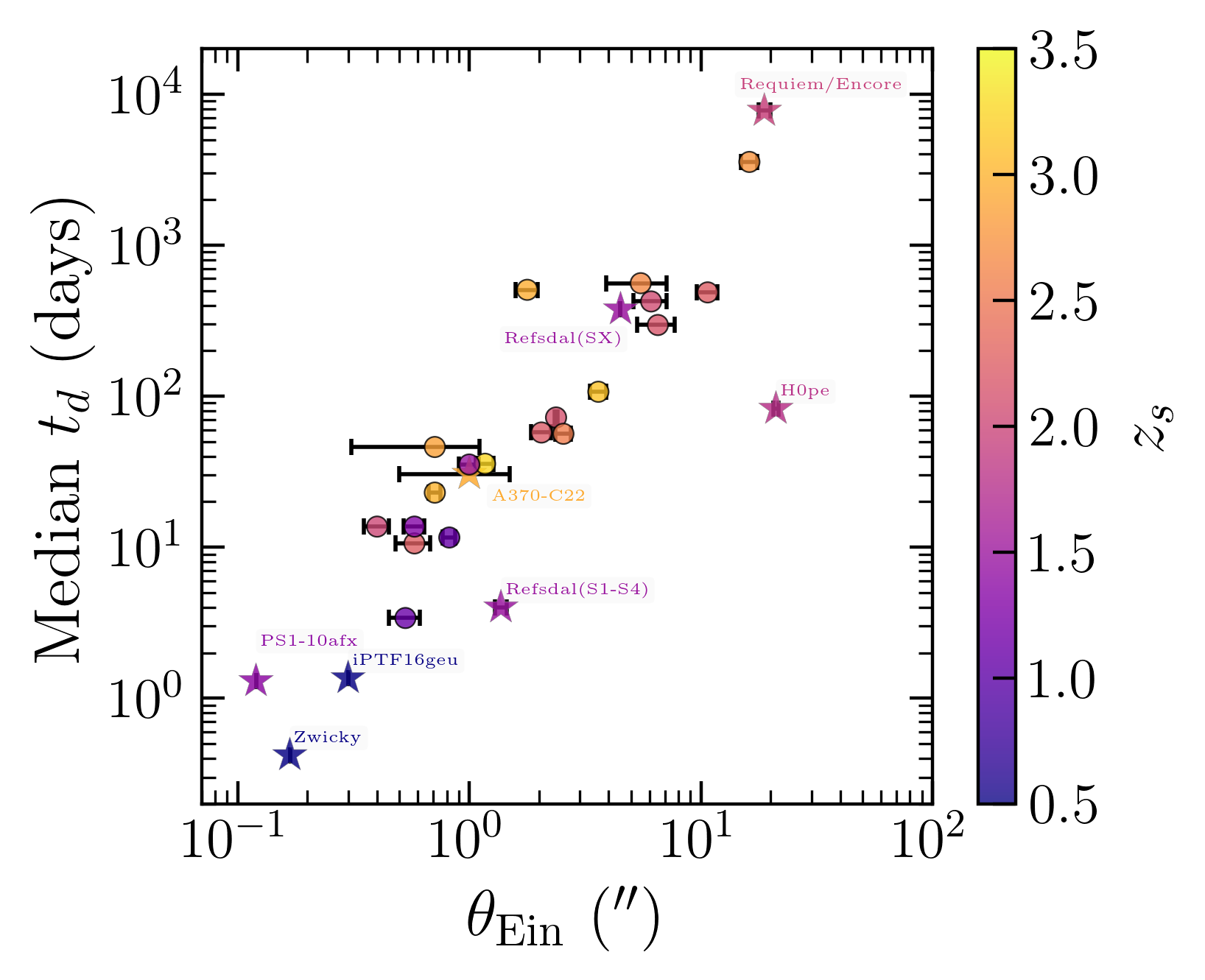}
\caption{
Median time delays $t_d$ (between the first image to arrive and the other multiple images) in days vs. the Einstein radius $\theta_E$ in arcsec; see Table~\ref{tab:td_Ein}.
Lensed SNe from the literature (and their time delay/Einstein radius measurements) shown as stars include
PSF1-10afx \citep{Chornock:2013aa, Quimby:2013aa, Quimby:2014aa},
SN Refsdal (for both the Einstein Cross images S1-S4 and the last image SX; \citealt{Kelly:2015aa, Kelly:2016aa, Kelly:2023aa, Rodney:2016aa}),
iPTF16geu \citep{Goobar:2017aa, Diego:2022ad},
SN Requiem \citep{Rodney:2021ab},
SN Zwicky \citep{Goobar:2022aa, Goobar:2023aa, Pierel:2023aa},
the $z=2.9$ SN discovered by \citet{Chen:2022ag},
and
SN H0pe \citep{Frye:2024aa, Pierel:2024aa};
see review by \citet{Suyu:2024aa}.
For SN H0pe and the SX image of SN Refsdal, the time delays are the photometric light curve-derived values; in all other cases, delays are lens model-predicted.
Despite the scatter, there is a statistically significant correlation between $\widetilde{t_d}$ and $\theta_E$, validating the assumption that larger image separations generally yield longer time delays (see Appendix \ref{sec:appendix_td_Re_corr}).
}
\label{fig:R_E}
\end{figure*}

Fig.~\ref{fig:R_E} shows the median lens model-predicted time delay ($\widetilde{t_d}$) vs. Einstein radius for the PASSAGES sample, along with the observed/predicted values for the set of known glSNe.
While there is a decent amount of scatter\textemdash especially at the high-$\theta_E$, high-$t_d$ end\textemdash there is a clear strong correlation
between $t_d$ and $\theta_E$ (Spearman's rank correlation coefficient $\rho = 0.89$, with high significance $p\ll 0.01$), such that $\theta_E$ may be used to select likely long-delay lenses as a rule-of-thumb.
As we discuss in Appendix \ref{sec:appendix_td_Re_corr}, this correlation has some theoretical underpinning, since time delays can be related to the image separations of arcs relative to the lens centroid.

\subsection{Optimizing the success of a random monitoring program}

\begin{figure*}[th]
\centering
\includegraphics[width=0.8\textwidth]{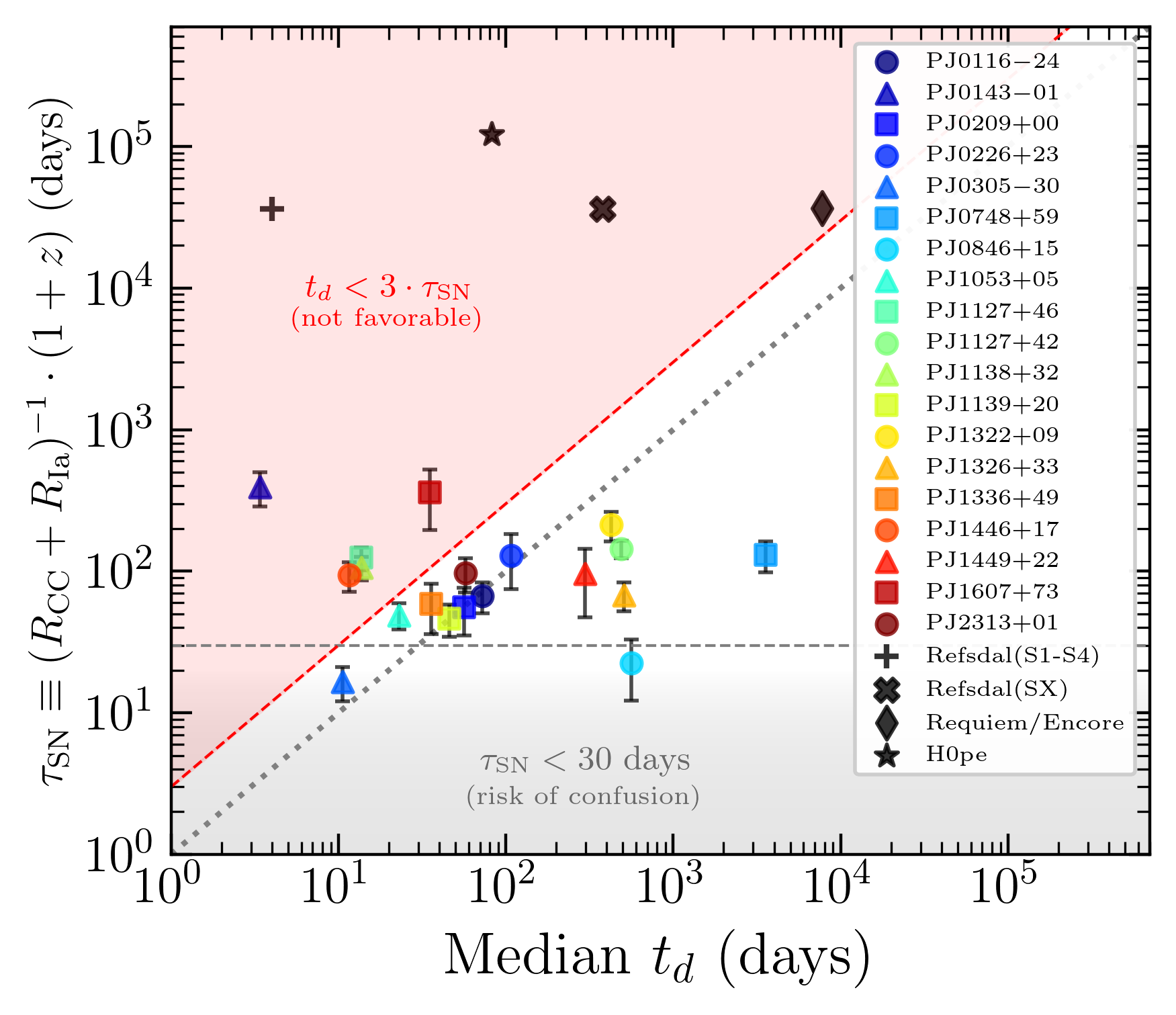}
\caption{
Characteristic timescale between supernovae in the observer-frame, $\tau_{\rm SN} \equiv (R_{\rm CC} + R_{\rm Ia})^{-1} \cdot (1+z)$, vs. the median time delays $t_d$ for PASSAGES and for glSN host galaxies in the literature.
A dotted line indicates $\widetilde{t_d} = \tau_{\rm SN}$, while the red shaded region marks $\widetilde{t_d} < 3 \cdot \tau_{\rm SN}$. We consider this regime not particularly favorable for a sparse monitoring cadence, as these target fields are likely to have long stretches with no images of glSNe arriving. On the other hand, targets with a very short $\tau_{\rm SN}$ (e.g. $<30$ days) carry a risk of confusion in ensuring that SN candidates identified in different arcs are in fact multiple images of the same SN. 
}
\label{fig:tau_SN}
\end{figure*}

Ideally, for a monitoring program to be successful, time delays between images should be relatively comparable to the average duration between supernovae in the observer-frame\textemdash i.e., $\tau_{\rm SN} \equiv (R_{\rm CC} + R_{\rm Ia})^{-1} \cdot (1+z)$.
If the median time delay is much less than this characteristic time between SNe, $\widetilde{t_d} \ll \tau_{\rm SN}$, then multiple imaging does not contribute very favorably to boosting the rate of observable SNe.
In this scenario, there would still be long durations where no SN images would be visible, presenting challenges for monitoring campaigns.
The PASSAGES targets and the hosts of some of the confirmed glSNe in the literature are shown in Fig.~\ref{fig:tau_SN}.
There are also unique challenges for fields with very short $\tau_{\rm SN}$ timescales (e.g. less than 1 month observed-frame), as it can then become difficult to confirm that transient events seen in multiple images of a galaxy are in fact images of the same supernova. 
This risk of confusion can certainly be mitigated: information on parity and relative location of the transient within the arc can be helpful in ruling out unrelated events. 
Very few targets are likely to lie in this regime, and a well-refined lens model can assist in assessing the reliability of glSN candidates. For PASSAGES, for example, PJ0846+15 nominally has a very short $\tau_{\rm SN}$, but in reality, the very high SN rates arise from a very high SFR from multiple galaxies in a candidate protocluster core \citep{Foo:2025aa}, in which case it is much easier to rule out unrelated transient events.

\startlongtable
\begin{deluxetable*}{lcccccc}
\tablecaption{Einstein radii and lens model-estimated median  time delays for PASSAGES targets and for glSNe drawn from the literature. 
\label{tab:td_Ein}}
\tablehead{
\colhead{ID} &  \colhead{$z_l$} & \colhead{$z_s$} & \colhead{$\theta_{\rm Ein}$} & \colhead{$\widetilde{t_d}^{\ \dagger}$} & \colhead{max $t_d^{\ \dagger}$} & \colhead{Refs.$^\ddagger$}  
\\[-2ex]
\colhead{} & \colhead{} & \colhead{} & \colhead{($\arcsec$)} & \colhead{(days)}& \colhead{(days)} & \colhead{}
}
\startdata
PJ0116$-$24& $0.555$ & $2.125$ & $2.37 \pm 0.04$ 	& $72$  & 201 & (1) \\
PJ0143$-$01& $0.594$ & $1.096$ & $0.53 \pm 0.08$ 	& $3.4$  & 10& (1) \\
PJ0209+00  & $0.202$ & $2.554$ & $2.6 \pm 0.2$ 	& $56$  & 60 & (1) \\
PJ0226+23  & $0.41$ & $3.120$ & $3.6 \pm 0.3$ 	& $110$  & 190 & (1) \\
PJ0305$-$30& $0.5$ & $2.263$ & $0.6 \pm 0.1$ 	& $11$  & 11 & (1) \\
PJ0748+59  & $0.402$ & $2.755$ & $16.2 \pm 1.3$ 	& $3600$  & 5400 & (2,3) \\
PJ0846+15  & $0.766$ & $2.664$ & {$5.5 \pm 1.6^{\dagger\dagger}$} 	& {560$^{\dagger\dagger}$}  & 780$^{\dagger\dagger}$ & (4) \\
PJ1053+60  & $0.837$ & $3.549$ & 5.9 	& \textemdash  & \textemdash & (5) \\
PJ1053+05  & $1.525$ & $3.005$ & $0.71 \pm 0.04$ 	& $23$  & 62 & (1) \\
PJ1127+46  & $0.42$ & $1.303$ & $0.58 \pm 0.06$ 	& $14$  & 14 & (1) \\
PJ1127+42  & $0.348$ & $2.236$ & $10.7 \pm 1.1$ 	& $470$  & 580 & (6,7) \\
PJ1138+32  & $0.52$ & $2.019$ & $0.40 \pm 0.05$ 	& $14$  & 16 & (1) \\
PJ1139+20  & $0.57$ & $2.858$ & $0.7 \pm 0.4$ 	& $46$  & 59 & (1) \\
PJ1322+09  & 0.6 & $2.068$ & $6.1\pm 1.0$   	& $420$  & 560 & (3) \\
PJ1326+33  & $0.786$ & $2.951$ & $1.8 \pm 0.2$ 	& $510$  & 580 & (1) \\
PJ1329+22  & $0.44$ & $2.040$ & $11.0 \pm 0.4$ 	& \textemdash  & \textemdash &  (8) \\
PJ1336+49  & $0.26$ & $3.254$ & $1.2 \pm 0.1$ 	& $36$  & 39 & (1) \\
PJ1446+17  & $0.493$ & $1.084$ & $0.82 \pm 0.05$ 	& $12$  & 14 &  (1) \\
PJ1449+22  & $0.4$ & $2.153$ & $6.5 \pm 1.2$ 	& $300$  & 460 & (1) \\
PJ1607+73  & $0.65$ & $1.482$ & $1.0 \pm 0.1$ 	& $35$  & 36 & (1) \\
PJ2313+01  & $0.560$ & $2.217$ & $2.1 \pm 0.2$ 	& $58$  & 94 & (1) \\
\hline
{\bf Literature} & & & & & \\
PS1-10afx   & $1.117$ & $1.388$ & $0.12$ 	& $1.3$  & \textemdash & (9,10) \\
SN Refsdal(S1-S4)  & $0.54$ & $1.49$ & $1.37 \pm 0.08^{\dagger\dagger}$ 	& $4$ & \textemdash & (11,12) \\
SN Refsdal(SX)  & \textemdash & \textemdash & $4.5^{\dagger\dagger}$ 	& $380$ & \textemdash & (11,13) \\
iPTF16geu   & $0.216$ & $0.409$ & $0.3$ 	& $1.4$  & \textemdash & (14,15,16) \\
SN Requiem/Encore  & $0.338$ & $1.95$ & $18.8 \pm 1.1$ 	& $7800$ & \textemdash & (17,18,19) \\
SN Zwicky   & $0.226$ & $0.354$ & $0.168$ 	& $0.42$ & \textemdash & (20,21) \\
SN A370-C22 & $0.375$ & $2.93$ & $1.0 \pm 0.5^{\dagger\dagger}$ 	& $31$ & \textemdash & (22) \\
SN H0pe     & $0.348$ & $1.78$ & ${21.1 \pm 0.5}^{\dagger\dagger}$ 	& $83$ & \textemdash & (6,23) \\
\enddata
\tablenotetext{^\dagger}{Median (and maximum) time delays for PASSAGES objects derived in this work (unless otherwise noted), using lens models from the supplied references.  The representative values are simply determined by taking the median/maximum of the delays from the first image in each lensed system to arrive. Estimates are not currently available for PJ1053+60 and PJ1329+22.}
\tablenotetext{^\ddagger}{References for table values, roughly in order of columns (e.g., SN discovery if applicable, followed by redshifts, then $\theta_{\rm Ein}$, and finally time delays). For SN Requiem/Encore, $\theta_{\rm Ein}$ is determined by rescaling $\theta_{\rm Ein}(z\to\infty)$ from \citet{Ertl:2025aa} to $z_s=1.95$.}
\tablenotetext{^{\dagger\dagger}}{Estimated in this work. The Einstein radii for the SNe in Abell 370 and MACS J1149 (Refsdal) are taken as the median $\theta_{\rm Ein}$ derived from the publicly available convergence maps from the Frontier Fields \citep{Lotz:2017aa} lens models at \url{https://archive.stsci.edu/prepds/frontier/lensmodels}\textemdash in particular, the CATS, GLAFIC, and Sharon models (given their similarity to the \lenstool\ models used for the PASSAGES objects).
The range of values are combined with
our own empirical measurements according to the method from \citet{Mork:2025aa}.
For PJ0846+15, $\theta_{\rm Ein}$ and $\widetilde{t_d}$ are estimated for the Einstein cross system, 12abcd \citep{Foo:2025aa}.
} 
\tablecomments{{\bf References}: 
(1) \citealt{Kamieneski:2024aa};
(2) \citealt{Amodeo:2018aa}; 
(3) P. Kamieneski et al., in prep.;
(4) \citealt{Foo:2025aa};
(5) \citealt{Wang:2024ab};
(6) \citealt{Frye:2024aa};
(7) \citealt{Kamieneski:2024ac}; 
(8) \citealt{Diaz-Sanchez:2017aa};
(9) \citealt{Chornock:2013aa};
(10) \citealt{Quimby:2014aa};
(11) \citealt{Kelly:2015aa};
(12) \citealt{Rodney:2016aa};
(13) \citealt{Kelly:2023aa}; 
(14) \citealt{Goobar:2017aa};
(15) \citealt{Diego:2022ad};
(16) \citealt{Dhawan:2020aa};
(17) \citealt{Rodney:2021ab};
(18) \citealt{Pierel:2024ab};
(19) \citealt{Ertl:2025aa};
(20) \citealt{Goobar:2022aa};
(21) \citealt{Pierel:2023aa};
(22) \citealt{Chen:2022ag};
(23) \citealt{Pierel:2024aa}.
}
\end{deluxetable*}

The sample members with the longest predicted median time delays $>100$ days include (in descending order)  
PJ0748+59, 
PJ0846+15 , 
PJ1326+33, 
PJ1127+42, 
PJ1322+09, 
PJ1449+22, 
and
PJ0226+23. 
These are also among the widest-separation targets, as expected.
Except for PJ0226+23, these all also have $\widetilde{t_d} > \tau_{\rm SN}$ and $\mathcal{R}_{\rm CC,obs} + \mathcal{R}_{\rm Ia,obs} > {1}$.
%
%
%
This upper echelon only considers the $\mathcal{R}$ rates and expected time delays\textemdash i.e., not magnification, dust attenuation, or other factors that influence the control time.

Lastly, we note that the vast majority ($90\%$) of objects in the full sample have an expected value of multiplicity, $\mathcal{N}_{\rm eff} > 2$.
This suggests that supernovae hosted by the primary DSFGs in these fields are likely to be multiply-imaged (and not just weakly lensed into a single image), and therefore hypothetically suitable for time-delay analyses\textemdash although $\mathcal{N}_{\rm eff} < 2$ certainly does not mean glSNe are not possible.
In the next section, we discuss contribution from other galaxies in the fields nearby to the DSFGs that might also be lensed into multiple images, but this generally requires a large lensing cross section that can only be attributed to galaxy cluster lenses.

\subsection{Contribution from galaxy-overdense regions}
\label{sec:environments}

Thus far, our estimated SN rates have been limited to the lensed dusty star-forming galaxies that were responsible for the submillimeter flux by which the PASSAGES sample was selected.
However, this does not account for the contributions from other galaxies that might also be gravitationally lensed, especially in the case of cluster-scale lenses with wide caustic networks.
Numerous recent works have revealed that DSFGs, like other very massive galaxies (\citealt{Greve:2005aa}), might be effective signposts of proto-clusters (or at least overdense regions; \citealt{Kauffmann:1999ab, Chiang:2013aa}; reviews by \citealt{Overzier:2016aa, Alberts:2022ab}).
This complements the established association of proto-clusters with over-densities of H$\alpha$ and Lyman-$\alpha$ emitters, High-$z$ Radio Galaxies, and Lyman Break Galaxies (e.g., \citealt{De-Breuck:2004aa, Venemans:2007aa, Miley:2008aa, Daddi:2009aa, Capak:2011ab, Drake:2020aa}).
The rich set of literature with evidence for DSFGs (or over-densities of DSFGs) as proto-cluster signposts includes 
\citet{Riechers:2010aa,  Dannerbauer:2014aa, Casey:2015aa, Umehata:2015aa, Casey:2016aa, Clements:2016aa, Hung:2016aa, Smolcic:2017ab, Lewis:2018aa, Miller:2018ab, Pavesi:2018aa,  Calvi:2023ab}.
Two members of the PASSAGES sample have themselves been highlighted as protocluster candidates (\citealt{Frye:2024aa, Foo:2025aa}).
On the other hand, DSFGs may not be efficient tracers of over-densities at all redshifts.
\citet{Miller:2015aa} suggest that since $z\lesssim 2.5$, the high-mass end of galaxies within the most massive dark matter halos have already quenched their star formation and ceased to be submillimeter-bright.

Semi-analytic modeling by 
\citet{Araya-Araya:2024aa} revealed an excess of SMGs in $z \gtrsim 2$ protocluster cores (relative to protocluster outskirts or field environments). 
The authors concluded that this was driven primarily by oversampling the high-mass end of the halo mass function. 
Fortunately, overdensities of starbursts in protocluster cores 
(\citealt{Oteo:2018aa, Long:2020aa, Champagne:2021aa, Arribas:2024aa})
suggest that they may be confined within regions of $\sim 100 - 500$ kpc in diameter ($\sim 0.2 - 1.0\arcmin$ at $z=3$). They thus can overlap more substantially with the multiply-imaged source-plane regions behind lensing clusters.
For example, \citet{Frye:2024aa} estimated an integrated SFR of  $100-500~{\rm M}_\odot~{\rm yr}^{-1}$ in two overdensities ($z\approx 1.8$ and $z\approx 2.2$) behind the G165 cluster, accounting for $\sim 1 - 3$ CCSNe yr$^{-1}$. 
Perhaps as a proof of concept, the triply-imaged SN H0pe was discovered in a massive host galaxy within the $z\sim 1.8$ overdensity behind G165 \citep{Frye:2023ab, Polletta:2023aa, Frye:2024aa},
through JWST/NIRCam imaging that was primarily targeting the PASSAGES DSFG that is nearby in projection.

\section{Summary}
\label{sec:summary}

So far, discoveries of gravitationally lensed supernovae that can be observed in multiple images have been very limited\textemdash unsurprisingly, as both lensing and supernovae are rare phenomena. 
They are, however, exceptionally useful for the purpose of time-delay cosmography, which depends upon the delays in arrival time between pairs of multiple images (induced by the difference both in path lengths of their respective geodesics and in the gravitational potential that they experience).
Considering the yet-unresolved Hubble tension in different measurements of the Universe's expansion rate, novel and independent methods to constrain $H_0$ are likely to be highly informative.
Yet, fewer than 10 multiply-imaged supernovae have ever been discovered, and only 2 of which have so far been employed successfully for TD cosmography (SN Refsdal and SN H0pe; \citealt{Kelly:2023ab, Pascale:2025aa}). 
There are several advantages to using supernovae for this measurement over, for example, lensed quasars
(e.g., reviews by \citealt{Treu:2022ac, Birrer:2024aa});
chief among them is that the light curve of a supernova is transient, not just variable like quasars, and is capable of being modeled.
For now, though, lensed quasars have a key advantage through being simply more numerous. 
A joint inference of $H_0$, combining information from several lensed supernovae, can be quite competitive with the level of precision achieved for other methods that have so far received much more attention. 
To get to this point in the next decade, however, requires many more discoveries, and it will likely not be sufficient to rely on luck alone.

To this end, we consider the feasibility of a targeted approach focused on galaxies that are expected to be uniquely fruitful for gravitationally lensed supernovae.
In this work, we have attempted to forecast the rates of glSNe that might be reasonably observed for a set of {21} dusty star-forming galaxies from the PASSAGES sample, spanning $z\approx 1.1 - 3.5$.
Using their magnification-corrected intrinsic star formation rates, derived in previous works, we tabulate their rates of core-collapse and Type Ia supernovae, $R_{\rm CC}$ and $R_{\rm Ia}$.
Given the expectation that only a fraction of this star formation is occurring in regions with low enough dust obscuration to ever be detectable, even with the likes of JWST, we apply an assumed unobscured fraction ($\sim 20\%$) to our supernova rates. 
The exact sensitivity and control time will depend on observing setup, but 
\citet{Mantynen:2025aa} found a detectable fraction of $\approx 34\%$ for local LIRGs in an experiment sensitive to $A_V < 16$.
We also include the reduction in observable rates due to the fact that time in these galaxies is slowed by a factor of $1+z$ from the perspective of Earth.
On the other hand, the observable rates are boosted by some factor \Neff\ (capturing the expected value of multiple images that a SN will appear in), based on existing gravitational lens models and the probabilistic distribution of star formation captured by dust and/or radio continuum.

As a result, we estimate a range of intrinsic rates 
$R_{\rm CC} \approx {1.8 - 65}~{\rm yr}^{-1}$ 
(core-collapse; median $11~{\rm yr}^{-1}$) and 
$R_{\rm Ia} \approx {0.2 - 6.4}~{\rm yr}^{-1}$ 
(Type Ia; median $1.1~{\rm yr}^{-1}$), 
with 
observable rates $\mathcal{R}_{\rm CC} \approx {0.4 - 8.0}~{\rm yr}^{-1}$ (median $2.3~{\rm yr}^{-1}$) and $\mathcal{R}_{\rm Ia} \approx {0.2 - 3.9}~{\rm yr}^{-1}$ (median $1.1~{\rm yr}^{-1}$).
As for the time-delayed rates of individual SNe, $R\cdot (1+z)^{-1}$, we find ranges of $0.8 - 20~{\rm yr}^{-1}$ (median $3.5~{\rm yr}^{-1}$) 
and
$0.1 - 2.0~{\rm yr}^{-1}$ (median $0.3~{\rm yr}^{-1}$)
for CCSNe and SNe Ia, respectively.
The number of supernovae that a monitoring campaign would yield is dependent on the control time of the observations, or the duration for which a given supernova would be above survey limits. 
In this regard, the biggest challenge for this particular sample would be ensuring adequate near-IR sensitivity to withstand dust attenuation, effectively necessitating JWST/NIRCam. While we have discussed some evidence that the attenuation might be inhomogeneous enough to allow some relatively unobscured sightlines, this is a source of large uncertainty at the moment. 

Despite this challenge, the population of lensed starbursts is a veritable wealth of supernovae, in the extremely rarefied stratum of galaxies with more than one supernova occurring per (rest-frame) year.
Notwithstanding sheer luck, for almost any other targeted strategy to discover glSNe, it would be necessary to monitor either a small sample for several years, or else a very large sample over a shorter timespan.
Whereas only 2 glSNe have yet delivered time-delay cosmographical measurements (SN Refsdal and SN H0pe, at time of writing), every new discovery has enormous potential to significantly improve the level of uncertainty on $H_0$ inferences and other cosmological parameters (e.g., \citealt{Grillo:2024aa}).
The ultimate goal of the next decade is to reach $\sim 1\%$ precision through joint inference \citep{Coe:2009aa, Treu:2016aa, Treu:2022ac}.

Fortunately, the PASSAGES objects typically have wider separations between lensed images; their Einstein radii satisfy $\theta_{\rm Ein} \gtrsim 0.5\arcsec$ in the vast majority of cases, with some even exceeding 10$\arcsec$. 
We find that a strong correlation between $\theta_{\rm Ein}$ and their median lens model-predicted time delays, which may even have some analytical basis connecting the maximum time delay that can be expected for an axisymmetric isothermal lens and its Einstein radius (Appendix \ref{sec:appendix_td_Re_corr}).
Finally, we compare the expected median time delays to the characteristic observed-frame timescale between supernova events (including both core-collapse and Type Ia). 
We find that PASSAGES occupies an ideal regime where the lensing time delays are not significantly shorter than the time between individual supernovae, such that multiple imaging from lensing helps improve the odds of catching one in a random monitoring program.

While this work forecasts the potential of the $\sim 30$ PASSAGES lensed DSFGs, in part because they account for a non-negligible fraction of the known HyLIRGs,
there are several hundred other candidate lensed DSFGs known currently, and perhaps many thousand more yet to be discovered at slightly lower intrinsic luminosities \citep{Sedgwick:2025aa}.
We contend that an optimal approach would involve targeted searches in tandem with (ongoing and planned) wide-area and blind surveys with {\it Euclid}, Rubin LSST, the {\it China Space Station Telescope}, and the {\it Roman Space Telescope}
(e.g., \citealt{Goldstein:2017aa, Goldstein:2019aa, Wojtak:2019aa, Pierel:2021ab, Ryczanowski:2023aa, Sainz-de-Murieta:2023aa, Arendse:2024ab, Dong:2024ab, Sainz-de-Murieta:2024aa, Bronikowski:2025aa}).

The level of precision in $H_0$ measurements from cluster-lensed supernovae has been of order $6-7\%$ for SN Refsdal \citep{Grillo:2018aa, Kelly:2023ab} and $7-13\%$ for SN H0pe \citep{Pascale:2025aa}.
While it will not be possible to replicate the fully-blinded process used in obtaining these results, there is a distinct possibility in refining these uncertainties. 
For example, the lens models for SN H0pe were constructed with only 5 spectroscopically-confirmed image system redshifts (out of 21), which can have a sizeable influence on model accuracy (e.g., \citealt{Johnson:2016aa}).
Similarly, the SN H0pe inference was subject to unique photometric uncertainties at the time of publication due to the lack of a ``template" image (where the supernova was not visible) of the host galaxy in most filters \citep{Pierel:2024aa}.
The images of SN Encore with the shortest time delays are similarly expected to yield a measurement of order $\approx 10\%$ precision, whereas the longest delays ($>3000$ days) could reduce this to $2-3\%$ (\citealt{Pierel:2024ab}; S. Suyu et al., in prep.; J. D. R. Pierel et al., in prep.).

Within PASSAGES, 5 cluster fields have estimated median time delays exceeding 100 days
and $\theta_{\rm Ein} > 5\arcsec$
(PJ0748+59, 
PJ0846+15 , 
PJ1127+42, 
PJ1322+09, 
and
PJ1449+22) 
and should in principle offer the greatest likelihood of a similar level of precision if glSNe were discovered in each.
A joint population inference for $\sim 20$ lensed supernovae has the potential to reach $\sim 1\%$ level precision on $H_0$ \citep{Suyu:2020aa}, but even a sample of only 6 lensed quasars has been able to reach $\sim 2\%$ relative uncertainty 
\citep{Wong:2020aa}.
The discovery of 5 lensed supernovae suitable for cosmography from a sample like PASSAGES would result in a competitive $H_0$ measurement, shedding much-needed light on the Hubble tension.

\begin{acknowledgments}

P.S.K. would like to thank Haojing Yan, Kate Whitaker, and Kevin Croker
for helpful discussions related to this work.
R.A.W., S.H.C., and R.A.J. acknowledge support from NASA JWST Interdisciplinary Scientist grants NAG5-12460, NNX14AN10G and 80NSSC18K0200 from GSFC.
T.C. acknowledges support from the Beus Center for Cosmic Foundations.
E.F.-J.A. acknowledges support from UNAM-PAPIIT project IA102023, and from CONAHCyT Ciencia de Frontera project ID:  CF-2023-I-506.

This paper makes use of the following ALMA data: 
ADS/JAO.ALMA \#2015.1.01518.S, 
\#2017.1.01214.S,
\#2016.1.00048.S,
and 
\#2019.1.01636.S.
ALMA is a partnership of ESO (representing its member states), NSF (USA) and NINS (Japan), together with NRC (Canada), MOST and ASIAA (Taiwan), and KASI (Republic of Korea), in cooperation with the Republic of Chile. The Joint ALMA Observatory is operated by ESO, AUI/NRAO and NAOJ. The National Radio Astronomy Observatory is a facility of the National Science Foundation operated under cooperative agreement by Associated Universities, Inc.
This research has made extensive use of NASA's Astrophysics Data System.

This work utilizes gravitational lensing models produced by PIs Brada\v{c}, Natarajan \& Kneib (CATS), Merten \& Zitrin, Sharon, Williams, Keeton, Bernstein and Diego, and the GLAFIC group. This lens modeling was partially funded by the HST Frontier Fields program conducted by STScI. STScI is operated by the Association of Universities for Research in Astronomy, Inc. under NASA contract NAS 5-26555. The lens models were obtained from the Mikulski Archive for Space Telescopes (MAST).

We also acknowledge the indigenous peoples of Arizona, including the Akimel O'odham (Pima) and Pee Posh (Maricopa) Indian Communities, whose care and keeping of the land has enabled us to be at ASU's Tempe campus in the Salt River Valley, where much of our work was conducted.

\end{acknowledgments}





%

\facilities{ALMA, VLA}


\software{
          {\sc APLpy} \citep{Robitaille:2012aa,Robitaille:2019aa},
          {\sc astropy} \citep{Astropy-Collaboration:2013aa, Astropy-Collaboration:2018aa, Astropy-Collaboration:2022aa},
          {\sc blobcat} \citep{Hales:2012aa},
          {\sc casa} \citep{McMullin:2007aa},
          \lenstool\ \citep{Kneib:1993aa, Kneib:1996aa, Jullo:2007aa, Jullo:2009aa}, 
          matplotlib \citep{Hunter:2007aa},
          pandas \citep{McKinney:2010aa, The-pandas-development-Team:2024aa},
          Ned Wright's Cosmology Calculator \citep{Wright:2006aa}
          }


\appendix

\section{Estimating Type Ia supernova rates robustly without stellar mass measurements}
\label{sec:appendix_RIa}

For a subset of the PASSAGES sample, \citet{Kamieneski:2024aa} made predictions for the sSFRs relative to the star-forming main sequence of galaxies between SFR and $M_\star$ (e.g., \citealt{Brinchmann:2004aa, Daddi:2007aa, Whitaker:2012aa, Speagle:2014aa}).
\citet{Kamieneski:2024aa} used the objects' gas depletion times ($\sim 150 - 350$ Myr; \citealt{Harrington:2016aa, Harrington:2021aa, Berman:2022aa}) to estimate that they occupied the regime of $\Delta{\rm MS} \equiv {\rm log}_{10}({\rm SFR}/ {\rm SFR}_{\rm MS}) \approx 0.0 - 0.9$, following the prescription of \citealt{Wang:2022ac} (see also \citealt{Genzel:2015aa, Tacconi:2018aa, Liu:2019ae}).
In an independent approach, \citet{Kamieneski:2024aa} used star formation rate surface densities $\Sigma_{\rm SFR}$ to estimate $\Delta{\rm MS}$, as per \citet{Jimenez-Andrade:2019aa}, resulting in a range of $\Delta{\rm MS} \approx -0.1 - 0.5$.

With this information in mind, a sensible upper limit for their sSFRs would be $\Delta{\rm MS} < 0.9$ dex, which happens to be $\approx 3$ times the $1\sigma$ scatter of the main sequence, and also 0.3 dex above the typical threshold for starburst galaxies \citep{Rodighiero:2011aa}.
To further motivate this, we consider some recent measurements of stellar masses in large samples of DSFGs. 
\citet{Miettinen:2017aa} found that $57\%$ of a sample of 124 mm-selected DSFGs in COSMOS were within a factor of 3 ($\approx 0.5$ dex) of the main sequence, and $89\%$ were below $\Delta{\rm MS} < 0.9$ dex.
Similarly, for 185 submm-selected DSFGs in the North Ecliptic Pole (NEP) field, \citet{Barrufet:2020aa} found 57\% within 0.5 dex of the main sequence.
More recently, in a large sample of $\approx$ 450 $z>1$ sources in the NEP selected at 850 $\mu$m, \citet{Shim:2022aa}
found 67\% with $\Delta{\rm MS} < 0.5$ dex and 90\% with $\Delta{\rm MS} < 0.9$ dex.
This finding that DSFGs are largely not dramatically above the main sequence\textemdash with apparently the majority even consistent with the SFMS\textemdash is not particularly surprising.
As discussed by \citet{Casey:2014aa}, 
the duty cycle of DSFGs in the starburst regime ($\Delta_{\rm MS}>0.6$ dex) might be quite short, so that they spend most of their time on the main sequence.

We can use this upper limit in sSFR of 0.9 dex above the main sequence (computed according to \citealt{Speagle:2014aa}),
along with the total SFRs inferred from the IR, 
in order to derive lower limits in $M_\star$. 
We also use a lower limit of sSFR 0.3 dex below the SFMS, giving the upper limits in $M_\star$.
By extension,
these provides lower and upper limits, respectively, in the Ia supernova rate, $R_{\rm Ia}$.
Fortunately, as illustrated in Equation \ref{eqn:RIa}, the dependence on $M_\star$ is quite weak relative to SFR for galaxies on or above the main sequence, so these approximate limits contribute extremely minimally to the error budget in comparison with the uncertainties on SFR.

\section{The connection between typical time delays and image separations}
\label{sec:appendix_td_Re_corr}

\begin{figure*}[th]
\centering
\includegraphics[width=0.495\textwidth]{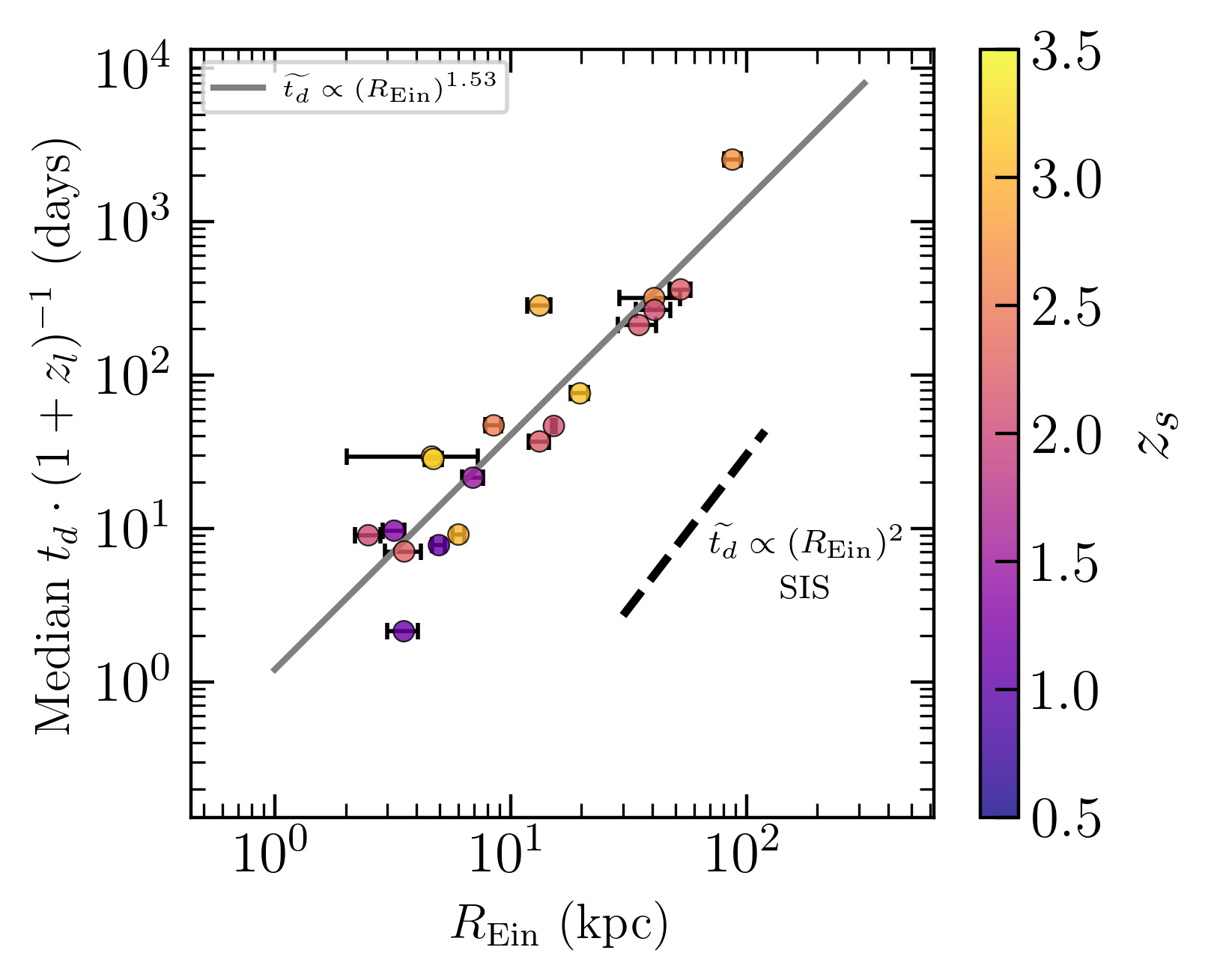}
\includegraphics[width=0.495\textwidth]{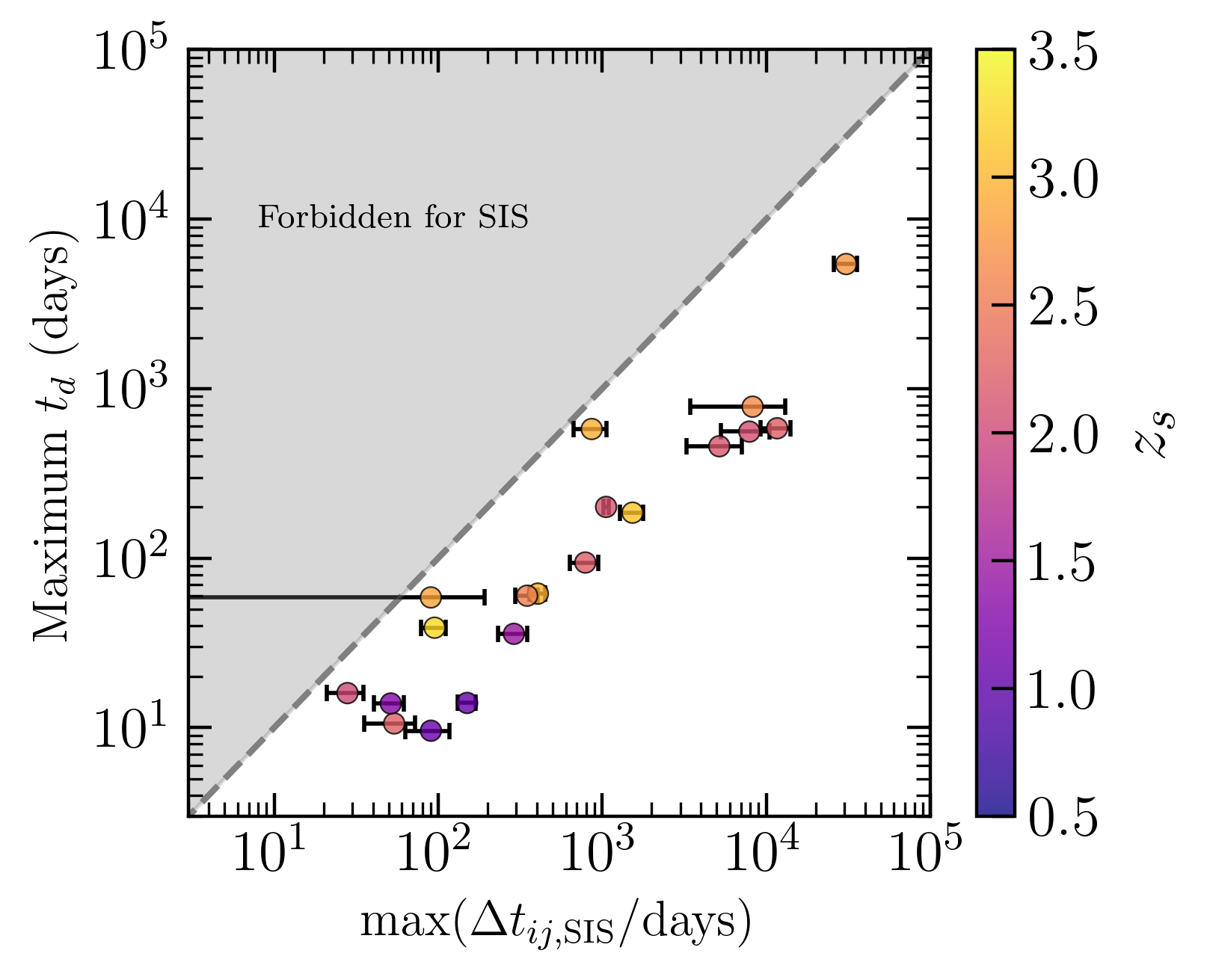}
\caption{
{\it (Left)} Fig.~\ref{fig:R_E} is modified to now plot the median time delays $t_d$ of PASSAGES at the redshift of the lens\textemdash i.e., including a dilation factor of $(1+z_l)^{-1}$\textemdash vs. the physical Einstein radius $R_{\rm Ein} = D_L \cdot \theta_{\rm Ein}$.
A gray line shows the best-fit power law, $\widetilde{t_d} \propto (R_{\rm Ein})^{1.53 \pm 0.16}$, while the dashed line segment shows the approximate scaling relation for an SIS with a power-law slope of 2.
The colorscale shows source-plane redshift.
{\it (Right)} Now, the {\it maximum} time delays vs. the predicted ${\rm max}(\Delta t_{\rm ij})$ computed based on the redshift geometries and $\theta_{\rm Ein}$ according to equation \ref{eqn:tij}.
The quasi-forbidden region (where lens model-predicted maximum time delays exceed the values allowed for an SIS lens at the respective lens/source redshifts with the respective Einstein radii), ${\rm max}(t_d) > {\rm max}(\Delta t_{ij, {\rm SIS}})$ is shown shaded in gray.
}
\label{fig:R_E_kpc}
\end{figure*}

As noted by \citet{Witt:2000ab} and \citet{Oguri:2007aa}, 
a lens with an isothermal potential $\phi(r,\theta) = r \cdot F(\theta)$
for generalized angular structure $F(\theta)$
will result in time delays between arcs $i$ and $j$
\begin{equation}
\Delta t_{ij} = \frac{1+z_l}{2c} \frac{D_L D_S}{D_{LS}} (r_j^2 - r_i^2)
\end{equation}
where $r_i$ is the angular distance from arc $i$ to the center of the lens.
This produces the result that $r_j$ and $r_i$ can be arbitrarily large\textemdash and therefore $\theta_{\rm Ein}$ as well\textemdash and still produce a small time delay, as long as $r_j \approx r_i$.
Also, at the redshift of the lens, the physical separation between deflected light rays is $R_{\rm Ein} = D_L \cdot \theta_{\rm Ein}$.
Considering the maximum possible time delays in this formulation (holding fixed the lens and source redshift),
\begin{equation}
{\rm max}(\Delta t_{ij}) \propto {\rm max}(r_j^2 - r_i^2) \propto (R_{\rm Ein})^2.
\end{equation}
In the left panel of Fig.~\ref{fig:R_E_kpc}, we modify Fig.~\ref{fig:R_E} to show $R_{\rm Ein}$ (in kpc) on the horizontal axis and the median time delay, dilated to the redshift of the lens with a $(1+z_l)^{-1}$ factor.
The best-fit power-law slope for the {\it median} time delays is $\widetilde{t_d} \propto (R_{\rm Ein})^{1.53 \pm 0.16}$, slightly shallower than the prediction for the {\it maximum} time delays.

In at least the specific case of a singular isothermal spheroid (SIS) lens, the two images satisfy 
\begin{equation}
r_j^2 - r_i^2 = (\theta_{\rm Ein} + \beta)^2 - (\theta_{\rm Ein} - \beta)^2 = 4 \theta_{\rm Ein} \beta
\end{equation}
for source-plane offset $\beta$ relative to the lens centroid \citep{Narayan:1996aa}. By necessity, $\beta < \theta_{\rm Ein}$, or else only a single image appears.
Therefore,
\begin{equation}
{\rm max}(r_j^2 - r_i^2) = 4\theta_{\rm Ein}^2
\end{equation}
and
\begin{equation}
\label{eqn:tij}
{\rm max}(\Delta t_{ij}) = \frac{1+z_l}{2c} \frac{D_L D_S}{D_{LS}} \cdot 4\theta_{\rm Ein}^2
\end{equation}
for an SIS lens, with $\theta_{\rm Ein}$ in radians.
The right panel of Fig.~\ref{fig:R_E_kpc} plots this term on the horizontal axis, with the lens model-predicted maximum time delays (Table~\ref{tab:td_Ein}) on the vertical axis. A quasi-forbidden region is shaded in gray, representing model-predicted time delays that exceed the maximum from the \citet{Witt:2000ab} and \citet{Oguri:2007aa} equation in the case of an SIS. 
We observe that the maximum model-predicted time delays $t_d$ remain below ${\rm max}(\Delta t_{ij})$.

Given the limiting assumptions implicit to this formulation (mainly that the lensing mass is consistent with a single SIS lens), this connection between time delays and Einstein radii will not hold in all cases.
Still, it is reasonably useful for our purposes in identifying lenses that are more likely to have time delays of order months to years. 
By equation \ref{eqn:tij}, for a fiducial SIS lens at $z_l=0.5$ and source at $z_s=2$, ${\rm max}(\Delta t_{ij}) < 30$ days for $\theta_{\rm Ein} \lesssim 0.4\arcsec$, while
${\rm max}(\Delta t_{ij}) < 100$ days for $\theta_{\rm Ein} \lesssim 0.8\arcsec$. 
Only $\sim 29\%$ of the PASSAGES subsample have $\theta_{\rm Ein} < 0.8\arcsec$ (Table~\ref{tab:td_Ein}), while none considered have $\theta_{\rm Ein} < 0.4\arcsec$, so their time delays are more likely to be longer, which generally yields improved $H_0$ precision.




\begin{thebibliography}{}
\expandafter\ifx\csname natexlab\endcsname\relax\def\natexlab#1{#1}\fi
\providecommand{\url}[1]{\href{#1}{#1}}
\providecommand{\dodoi}[1]{doi:~\href{http://doi.org/#1}{\nolinkurl{#1}}}
\providecommand{\doeprint}[1]{\href{http://ascl.net/#1}{\nolinkurl{http://ascl.net/#1}}}
\providecommand{\doarXiv}[1]{\href{https://arxiv.org/abs/#1}{\nolinkurl{https://arxiv.org/abs/#1}}}

\bibitem[{S. {Aalto} {et~al.}(2015){Aalto}, {Mart{\'\i}n}, {Costagliola},
  {Gonz{\'a}lez-Alfonso}, {Muller}, {Sakamoto}, {Fuller},
  {Garc{\'\i}a-Burillo}, {van der Werf}, {Neri}, {Spaans}, {Combes}, {Viti},
  {M{\"u}hle}, {Armus}, {Evans}, {Sturm}, {Cernicharo}, {Henkel}, \&
  {Greve}}]{Aalto:2015aa}
{Aalto}, S., {Mart{\'\i}n}, S., {Costagliola}, F., {et~al.} 2015,
  \bibinfo{title}{{Probing highly obscured, self-absorbed galaxy nuclei with
  vibrationally excited HCN},} \aap, 584, A42,
  \dodoi{10.1051/0004-6361/201526410}

\bibitem[{ {Abdurro'uf} {et~al.}(2023){Abdurro'uf}, {Coe}, {Jung}, {Ferguson},
  {Brammer}, {Iyer}, {Bradley}, {Dayal}, {Windhorst}, {Zitrin}, {Meena},
  {Oguri}, {Diego}, {Kokorev}, {Dimauro}, {Adamo}, {Conselice}, {Welch},
  {Vanzella}, {Hsiao}, {Xu}, {Roy}, \& {Mulcahey}}]{Abdurrouf:2023aa}
{Abdurro'uf}, {Coe}, D., {Jung}, I., {et~al.} 2023, \bibinfo{title}{{Spatially
  Resolved Stellar Populations of 0.3 < z < 6.0 Galaxies in WHL 0137-08 and
  MACS 0647+70 Clusters as Revealed by JWST: How Do Galaxies Grow and Quench
  over Cosmic Time?},} \apj, 945, 117, \dodoi{10.3847/1538-4357/acba06}

\bibitem[{S. {Aiola} {et~al.}(2020){Aiola}, {Calabrese}, {Maurin}, {Naess},
  {Schmitt}, {Abitbol}, {Addison}, {Ade}, {Alonso}, {Amiri}, {Amodeo},
  {Angile}, {Austermann}, {Baildon}, {Battaglia}, {Beall}, {Bean}, {Becker},
  {Bond}, {Bruno}, {Calafut}, {Campusano}, {Carrero}, {Chesmore}, {Cho},
  {Choi}, {Clark}, {Cothard}, {Crichton}, {Crowley}, {Darwish}, {Datta},
  {Denison}, {Devlin}, {Duell}, {Duff}, {Duivenvoorden}, {Dunkley},
  {D{\"u}nner}, {Essinger-Hileman}, {Fankhanel}, {Ferraro}, {Fox}, {Fuzia},
  {Gallardo}, {Gluscevic}, {Golec}, {Grace}, {Gralla}, {Guan}, {Hall},
  {Halpern}, {Han}, {Hargrave}, {Hasselfield}, {Helton}, {Henderson},
  {Hensley}, {Hill}, {Hilton}, {Hilton}, {Hincks}, {Hlo{\v{z}}ek}, {Ho},
  {Hubmayr}, {Huffenberger}, {Hughes}, {Infante}, {Irwin}, {Jackson}, {Klein},
  {Knowles}, {Koopman}, {Kosowsky}, {Lakey}, {Li}, {Li}, {Li}, {Lokken},
  {Louis}, {Lungu}, {MacInnis}, {Madhavacheril}, {Maldonado}, {Mallaby-Kay},
  {Marsden}, {McMahon}, {Menanteau}, {Moodley}, {Morton}, {Namikawa}, {Nati},
  {Newburgh}, {Nibarger}, {Nicola}, {Niemack}, {Nolta}, {Orlowski-Sherer},
  {Page}, {Pappas}, {Partridge}, {Phakathi}, {Pisano}, {Prince}, {Puddu}, {Qu},
  {Rivera}, {Robertson}, {Rojas}, {Salatino}, {Schaan}, {Schillaci}, {Sehgal},
  {Sherwin}, {Sierra}, {Sievers}, {Sifon}, {Sikhosana}, {Simon}, {Spergel},
  {Staggs}, {Stevens}, {Storer}, {Sunder}, {Switzer}, {Thorne}, {Thornton},
  {Trac}, {Treu}, {Tucker}, {Vale}, {Van Engelen}, {Van Lanen}, {Vavagiakis},
  {Wagoner}, {Wang}, {Ward}, {Wollack}, {Xu}, {Zago}, \& {Zhu}}]{Aiola:2020aa}
{Aiola}, S., {Calabrese}, E., {Maurin}, L., {et~al.} 2020, \bibinfo{title}{{The
  Atacama Cosmology Telescope: DR4 maps and cosmological parameters},} \jcap,
  2020, 047, \dodoi{10.1088/1475-7516/2020/12/047}

\bibitem[{H.~B. {Akins} {et~al.}(2025){Akins}, {Casey}, {Champagne}, {Cooper},
  {Franco}, {Fujimoto}, {Knudsen}, {Koekemoer}, {Long}, {Man}, {Manning},
  {McKinney}, {Zavala}, {Arrabal Haro}, {Dickinson}, {Kokorev}, \&
  {Taylor}}]{Akins:2025ab}
{Akins}, H.~B., {Casey}, C.~M., {Champagne}, J.~B., {et~al.} 2025,
  \bibinfo{title}{{JWST+ALMA reveal the ISM kinematics and stellar structure of
  MAMBO-9, a merging pair of DSFGs in an overdense environment at $z=5.85$},}
  arXiv e-prints, arXiv:2508.06607.
\newblock \doarXiv{2508.06607}

\bibitem[{S. {Alberts} \& A. {Noble}(2022){Alberts} \&
  {Noble}}]{Alberts:2022ab}
{Alberts}, S., \& {Noble}, A. 2022, \bibinfo{title}{{From Clusters to
  Proto-Clusters: The Infrared Perspective on Environmental Galaxy Evolution},}
  Universe, 8, 554, \dodoi{10.3390/universe8110554}

\bibitem[{S. {Amodeo} {et~al.}(2018){Amodeo}, {Mei}, {Stanford}, {Lawrence},
  {Bartlett}, {Stern}, {Chary}, {Shim}, {Marleau}, {Melin}, \&
  {Rodr{\'\i}guez-Gonz{\'a}lvez}}]{Amodeo:2018aa}
{Amodeo}, S., {Mei}, S., {Stanford}, S.~A., {et~al.} 2018,
  \bibinfo{title}{{Spectroscopic Confirmation and Velocity Dispersions for 20
  Planck Galaxy Clusters at 0.16 < z < 0.78},} \apj, 853, 36,
  \dodoi{10.3847/1538-4357/aa98dd}

\bibitem[{G.~S. {Anand} {et~al.}(2024){Anand}, {Riess}, {Yuan}, {Beaton},
  {Casertano}, {Li}, {Makarov}, {Makarova}, {Tully}, {Anderson}, {Breuval},
  {Dolphin}, {Karachentsev}, {Macri}, \& {Scolnic}}]{Anand:2024aa}
{Anand}, G.~S., {Riess}, A.~G., {Yuan}, W., {et~al.} 2024, \bibinfo{title}{{Tip
  of the Red Giant Branch Distances with JWST: An Absolute Calibration in NGC
  4258 and First Applications to Type Ia Supernova Hosts},} \apj, 966, 89,
  \dodoi{10.3847/1538-4357/ad2e0a}

\bibitem[{P. {Andersen} \& J. {Hjorth}(2018){Andersen} \&
  {Hjorth}}]{Andersen:2018aa}
{Andersen}, P., \& {Hjorth}, J. 2018, \bibinfo{title}{{Reconciling volumetric
  and individual galaxy type Ia supernova rates},} \mnras, 480, 68,
  \dodoi{10.1093/mnras/sty1837}

\bibitem[{P. {Araya-Araya} {et~al.}(2024){Araya-Araya}, {Cochrane}, {Hayward},
  {Yates}, {Sodr{\'e}}, {Vicentin}, {Rennehan}, {Overzier}, \& {van
  Daalen}}]{Araya-Araya:2024aa}
{Araya-Araya}, P., {Cochrane}, R.~K., {Hayward}, C.~C., {et~al.} 2024,
  \bibinfo{title}{{Modeling the Multiwavelength Detection of Protoclusters. I.
  An Excess of Submillimeter Galaxies in Protocluster Cores},} \apj, 977, 204,
  \dodoi{10.3847/1538-4357/ad90ae}

\bibitem[{N. {Arendse} {et~al.}(2024){Arendse}, {Dhawan}, {Sagu{\'e}s
  Carracedo}, {Peiris}, {Goobar}, {Wojtak}, {Alves}, {Biswas}, {Huber},
  {Birrer}, \& {The LSST Dark Enrgy Science Collaboration}}]{Arendse:2024ab}
{Arendse}, N., {Dhawan}, S., {Sagu{\'e}s Carracedo}, A., {et~al.} 2024,
  \bibinfo{title}{{Detecting strongly lensed type Ia supernovae with LSST},}
  \mnras, 531, 3509, \dodoi{10.1093/mnras/stae1356}

\bibitem[{S. {Arribas} {et~al.}(2024){Arribas}, {Perna}, {Rodr{\'\i}guez Del
  Pino}, {Lamperti}, {D'Eugenio}, {P{\'e}rez-Gonz{\'a}lez}, {Jones}, {Crespo
  G{\'o}mez}, {Curti}, {Lim}, {{\'A}lvarez-M{\'a}rquez}, {Bunker}, {Carniani},
  {Charlot}, {Jakobsen}, {Maiolino}, {{\"U}bler}, {Willott}, {B{\"o}ker},
  {Chevallard}, {Circosta}, {Cresci}, {Kumari}, {Parlanti}, {Scholtz},
  {Venturi}, \& {Witstok}}]{Arribas:2024aa}
{Arribas}, S., {Perna}, M., {Rodr{\'\i}guez Del Pino}, B., {et~al.} 2024,
  \bibinfo{title}{{GA-NIFS: The core of an extremely massive protocluster at
  the epoch of reionisation probed with JWST/NIRSpec},} \aap, 688, A146,
  \dodoi{10.1051/0004-6361/202348824}

\bibitem[{ {Astropy Collaboration} {et~al.}(2013){Astropy Collaboration},
  {Robitaille}, {Tollerud}, {Greenfield}, {Droettboom}, {Bray}, {Aldcroft},
  {Davis}, {Ginsburg}, {Price-Whelan}, {Kerzendorf}, {Conley}, {Crighton},
  {Barbary}, {Muna}, {Ferguson}, {Grollier}, {Parikh}, {Nair}, {Unther},
  {Deil}, {Woillez}, {Conseil}, {Kramer}, {Turner}, {Singer}, {Fox}, {Weaver},
  {Zabalza}, {Edwards}, {Azalee Bostroem}, {Burke}, {Casey}, {Crawford},
  {Dencheva}, {Ely}, {Jenness}, {Labrie}, {Lim}, {Pierfederici}, {Pontzen},
  {Ptak}, {Refsdal}, {Servillat}, \&
  {Streicher}}]{Astropy-Collaboration:2013aa}
{Astropy Collaboration}, {Robitaille}, T.~P., {Tollerud}, E.~J., {et~al.} 2013,
  \bibinfo{title}{{Astropy: A community Python package for astronomy},} \aap,
  558, A33, \dodoi{10.1051/0004-6361/201322068}

\bibitem[{ {Astropy Collaboration} {et~al.}(2018){Astropy Collaboration},
  {Price-Whelan}, {Sip{\H{o}}cz}, {G{\"u}nther}, {Lim}, {Crawford}, {Conseil},
  {Shupe}, {Craig}, {Dencheva}, {Ginsburg}, {VanderPlas}, {Bradley},
  {P{\'e}rez-Su{\'a}rez}, {de Val-Borro}, {Aldcroft}, {Cruz}, {Robitaille},
  {Tollerud}, {Ardelean}, {Babej}, {Bach}, {Bachetti}, {Bakanov}, {Bamford},
  {Barentsen}, {Barmby}, {Baumbach}, {Berry}, {Biscani}, {Boquien}, {Bostroem},
  {Bouma}, {Brammer}, {Bray}, {Breytenbach}, {Buddelmeijer}, {Burke},
  {Calderone}, {Cano Rodr{\'\i}guez}, {Cara}, {Cardoso}, {Cheedella}, {Copin},
  {Corrales}, {Crichton}, {D'Avella}, {Deil}, {Depagne}, {Dietrich}, {Donath},
  {Droettboom}, {Earl}, {Erben}, {Fabbro}, {Ferreira}, {Finethy}, {Fox},
  {Garrison}, {Gibbons}, {Goldstein}, {Gommers}, {Greco}, {Greenfield},
  {Groener}, {Grollier}, {Hagen}, {Hirst}, {Homeier}, {Horton}, {Hosseinzadeh},
  {Hu}, {Hunkeler}, {Ivezi{\'c}}, {Jain}, {Jenness}, {Kanarek}, {Kendrew},
  {Kern}, {Kerzendorf}, {Khvalko}, {King}, {Kirkby}, {Kulkarni}, {Kumar},
  {Lee}, {Lenz}, {Littlefair}, {Ma}, {Macleod}, {Mastropietro}, {McCully},
  {Montagnac}, {Morris}, {Mueller}, {Mumford}, {Muna}, {Murphy}, {Nelson},
  {Nguyen}, {Ninan}, {N{\"o}the}, {Ogaz}, {Oh}, {Parejko}, {Parley}, {Pascual},
  {Patil}, {Patil}, {Plunkett}, {Prochaska}, {Rastogi}, {Reddy Janga},
  {Sabater}, {Sakurikar}, {Seifert}, {Sherbert}, {Sherwood-Taylor}, {Shih},
  {Sick}, {Silbiger}, {Singanamalla}, {Singer}, {Sladen}, {Sooley},
  {Sornarajah}, {Streicher}, {Teuben}, {Thomas}, {Tremblay}, {Turner},
  {Terr{\'o}n}, {van Kerkwijk}, {de la Vega}, {Watkins}, {Weaver}, {Whitmore},
  {Woillez}, {Zabalza}, \& {Astropy
  Contributors}}]{Astropy-Collaboration:2018aa}
{Astropy Collaboration}, {Price-Whelan}, A.~M., {Sip{\H{o}}cz}, B.~M., {et~al.}
  2018, \bibinfo{title}{{The Astropy Project: Building an Open-science Project
  and Status of the v2.0 Core Package},} \aj, 156, 123,
  \dodoi{10.3847/1538-3881/aabc4f}

\bibitem[{ {Astropy Collaboration} {et~al.}(2022){Astropy Collaboration},
  {Price-Whelan}, {Lim}, {Earl}, {Starkman}, {Bradley}, {Shupe}, {Patil},
  {Corrales}, {Brasseur}, {N{\"o}the}, {Donath}, {Tollerud}, {Morris},
  {Ginsburg}, {Vaher}, {Weaver}, {Tocknell}, {Jamieson}, {van Kerkwijk},
  {Robitaille}, {Merry}, {Bachetti}, {G{\"u}nther}, {Aldcroft},
  {Alvarado-Montes}, {Archibald}, {B{\'o}di}, {Bapat}, {Barentsen},
  {Baz{\'a}n}, {Biswas}, {Boquien}, {Burke}, {Cara}, {Cara}, {Conroy},
  {Conseil}, {Craig}, {Cross}, {Cruz}, {D'Eugenio}, {Dencheva}, {Devillepoix},
  {Dietrich}, {Eigenbrot}, {Erben}, {Ferreira}, {Foreman-Mackey}, {Fox},
  {Freij}, {Garg}, {Geda}, {Glattly}, {Gondhalekar}, {Gordon}, {Grant},
  {Greenfield}, {Groener}, {Guest}, {Gurovich}, {Handberg}, {Hart},
  {Hatfield-Dodds}, {Homeier}, {Hosseinzadeh}, {Jenness}, {Jones}, {Joseph},
  {Kalmbach}, {Karamehmetoglu}, {Ka{\l}uszy{\'n}ski}, {Kelley}, {Kern},
  {Kerzendorf}, {Koch}, {Kulumani}, {Lee}, {Ly}, {Ma}, {MacBride}, {Maljaars},
  {Muna}, {Murphy}, {Norman}, {O'Steen}, {Oman}, {Pacifici}, {Pascual},
  {Pascual-Granado}, {Patil}, {Perren}, {Pickering}, {Rastogi}, {Roulston},
  {Ryan}, {Rykoff}, {Sabater}, {Sakurikar}, {Salgado}, {Sanghi}, {Saunders},
  {Savchenko}, {Schwardt}, {Seifert-Eckert}, {Shih}, {Jain}, {Shukla}, {Sick},
  {Simpson}, {Singanamalla}, {Singer}, {Singhal}, {Sinha}, {Sip{\H{o}}cz},
  {Spitler}, {Stansby}, {Streicher}, {{\v{S}}umak}, {Swinbank}, {Taranu},
  {Tewary}, {Tremblay}, {de Val-Borro}, {Van Kooten}, {Vasovi{\'c}}, {Verma},
  {de Miranda Cardoso}, {Williams}, {Wilson}, {Winkel}, {Wood-Vasey}, {Xue},
  {Yoachim}, {Zhang}, {Zonca}, \& {Astropy Project
  Contributors}}]{Astropy-Collaboration:2022aa}
{Astropy Collaboration}, {Price-Whelan}, A.~M., {Lim}, P.~L., {et~al.} 2022,
  \bibinfo{title}{{The Astropy Project: Sustaining and Growing a
  Community-oriented Open-source Project and the Latest Major Release (v5.0) of
  the Core Package},} \apj, 935, 167, \dodoi{10.3847/1538-4357/ac7c74}

\bibitem[{S. {Bag} {et~al.}(2024){Bag}, {Huber}, {Suyu}, {Arendse}, {Andika},
  {Ca{\~n}ameras}, {Kim}, {Linder}, {Lodha}, {Melo}, {More}, {Schuldt}, \&
  {Shafieloo}}]{Bag:2024aa}
{Bag}, S., {Huber}, S., {Suyu}, S.~H., {et~al.} 2024,
  \bibinfo{title}{{Detecting unresolved lensed SNe Ia in LSST using blended
  light curves},} \aap, 691, A100, \dodoi{10.1051/0004-6361/202450485}

\bibitem[{W.~M. {Baker} {et~al.}(2022){Baker}, {Maiolino}, {Bluck}, {Lin},
  {Ellison}, {Belfiore}, {Pan}, \& {Thorp}}]{Baker:2022aa}
{Baker}, W.~M., {Maiolino}, R., {Bluck}, A. F.~L., {et~al.} 2022,
  \bibinfo{title}{{The ALMaQUEST survey IX: the nature of the resolved star
  forming main sequence},} \mnras, 510, 3622, \dodoi{10.1093/mnras/stab3672}

\bibitem[{T.~J.~L.~C. {Bakx} {et~al.}(2024{\natexlab{a}}){Bakx}, {Gray},
  {Gonz{\'a}lez-Nuevo}, {Bonavera}, {Amvrosiadis}, {Eales}, {Hagimoto}, \&
  {Serjeant}}]{Bakx:2024aa}
{Bakx}, T. J.~L.~C., {Gray}, B.~S., {Gonz{\'a}lez-Nuevo}, J., {et~al.}
  2024{\natexlab{a}}, \bibinfo{title}{{FLASH: Faint Lenses from Associated
  Selection with Herschel},} \mnras, 527, 8865, \dodoi{10.1093/mnras/stad3759}

\bibitem[{T.~J.~L.~C. {Bakx} {et~al.}(2018){Bakx}, {Eales}, {Negrello},
  {Smith}, {Valiante}, {Holland}, {Baes}, {Bourne}, {Clements}, {Dannerbauer},
  {De Zotti}, {Dunne}, {Dye}, {Furlanetto}, {Ivison}, {Maddox}, {Marchetti},
  {Micha{\l}owski}, {Omont}, {Oteo}, {Wardlow}, {van der Werf}, \&
  {Yang}}]{Bakx:2018aa}
{Bakx}, T. J.~L.~C., {Eales}, S.~A., {Negrello}, M., {et~al.} 2018,
  \bibinfo{title}{{The Herschel Bright Sources (HerBS): sample definition and
  SCUBA-2 observations},} \mnras, 473, 1751, \dodoi{10.1093/mnras/stx2267}

\bibitem[{T.~J.~L.~C. {Bakx} {et~al.}(2024{\natexlab{b}}){Bakx}, {Amvrosiadis},
  {Bendo}, {Algera}, {Serjeant}, {Bonavera}, {Borsato}, {Chen}, {Cox},
  {Gonz{\'a}lez-Nuevo}, {Hagimoto}, {Harrington}, {Ivison}, {Kamieneski},
  {Marchetti}, {Riechers}, {Tsukui}, {van der Werf}, {Yang}, {Zavala},
  {Andreani}, {Berta}, {Cooray}, {De Zotti}, {Eales}, {Ikeda}, {Knudsen},
  {Mitsuhashi}, {Negrello}, {Neri}, {Omont}, {Scott}, {Tamura}, {Temi}, \&
  {Urquhart}}]{Bakx:2024ad}
{Bakx}, T.~J.~L.~C., {Amvrosiadis}, A., {Bendo}, G.~J., {et~al.}
  2024{\natexlab{b}}, \bibinfo{title}{{A novel high-z submm galaxy efficient
  line survey in ALMA Bands 3 through 8 - an ANGELS pilot},} \mnras, 535, 1533,
  \dodoi{10.1093/mnras/stae2409}

\bibitem[{L. {Barrufet} {et~al.}(2020){Barrufet}, {Pearson}, {Serjeant},
  {Ma{\l}ek}, {Baronchelli}, {Campos-Varillas}, {White}, {Valtchanov},
  {Matsuhara}, {Conversi}, {Kim}, {Goto}, {Oi}, {Malkan}, {Kim}, {Ikeda},
  {Takagi}, {Toba}, \& {Miyaji}}]{Barrufet:2020aa}
{Barrufet}, L., {Pearson}, C., {Serjeant}, S., {et~al.} 2020,
  \bibinfo{title}{{A high redshift population of galaxies at the North Ecliptic
  Pole. Unveiling the main sequence of dusty galaxies},} \aap, 641, A129,
  \dodoi{10.1051/0004-6361/202037838}

\bibitem[{E. {Bellm}(2014){Bellm}}]{Bellm:2014aa}
{Bellm}, E. 2014, \bibinfo{title}{{The Zwicky Transient Facility},} in The
  Third Hot-wiring the Transient Universe Workshop, ed. P.~R. {Wozniak}, M.~J.
  {Graham}, A.~A. {Mahabal}, \& R.~{Seaman}, 27--33,
  \dodoi{10.48550/arXiv.1410.8185}

\bibitem[{D.~A. {Berman} {et~al.}(2022){Berman}, {Yun}, {Harrington},
  {Kamieneski}, {Lowenthal}, {Frye}, {Wang}, {Wilson}, {Aretxaga}, {Chavez},
  {Cybulski}, {De la Luz}, {Erickson}, {Ferrusca}, {Hughes}, {Monta{\~n}a},
  {Narayanan}, {S{\'a}nchez-Arg{\"u}elles}, {Schloerb}, {Souccar}, {Terlevich},
  {Terlevich}, \& {Zavala}}]{Berman:2022aa}
{Berman}, D.~A., {Yun}, M.~S., {Harrington}, K.~C., {et~al.} 2022,
  \bibinfo{title}{{PASSAGES: the Large Millimeter Telescope and ALMA
  observations of extremely luminous high-redshift galaxies identified by the
  Planck},} \mnras, 515, 3911, \dodoi{10.1093/mnras/stac1494}

\bibitem[{A. {Bik} {et~al.}(2018){Bik}, {{\"O}stlin}, {Menacho}, {Adamo},
  {Hayes}, {Herenz}, \& {Melinder}}]{Bik:2018aa}
{Bik}, A., {{\"O}stlin}, G., {Menacho}, V., {et~al.} 2018,
  \bibinfo{title}{{Super star cluster feedback driving ionization, shocks and
  outflows in the halo of the nearby starburst ESO 338-IG04},} \aap, 619, A131,
  \dodoi{10.1051/0004-6361/201833916}

\bibitem[{S. {Birrer} {et~al.}(2020){Birrer}, {Shajib}, {Galan}, {Millon},
  {Treu}, {Agnello}, {Auger}, {Chen}, {Christensen}, {Collett}, {Courbin},
  {Fassnacht}, {Koopmans}, {Marshall}, {Park}, {Rusu}, {Sluse}, {Spiniello},
  {Suyu}, {Wagner-Carena}, {Wong}, {Barnab{\`e}}, {Bolton}, {Czoske}, {Ding},
  {Frieman}, \& {Van de Vyvere}}]{Birrer:2020ab}
{Birrer}, S., {Shajib}, A.~J., {Galan}, A., {et~al.} 2020,
  \bibinfo{title}{{TDCOSMO. IV. Hierarchical time-delay cosmography - joint
  inference of the Hubble constant and galaxy density profiles},} \aap, 643,
  A165, \dodoi{10.1051/0004-6361/202038861}

\bibitem[{S. {Birrer} {et~al.}(2024){Birrer}, {Millon}, {Sluse}, {Shajib},
  {Courbin}, {Erickson}, {Koopmans}, {Suyu}, \& {Treu}}]{Birrer:2024aa}
{Birrer}, S., {Millon}, M., {Sluse}, D., {et~al.} 2024,
  \bibinfo{title}{{Time-Delay Cosmography: Measuring the Hubble Constant and
  Other Cosmological Parameters with Strong Gravitational Lensing},} \ssr, 220,
  48, \dodoi{10.1007/s11214-024-01079-w}

\bibitem[{A.~W. {Blain}(1996){Blain}}]{Blain:1996aa}
{Blain}, A.~W. 1996, \bibinfo{title}{{Galaxy-galaxy gravitational lensing in
  the millimetre/submillimetre waveband},} \mnras, 283, 1340,
  \dodoi{10.1093/mnras/283.4.1340}

\bibitem[{R.~D. {Blandford} \& R. {Narayan}(1992){Blandford} \&
  {Narayan}}]{Blandford:1992aa}
{Blandford}, R.~D., \& {Narayan}, R. 1992, \bibinfo{title}{{Cosmological
  applications of gravitational lensing},} \araa, 30, 311,
  \dodoi{10.1146/annurev.aa.30.090192.001523}

\bibitem[{M. {Bondi} {et~al.}(2012){Bondi}, {P{\'e}rez-Torres},
  {Herrero-Illana}, \& {Alberdi}}]{Bondi:2012aa}
{Bondi}, M., {P{\'e}rez-Torres}, M.~A., {Herrero-Illana}, R., \& {Alberdi}, A.
  2012, \bibinfo{title}{{The nuclear starburst in Arp 299-A: from the 5.0 GHz
  VLBI radio light-curves to its core-collapse supernova rate},} \aap, 539,
  A134, \dodoi{10.1051/0004-6361/201118446}

\bibitem[{V. {Bonvin} {et~al.}(2017){Bonvin}, {Courbin}, {Suyu}, {Marshall},
  {Rusu}, {Sluse}, {Tewes}, {Wong}, {Collett}, {Fassnacht}, {Treu}, {Auger},
  {Hilbert}, {Koopmans}, {Meylan}, {Rumbaugh}, {Sonnenfeld}, \&
  {Spiniello}}]{Bonvin:2017aa}
{Bonvin}, V., {Courbin}, F., {Suyu}, S.~H., {et~al.} 2017,
  \bibinfo{title}{{H0LiCOW - V. New COSMOGRAIL time delays of HE 0435-1223:
  H$_{0}$ to 3.8 per cent precision from strong lensing in a flat
  {$\Lambda$}CDM model},} \mnras, 465, 4914, \dodoi{10.1093/mnras/stw3006}

\bibitem[{M. {Boquien} {et~al.}(2009){Boquien}, {Calzetti}, {Kennicutt},
  {Dale}, {Engelbracht}, {Gordon}, {Hong}, {Lee}, \&
  {Portouw}}]{Boquien:2009aa}
{Boquien}, M., {Calzetti}, D., {Kennicutt}, R., {et~al.} 2009,
  \bibinfo{title}{{Star-Forming or Starbursting? The Ultraviolet Conundrum},}
  \apj, 706, 553, \dodoi{10.1088/0004-637X/706/1/553}

\bibitem[{E. {Borsato} {et~al.}(2024){Borsato}, {Marchetti}, {Negrello},
  {Corsini}, {Wake}, {Amvrosiadis}, {Baker}, {Bakx}, {Beelen}, {Berta},
  {Beyer}, {Clements}, {Cooray}, {Cox}, {Dannerbauer}, {de Zotti}, {Dye},
  {Eales}, {Enia}, {Farrah}, {Gonzalez-Nuevo}, {Hughes}, {Ismail}, {Jin},
  {Lapi}, {Lehnert}, {Neri}, {P{\'e}rez-Fournon}, {Riechers}, {Rodighiero},
  {Scott}, {Serjeant}, {Stanley}, {Urquhart}, {van der Werf}, {Vaccari},
  {Wang}, {Yang}, \& {Young}}]{Borsato:2024aa}
{Borsato}, E., {Marchetti}, L., {Negrello}, M., {et~al.} 2024,
  \bibinfo{title}{{Characterization of Herschel-selected strong lens candidates
  through HST and sub-mm/mm observations},} \mnras, 528, 6222,
  \dodoi{10.1093/mnras/stad3381}

\bibitem[{F. {Bournaud} {et~al.}(2007){Bournaud}, {Elmegreen}, \&
  {Elmegreen}}]{Bournaud:2007aa}
{Bournaud}, F., {Elmegreen}, B.~G., \& {Elmegreen}, D.~M. 2007,
  \bibinfo{title}{{Rapid Formation of Exponential Disks and Bulges at High
  Redshift from the Dynamical Evolution of Clump-Cluster and Chain Galaxies},}
  \apj, 670, 237, \dodoi{10.1086/522077}

\bibitem[{J. {Brinchmann} {et~al.}(2004){Brinchmann}, {Charlot}, {White},
  {Tremonti}, {Kauffmann}, {Heckman}, \& {Brinkmann}}]{Brinchmann:2004aa}
{Brinchmann}, J., {Charlot}, S., {White}, S.~D.~M., {et~al.} 2004,
  \bibinfo{title}{{The physical properties of star-forming galaxies in the
  low-redshift Universe},} \mnras, 351, 1151,
  \dodoi{10.1111/j.1365-2966.2004.07881.x}

\bibitem[{M. {Bronikowski} {et~al.}(2025){Bronikowski}, {Petrushevska},
  {Pierel}, {Acebron}, {Donevski}, {Apostolova}, {Blagorodnova}, \&
  {Jankovi{\v{c}}}}]{Bronikowski:2025aa}
{Bronikowski}, M., {Petrushevska}, T., {Pierel}, J.~D.~R., {et~al.} 2025,
  \bibinfo{title}{{Cluster-lensed supernova yields from the Vera C. Rubin
  Observatory and Nancy Grace Roman Space Telescope},} arXiv e-prints,
  arXiv:2504.01068.
\newblock \doarXiv{2504.01068}

\bibitem[{R.~S. {Bussmann} {et~al.}(2013){Bussmann}, {P{\'e}rez-Fournon},
  {Amber}, {Calanog}, {Gurwell}, {Dannerbauer}, {De Bernardis}, {Fu}, {Harris},
  {Krips}, {Lapi}, {Maiolino}, {Omont}, {Riechers}, {Wardlow}, {Baker},
  {Birkinshaw}, {Bock}, {Bourne}, {Clements}, {Cooray}, {De Zotti}, {Dunne},
  {Dye}, {Eales}, {Farrah}, {Gavazzi}, {Gonz{\'a}lez Nuevo}, {Hopwood}, {Ibar},
  {Ivison}, {Laporte}, {Maddox}, {Mart{\'{\i}}nez-Navajas}, {Michalowski},
  {Negrello}, {Oliver}, {Roseboom}, {Scott}, {Serjeant}, {Smith}, {Smith},
  {Streblyanska}, {Valiante}, {van der Werf}, {Verma}, {Vieira}, {Wang}, \&
  {Wilner}}]{Bussmann:2013aa}
{Bussmann}, R.~S., {P{\'e}rez-Fournon}, I., {Amber}, S., {et~al.} 2013,
  \bibinfo{title}{{Gravitational Lens Models Based on Submillimeter Array
  Imaging of Herschel-selected Strongly Lensed Sub-millimeter Galaxies at z
  $\gt$ 1.5},} \apj, 779, 25, \dodoi{10.1088/0004-637X/779/1/25}

\bibitem[{R. {Ca{\~n}ameras} {et~al.}(2015){Ca{\~n}ameras}, {Nesvadba},
  {Guery}, {McKenzie}, {K{\"o}nig}, {Petitpas}, {Dole}, {Frye}, {Flores-Cacho},
  {Montier}, {Negrello}, {Beelen}, {Boone}, {Dicken}, {Lagache}, {Le Floc'h},
  {Altieri}, {B{\'e}thermin}, {Chary}, {de Zotti}, {Giard}, {Kneissl}, {Krips},
  {Malhotra}, {Martinache}, {Omont}, {Pointecouteau}, {Puget}, {Scott},
  {Soucail}, {Valtchanov}, {Welikala}, \& {Yan}}]{Canameras:2015aa}
{Ca{\~n}ameras}, R., {Nesvadba}, N.~P.~H., {Guery}, D., {et~al.} 2015,
  \bibinfo{title}{{Planck's dusty GEMS: The brightest gravitationally lensed
  galaxies discovered with the Planck all-sky survey},} \aap, 581, A105,
  \dodoi{10.1051/0004-6361/201425128}

\bibitem[{R. {Ca{\~n}ameras} {et~al.}(2018){Ca{\~n}ameras}, {Nesvadba},
  {Limousin}, {Dole}, {Kneissl}, {Koenig}, {Le Floc'h}, {Petitpas}, \&
  {Scott}}]{Canameras:2018aa}
{Ca{\~n}ameras}, R., {Nesvadba}, N.~P.~H., {Limousin}, M., {et~al.} 2018,
  \bibinfo{title}{{Planck's dusty GEMS. V. Molecular wind and clump stability
  in a strongly lensed star-forming galaxy at z = 2.2},} \aap, 620, A60,
  \dodoi{10.1051/0004-6361/201833679}

\bibitem[{R. {Calvi} {et~al.}(2023){Calvi}, {Castignani}, \&
  {Dannerbauer}}]{Calvi:2023ab}
{Calvi}, R., {Castignani}, G., \& {Dannerbauer}, H. 2023,
  \bibinfo{title}{{Bright submillimeter galaxies do trace galaxy
  protoclusters},} \aap, 678, A15, \dodoi{10.1051/0004-6361/202346200}

\bibitem[{D. {Calzetti}(1997){Calzetti}}]{Calzetti:1997ab}
{Calzetti}, D. 1997, \bibinfo{title}{{UV opacity in nearby galaxies and
  application to distant galaxies},} in American Institute of Physics
  Conference Series, Vol. 408, The ultraviolet universe at low and High
  redshift, ed. W.~H. {Waller}, 403--412, \dodoi{10.1063/1.53764}

\bibitem[{D. {Calzetti} {et~al.}(2000){Calzetti}, {Armus}, {Bohlin}, {Kinney},
  {Koornneef}, \& {Storchi-Bergmann}}]{Calzetti:2000aa}
{Calzetti}, D., {Armus}, L., {Bohlin}, R.~C., {et~al.} 2000,
  \bibinfo{title}{{The Dust Content and Opacity of Actively Star-forming
  Galaxies},} \apj, 533, 682, \dodoi{10.1086/308692}

\bibitem[{P.~L. {Capak} {et~al.}(2011){Capak}, {Riechers}, {Scoville},
  {Carilli}, {Cox}, {Neri}, {Robertson}, {Salvato}, {Schinnerer}, {Yan},
  {Wilson}, {Yun}, {Civano}, {Elvis}, {Karim}, {Mobasher}, \&
  {Staguhn}}]{Capak:2011ab}
{Capak}, P.~L., {Riechers}, D., {Scoville}, N.~Z., {et~al.} 2011,
  \bibinfo{title}{{A massive protocluster of galaxies at a redshift of
  z\raisebox{-0.5ex}\textasciitilde5.3},} \nat, 470, 233,
  \dodoi{10.1038/nature09681}

\bibitem[{C.~M. {Casey}(2016){Casey}}]{Casey:2016aa}
{Casey}, C.~M. 2016, \bibinfo{title}{{The Ubiquity of Coeval Starbursts in
  Massive Galaxy Cluster Progenitors},} \apj, 824, 36,
  \dodoi{10.3847/0004-637X/824/1/36}

\bibitem[{C.~M. {Casey} {et~al.}(2014{\natexlab{a}}){Casey}, {Narayanan}, \&
  {Cooray}}]{Casey:2014aa}
{Casey}, C.~M., {Narayanan}, D., \& {Cooray}, A. 2014{\natexlab{a}},
  \bibinfo{title}{{Dusty star-forming galaxies at high redshift},} \physrep,
  541, 45, \dodoi{10.1016/j.physrep.2014.02.009}

\bibitem[{C.~M. {Casey} {et~al.}(2014{\natexlab{b}}){Casey}, {Scoville},
  {Sanders}, {Lee}, {Cooray}, {Finkelstein}, {Capak}, {Conley}, {De Zotti},
  {Farrah}, {Fu}, {Le Floc'h}, {Ilbert}, {Ivison}, \&
  {Takeuchi}}]{Casey:2014ab}
{Casey}, C.~M., {Scoville}, N.~Z., {Sanders}, D.~B., {et~al.}
  2014{\natexlab{b}}, \bibinfo{title}{{Are Dusty Galaxies Blue? Insights on UV
  Attenuation from Dust-selected Galaxies},} \apj, 796, 95,
  \dodoi{10.1088/0004-637X/796/2/95}

\bibitem[{C.~M. {Casey} {et~al.}(2015){Casey}, {Cooray}, {Capak}, {Fu},
  {Kovac}, {Lilly}, {Sanders}, {Scoville}, \& {Treister}}]{Casey:2015aa}
{Casey}, C.~M., {Cooray}, A., {Capak}, P., {et~al.} 2015, \bibinfo{title}{{A
  Massive, Distant Proto-cluster at z = 2.47 Caught in a Phase of Rapid
  Formation?},} \apjl, 808, L33, \dodoi{10.1088/2041-8205/808/2/L33}

\bibitem[{C.~M. {Casey} {et~al.}(2017){Casey}, {Cooray}, {Killi}, {Capak},
  {Chen}, {Hung}, {Kartaltepe}, {Sanders}, \& {Scoville}}]{Casey:2017aa}
{Casey}, C.~M., {Cooray}, A., {Killi}, M., {et~al.} 2017,
  \bibinfo{title}{{Near-infrared MOSFIRE Spectra of Dusty Star-forming Galaxies
  at 0.2 < z < 4},} \apj, 840, 101, \dodoi{10.3847/1538-4357/aa6cb1}

\bibitem[{G. {Chabrier}(2003){Chabrier}}]{Chabrier:2003aa}
{Chabrier}, G. 2003, \bibinfo{title}{{Galactic Stellar and Substellar Initial
  Mass Function},} \pasp, 115, 763, \dodoi{10.1086/376392}

\bibitem[{J.~B. {Champagne} {et~al.}(2021){Champagne}, {Casey}, {Zavala},
  {Cooray}, {Dannerbauer}, {Fabian}, {Hayward}, {Long}, \&
  {Spilker}}]{Champagne:2021aa}
{Champagne}, J.~B., {Casey}, C.~M., {Zavala}, J.~A., {et~al.} 2021,
  \bibinfo{title}{{Comprehensive Gas Characterization of a z = 2.5
  Protocluster: A Cluster Core Caught in the Beginning of Virialization?},}
  \apj, 913, 110, \dodoi{10.3847/1538-4357/abf4e6}

\bibitem[{C.-C. {Chen} {et~al.}(2020){Chen}, {Harrison}, {Smail}, {Swinbank},
  {Turner}, {Wardlow}, {Brandt}, {Calistro Rivera}, {Chapman}, {Cooke},
  {Dannerbauer}, {Dunlop}, {Farrah}, {Micha{\l}owski}, {Schinnerer}, {Simpson},
  {Thomson}, \& {van der Werf}}]{Chen:2020aa}
{Chen}, C.-C., {Harrison}, C.~M., {Smail}, I., {et~al.} 2020,
  \bibinfo{title}{{Extended H{\ensuremath{\alpha}} over compact far-infrared
  continuum in dusty submillimeter galaxies. Insights into dust distributions
  and star-formation rates at z {\ensuremath{\sim}} 2},} \aap, 635, A119,
  \dodoi{10.1051/0004-6361/201936286}

\bibitem[{W. {Chen} {et~al.}(2022){Chen}, {Kelly}, {Oguri}, {Broadhurst},
  {Diego}, {Emami}, {Filippenko}, {Treu}, \& {Zitrin}}]{Chen:2022ag}
{Chen}, W., {Kelly}, P.~L., {Oguri}, M., {et~al.} 2022, \bibinfo{title}{{Shock
  cooling of a red-supergiant supernova at redshift 3 in lensed images},} \nat,
  611, 256, \dodoi{10.1038/s41586-022-05252-5}

\bibitem[{W. {Chen} {et~al.}(2024){Chen}, {Kelly}, {Frye}, {Pierel}, {Willner},
  {Pascale}, {Cohen}, {Conselice}, {Engesser}, {Furtak}, {Gilman}, {Grogin},
  {Huber}, {Jha}, {Johansson}, {Koekemoer}, {Larison}, {Meena}, {Siebert},
  {Windhorst}, {Yan}, \& {Zitrin}}]{Chen:2024ab}
{Chen}, W., {Kelly}, P.~L., {Frye}, B.~L., {et~al.} 2024, \bibinfo{title}{{JWST
  Spectroscopy of SN H0pe: Classification and Time Delays of a Triply Imaged
  Type Ia Supernova at z = 1.78},} \apj, 970, 102,
  \dodoi{10.3847/1538-4357/ad50a5}

\bibitem[{C. {Cheng} {et~al.}(2020){Cheng}, {Ibar}, {Smail}, {Molina},
  {Sobral}, {Escala}, {Best}, {Cochrane}, {Gillman}, {Swinbank}, {Ivison},
  {Huang}, {Hughes}, {Villard}, \& {Cirasuolo}}]{Cheng:2020ac}
{Cheng}, C., {Ibar}, E., {Smail}, I., {et~al.} 2020, \bibinfo{title}{{A
  kpc-scale-resolved study of unobscured and obscured star formation activity
  in normal galaxies at z = 1.5 and 2.2 from ALMA and HiZELS},} \mnras, 499,
  5241, \dodoi{10.1093/mnras/staa3036}

\bibitem[{M. {Chevance} {et~al.}(2022){Chevance}, {Kruijssen}, {Krumholz},
  {Groves}, {Keller}, {Hughes}, {Glover}, {Henshaw}, {Herrera}, {Kim}, {Leroy},
  {Pety}, {Razza}, {Rosolowsky}, {Schinnerer}, {Schruba}, {Barnes}, {Bigiel},
  {Blanc}, {Dale}, {Emsellem}, {Faesi}, {Grasha}, {Klessen}, {Kreckel}, {Liu},
  {Longmore}, {Meidt}, {Querejeta}, {Saito}, {Sun}, \&
  {Usero}}]{Chevance:2022ab}
{Chevance}, M., {Kruijssen}, J.~M.~D., {Krumholz}, M.~R., {et~al.} 2022,
  \bibinfo{title}{{Pre-supernova feedback mechanisms drive the destruction of
  molecular clouds in nearby star-forming disc galaxies},} \mnras, 509, 272,
  \dodoi{10.1093/mnras/stab2938}

\bibitem[{Y.-K. {Chiang} {et~al.}(2013){Chiang}, {Overzier}, \&
  {Gebhardt}}]{Chiang:2013aa}
{Chiang}, Y.-K., {Overzier}, R., \& {Gebhardt}, K. 2013,
  \bibinfo{title}{{Ancient Light from Young Cosmic Cities: Physical and
  Observational Signatures of Galaxy Proto-clusters},} \apj, 779, 127,
  \dodoi{10.1088/0004-637X/779/2/127}

\bibitem[{R. {Chornock} {et~al.}(2013){Chornock}, {Berger}, {Rest},
  {Milisavljevic}, {Lunnan}, {Foley}, {Soderberg}, {Smartt}, {Burgasser},
  {Challis}, {Chomiuk}, {Czekala}, {Drout}, {Fong}, {Huber}, {Kirshner},
  {Leibler}, {McLeod}, {Marion}, {Narayan}, {Riess}, {Roth}, {Sanders},
  {Scolnic}, {Smith}, {Stubbs}, {Tonry}, {Valenti}, {Burgett}, {Chambers},
  {Hodapp}, {Kaiser}, {Kudritzki}, {Magnier}, \& {Price}}]{Chornock:2013aa}
{Chornock}, R., {Berger}, E., {Rest}, A., {et~al.} 2013,
  \bibinfo{title}{{PS1-10afx at z = 1.388: Pan-STARRS1 Discovery of a New Type
  of Superluminous Supernova},} \apj, 767, 162,
  \dodoi{10.1088/0004-637X/767/2/162}

\bibitem[{C. {Clarke} \& M.~S. {Oey}(2002){Clarke} \& {Oey}}]{Clarke:2002aa}
{Clarke}, C., \& {Oey}, M.~S. 2002, \bibinfo{title}{{Galactic porosity and a
  star formation threshold for the escape of ionizing radiation from
  galaxies},} \mnras, 337, 1299, \dodoi{10.1046/j.1365-8711.2002.05976.x}

\bibitem[{D.~L. {Clements} {et~al.}(2016){Clements}, {Braglia}, {Petitpas},
  {Greenslade}, {Cooray}, {Valiante}, {De Zotti}, {O'Halloran}, {Holdship},
  {Morris}, {P{\'e}rez-Fournon}, {Herranz}, {Riechers}, {Baes}, {Bremer},
  {Bourne}, {Dannerbauer}, {Dariush}, {Dunne}, {Eales}, {Fritz},
  {Gonzalez-Nuevo}, {Hopwood}, {Ibar}, {Ivison}, {Leeuw}, {Maddox},
  {Micha{\l}owski}, {Negrello}, {Omont}, {Oteo}, {Serjeant}, {Valtchanov},
  {Vieira}, {Wardlow}, \& {van der Werf}}]{Clements:2016aa}
{Clements}, D.~L., {Braglia}, F., {Petitpas}, G., {et~al.} 2016,
  \bibinfo{title}{{H-ATLAS: a candidate high redshift cluster/protocluster of
  star-forming galaxies},} \mnras, 461, 1719, \dodoi{10.1093/mnras/stw1224}

\bibitem[{D. {Coe} \& L.~A. {Moustakas}(2009){Coe} \& {Moustakas}}]{Coe:2009aa}
{Coe}, D., \& {Moustakas}, L.~A. 2009, \bibinfo{title}{{Cosmological
  Constraints from Gravitational Lens Time Delays},} \apj, 706, 45,
  \dodoi{10.1088/0004-637X/706/1/45}

\bibitem[{J.~J. {Condon}(1992){Condon}}]{Condon:1992aa}
{Condon}, J.~J. 1992, \bibinfo{title}{{Radio emission from normal galaxies},}
  \araa, 30, 575, \dodoi{10.1146/annurev.aa.30.090192.003043}

\bibitem[{O.~R. {Cooper} {et~al.}(2025){Cooper}, {Brammer}, {Heintz}, {Toft},
  {Casey}, {Setton}, {de Graaff}, {Boogaard}, {Cleri}, {Gillman},
  {Gottumukkala}, {Greene}, {Gullberg}, {Hirschmann}, {Hviding}, {Lambrides},
  {Leja}, {Long}, {Manning}, {Maseda}, {McConachie}, {McKinney}, {Narayanan},
  {Price}, {Strait}, {Suess}, {Weibel}, \& {Williams}}]{Cooper:2025aa}
{Cooper}, O.~R., {Brammer}, G., {Heintz}, K.~E., {et~al.} 2025,
  \bibinfo{title}{{RUBIES: JWST/NIRSpec Resolves Evolutionary Phases of Dusty
  Star-forming Galaxies at z {\ensuremath{\sim}} 2},} \apj, 982, 125,
  \dodoi{10.3847/1538-4357/adb8e1}

\bibitem[{L. {Cortese} {et~al.}(2006){Cortese}, {Boselli}, {Buat}, {Gavazzi},
  {Boissier}, {Gil de Paz}, {Seibert}, {Madore}, \& {Martin}}]{Cortese:2006ab}
{Cortese}, L., {Boselli}, A., {Buat}, V., {et~al.} 2006, \bibinfo{title}{{UV
  Dust Attenuation in Normal Star-Forming Galaxies. I. Estimating the
  L$_{TIR}$/L$_{FUV}$ Ratio},} \apj, 637, 242, \dodoi{10.1086/498296}

\bibitem[{F. {Courbin} {et~al.}(2005){Courbin}, {Eigenbrod}, {Vuissoz},
  {Meylan}, \& {Magain}}]{Courbin:2005aa}
{Courbin}, F., {Eigenbrod}, A., {Vuissoz}, C., {Meylan}, G., \& {Magain}, P.
  2005, \bibinfo{title}{{COSMOGRAIL: the COSmological MOnitoring of
  GRAvItational Lenses},} in Gravitational Lensing Impact on Cosmology, ed.
  Y.~{Mellier} \& G.~{Meylan}, Vol. 225, 297--303,
  \dodoi{10.1017/S1743921305002097}

\bibitem[{P. {Craig} {et~al.}(2024){Craig}, {O'Connor}, {Chakrabarti},
  {Rodney}, {Pierel}, {McCully}, \& {Perez-Fournon}}]{Craig:2024aa}
{Craig}, P., {O'Connor}, K., {Chakrabarti}, S., {et~al.} 2024,
  \bibinfo{title}{{A targeted search for strongly lensed supernovae with the
  Las Cumbres Observatory},} \mnras, 534, 1077, \dodoi{10.1093/mnras/stae2103}

\bibitem[{G. {Cresci} {et~al.}(2007){Cresci}, {Mannucci}, {Della Valle}, \&
  {Maiolino}}]{Cresci:2007aa}
{Cresci}, G., {Mannucci}, F., {Della Valle}, M., \& {Maiolino}, R. 2007,
  \bibinfo{title}{{A NICMOS search for obscured supernovae in starburst
  galaxies},} \aap, 462, 927, \dodoi{10.1051/0004-6361:20065364}

\bibitem[{E. {Daddi} {et~al.}(2007){Daddi}, {Dickinson}, {Morrison}, {Chary},
  {Cimatti}, {Elbaz}, {Frayer}, {Renzini}, {Pope}, {Alexander}, {Bauer},
  {Giavalisco}, {Huynh}, {Kurk}, \& {Mignoli}}]{Daddi:2007aa}
{Daddi}, E., {Dickinson}, M., {Morrison}, G., {et~al.} 2007,
  \bibinfo{title}{{Multiwavelength Study of Massive Galaxies at z\~{}2. I. Star
  Formation and Galaxy Growth},} \apj, 670, 156, \dodoi{10.1086/521818}

\bibitem[{E. {Daddi} {et~al.}(2009){Daddi}, {Dannerbauer}, {Stern},
  {Dickinson}, {Morrison}, {Elbaz}, {Giavalisco}, {Mancini}, {Pope}, \&
  {Spinrad}}]{Daddi:2009aa}
{Daddi}, E., {Dannerbauer}, H., {Stern}, D., {et~al.} 2009,
  \bibinfo{title}{{Two Bright Submillimeter Galaxies in a z = 4.05 Protocluster
  in Goods-North, and Accurate Radio-Infrared Photometric Redshifts},} \apj,
  694, 1517, \dodoi{10.1088/0004-637X/694/2/1517}

\bibitem[{H. {Dahle} {et~al.}(2015){Dahle}, {Gladders}, {Sharon}, {Bayliss}, \&
  {Rigby}}]{Dahle:2015aa}
{Dahle}, H., {Gladders}, M.~D., {Sharon}, K., {Bayliss}, M.~B., \& {Rigby},
  J.~R. 2015, \bibinfo{title}{{Time Delay Measurements for the Cluster-lensed
  Sextuple Quasar SDSS J2222+2745},} \apj, 813, 67,
  \dodoi{10.1088/0004-637X/813/1/67}

\bibitem[{T. {Dahlen} {et~al.}(2012){Dahlen}, {Strolger}, {Riess}, {Mattila},
  {Kankare}, \& {Mobasher}}]{Dahlen:2012aa}
{Dahlen}, T., {Strolger}, L.-G., {Riess}, A.~G., {et~al.} 2012,
  \bibinfo{title}{{The Extended Hubble Space Telescope Supernova Survey: The
  Rate of Core Collapse Supernovae to z \raisebox{-0.5ex}\textasciitilde 1},}
  \apj, 757, 70, \dodoi{10.1088/0004-637X/757/1/70}

\bibitem[{H. {Dannerbauer} {et~al.}(2014){Dannerbauer}, {Kurk}, {De Breuck},
  {Wylezalek}, {Santos}, {Koyama}, {Seymour}, {Tanaka}, {Hatch}, {Altieri},
  {Coia}, {Galametz}, {Kodama}, {Miley}, {R{\"o}ttgering}, {Sanchez-Portal},
  {Valtchanov}, {Venemans}, \& {Ziegler}}]{Dannerbauer:2014aa}
{Dannerbauer}, H., {Kurk}, J.~D., {De Breuck}, C., {et~al.} 2014,
  \bibinfo{title}{{An excess of dusty starbursts related to the Spiderweb
  galaxy},} \aap, 570, A55, \dodoi{10.1051/0004-6361/201423771}

\bibitem[{C. {De Breuck} {et~al.}(2004){De Breuck}, {Bertoldi}, {Carilli},
  {Omont}, {Venemans}, {R{\"o}ttgering}, {Overzier}, {Reuland}, {Miley},
  {Ivison}, \& {van Breugel}}]{De-Breuck:2004aa}
{De Breuck}, C., {Bertoldi}, F., {Carilli}, C., {et~al.} 2004,
  \bibinfo{title}{{A multi-wavelength study of the proto-cluster surrounding
  the z = 4.1 radio galaxy TN J1338-1942},} \aap, 424, 1,
  \dodoi{10.1051/0004-6361:20035885}

\bibitem[{R. {de Grijs} {et~al.}(2003){de Grijs}, {Anders}, {Bastian}, {Lynds},
  {Lamers}, \& {O'Neil}}]{de-Grijs:2003aa}
{de Grijs}, R., {Anders}, P., {Bastian}, N., {et~al.} 2003,
  \bibinfo{title}{{Star cluster formation and evolution in nearby starburst
  galaxies - II. Initial conditions},} \mnras, 343, 1285,
  \dodoi{10.1046/j.1365-8711.2003.06777.x}

\bibitem[{A. {Dekel} {et~al.}(2009){Dekel}, {Sari}, \&
  {Ceverino}}]{Dekel:2009aa}
{Dekel}, A., {Sari}, R., \& {Ceverino}, D. 2009, \bibinfo{title}{{Formation of
  Massive Galaxies at High Redshift: Cold Streams, Clumpy Disks, and Compact
  Spheroids},} \apj, 703, 785, \dodoi{10.1088/0004-637X/703/1/785}

\bibitem[{A. {Del Moro} {et~al.}(2013){Del Moro}, {Alexander}, {Mullaney},
  {Daddi}, {Pannella}, {Bauer}, {Pope}, {Dickinson}, {Elbaz}, {Barthel},
  {Garrett}, {Brandt}, {Charmandaris}, {Chary}, {Dasyra}, {Gilli}, {Hickox},
  {Hwang}, {Ivison}, {Juneau}, {Le Floc'h}, {Luo}, {Morrison}, {Rovilos},
  {Sargent}, \& {Xue}}]{Del-Moro:2013aa}
{Del Moro}, A., {Alexander}, D.~M., {Mullaney}, J.~R., {et~al.} 2013,
  \bibinfo{title}{{GOODS-Herschel: radio-excess signature of hidden AGN
  activity in distant star-forming galaxies},} \aap, 549, A59,
  \dodoi{10.1051/0004-6361/201219880}

\bibitem[{I. {Delvecchio} {et~al.}(2017){Delvecchio}, {Smol{\v{c}}i{\'c}},
  {Zamorani}, {Lagos}, {Berta}, {Delhaize}, {Baran}, {Alexander}, {Rosario},
  {Gonzalez-Perez}, {Ilbert}, {Lacey}, {Le F{\`e}vre}, {Miettinen}, {Aravena},
  {Bondi}, {Carilli}, {Ciliegi}, {Mooley}, {Novak}, {Schinnerer}, {Capak},
  {Civano}, {Fanidakis}, {Herrera Ruiz}, {Karim}, {Laigle}, {Marchesi},
  {McCracken}, {Middleberg}, {Salvato}, \& {Tasca}}]{Delvecchio:2017aa}
{Delvecchio}, I., {Smol{\v{c}}i{\'c}}, V., {Zamorani}, G., {et~al.} 2017,
  \bibinfo{title}{{The VLA-COSMOS 3 GHz Large Project: AGN and host-galaxy
  properties out to z {\ensuremath{\lesssim}} 6},} \aap, 602, A3,
  \dodoi{10.1051/0004-6361/201629367}

\bibitem[{S. {Dhawan} {et~al.}(2020){Dhawan}, {Johansson}, {Goobar},
  {Amanullah}, {M{\"o}rtsell}, {Cenko}, {Cooray}, {Fox}, {Goldstein},
  {Kalender}, {Kasliwal}, {Kulkarni}, {Lee}, {Nayyeri}, {Nugent}, {Ofek}, \&
  {Quimby}}]{Dhawan:2020aa}
{Dhawan}, S., {Johansson}, J., {Goobar}, A., {et~al.} 2020,
  \bibinfo{title}{{Magnification, dust, and time-delay constraints from the
  first resolved strongly lensed Type Ia supernova iPTF16geu},} \mnras, 491,
  2639, \dodoi{10.1093/mnras/stz2965}

\bibitem[{A. {D{\'{\i}}az-S{\'a}nchez} {et~al.}(2017){D{\'{\i}}az-S{\'a}nchez},
  {Iglesias-Groth}, {Rebolo}, \& {Dannerbauer}}]{Diaz-Sanchez:2017aa}
{D{\'{\i}}az-S{\'a}nchez}, A., {Iglesias-Groth}, S., {Rebolo}, R., \&
  {Dannerbauer}, H. 2017, \bibinfo{title}{{Discovery of a Lensed Ultrabright
  Submillimeter Galaxy at z = 2.0439},} \apjl, 843, L22,
  \dodoi{10.3847/2041-8213/aa79ef}

\bibitem[{J.~M. {Diego} {et~al.}(2022){Diego}, {Bernstein}, {Chen}, {Goobar},
  {Johansson}, {Kelly}, {M{\"o}rtsell}, \& {Nightingale}}]{Diego:2022ad}
{Diego}, J.~M., {Bernstein}, G., {Chen}, W., {et~al.} 2022,
  \bibinfo{title}{{Microlensing and the type Ia supernova iPTF16geu},} \aap,
  662, A34, \dodoi{10.1051/0004-6361/202143009}

\bibitem[{H. {Dom{\'\i}nguez S{\'a}nchez} {et~al.}(2012){Dom{\'\i}nguez
  S{\'a}nchez}, {Mignoli}, {Pozzi}, {Calura}, {Cimatti}, {Gruppioni}, {Cepa},
  {S{\'a}nchez Portal}, {Zamorani}, {Berta}, {Elbaz}, {Le Floc'h}, {Granato},
  {Lutz}, {Maiolino}, {Matteucci}, {Nair}, {Nordon}, {Pozzetti}, {Silva},
  {Silverman}, {Wuyts}, {Carollo}, {Contini}, {Kneib}, {Le F{\`e}vre}, {Lilly},
  {Mainieri}, {Renzini}, {Scodeggio}, {Bardelli}, {Bolzonella}, {Bongiorno},
  {Caputi}, {Coppa}, {Cucciati}, {de la Torre}, {de Ravel}, {Franzetti},
  {Garilli}, {Iovino}, {Kampczyk}, {Knobel}, {Kova{\v{c}}}, {Lamareille}, {Le
  Borgne}, {Le Brun}, {Maier}, {Magnelli}, {Pell{\'o}}, {Peng},
  {Perez-Montero}, {Ricciardelli}, {Riguccini}, {Tanaka}, {Tasca}, {Tresse},
  {Vergani}, \& {Zucca}}]{Dominguez-Sanchez:2012aa}
{Dom{\'\i}nguez S{\'a}nchez}, H., {Mignoli}, M., {Pozzi}, F., {et~al.} 2012,
  \bibinfo{title}{{Comparison of star formation rates from
  H{\ensuremath{\alpha}} and infrared luminosity as seen by Herschel},} \mnras,
  426, 330, \dodoi{10.1111/j.1365-2966.2012.21710.x}

\bibitem[{J. {Dong} {et~al.}(2024){Dong}, {Shu}, {Li}, {Er}, {Hu}, \&
  {Xu}}]{Dong:2024ab}
{Dong}, J., {Shu}, Y., {Li}, G., {et~al.} 2024, \bibinfo{title}{{Forecast of
  strongly lensed supernovae rates in the China Space Station Telescope
  surveys},} \aap, 689, A192, \dodoi{10.1051/0004-6361/202450838}

\bibitem[{J.~L. {Donley} {et~al.}(2005){Donley}, {Rieke}, {Rigby}, \&
  {P{\'e}rez-Gonz{\'a}lez}}]{Donley:2005aa}
{Donley}, J.~L., {Rieke}, G.~H., {Rigby}, J.~R., \& {P{\'e}rez-Gonz{\'a}lez},
  P.~G. 2005, \bibinfo{title}{{Unveiling a Population of AGNs Not Detected in
  X-Rays},} \apj, 634, 169, \dodoi{10.1086/491668}

\bibitem[{A.~B. {Drake} {et~al.}(2020){Drake}, {Walter}, {Novak}, {Farina},
  {Neeleman}, {Riechers}, {Carilli}, {Decarli}, {Mazzucchelli}, \&
  {Onoue}}]{Drake:2020aa}
{Drake}, A.~B., {Walter}, F., {Novak}, M., {et~al.} 2020, \bibinfo{title}{{The
  Ionized- and Cool-gas Content of the BR1202-0725 System as Seen by MUSE and
  ALMA},} \apj, 902, 37, \dodoi{10.3847/1538-4357/aba832}

\bibitem[{K.~L. {Emig} {et~al.}(2020){Emig}, {Bolatto}, {Leroy}, {Mills},
  {Jim{\'e}nez Donaire}, {Tielens}, {Ginsburg}, {Gorski}, {Krieger}, {Levy},
  {Meier}, {Ott}, {Rosolowsky}, {Thompson}, \& {Veilleux}}]{Emig:2020aa}
{Emig}, K.~L., {Bolatto}, A.~D., {Leroy}, A.~K., {et~al.} 2020,
  \bibinfo{title}{{Super Star Clusters in the Central Starburst of NGC 4945},}
  \apj, 903, 50, \dodoi{10.3847/1538-4357/abb67d}

\bibitem[{S. {Ertl} {et~al.}(2025){Ertl}, {Suyu}, {Schuldt}, {Granata},
  {Grillo}, {Caminha}, {Acebron}, {Bergamini}, {Ca{\~n}ameras}, {Cha}, {Diego},
  {Foo}, {Frye}, {Fudamoto}, {Halkola}, {Jee}, {Kamieneski}, {Koekemoer},
  {Meena}, {Nishida}, {Oguri}, {Pierel}, {Rosati}, {Tortorelli}, {Wang}, \&
  {Zitrin}}]{Ertl:2025aa}
{Ertl}, S., {Suyu}, S.~H., {Schuldt}, S., {et~al.} 2025,
  \bibinfo{title}{{Cosmology with Supernova Encore in the strong lensing
  cluster MACS J0138$-$2155: photometry, cluster members, and lens mass
  model},} arXiv e-prints, arXiv:2503.09718.
\newblock \doarXiv{2503.09718}

\bibitem[{ {Euclid Collaboration} {et~al.}(2022){Euclid Collaboration},
  {Scaramella}, {Amiaux}, {Mellier}, {Burigana}, {Carvalho}, {Cuillandre}, {Da
  Silva}, {Derosa}, {Dinis}, {Maiorano}, {Maris}, {Tereno}, {Laureijs},
  {Boenke}, {Buenadicha}, {Dupac}, {Gaspar Venancio}, {G{\'o}mez-{\'A}lvarez},
  {Hoar}, {Lorenzo Alvarez}, {Racca}, {Saavedra-Criado}, {Schwartz}, {Vavrek},
  {Schirmer}, {Aussel}, {Azzollini}, {Cardone}, {Cropper}, {Ealet}, {Garilli},
  {Gillard}, {Granett}, {Guzzo}, {Hoekstra}, {Jahnke}, {Kitching}, {Maciaszek},
  {Meneghetti}, {Miller}, {Nakajima}, {Niemi}, {Pasian}, {Percival},
  {Pottinger}, {Sauvage}, {Scodeggio}, {Wachter}, {Zacchei}, {Aghanim},
  {Amara}, {Auphan}, {Auricchio}, {Awan}, {Balestra}, {Bender}, {Bodendorf},
  {Bonino}, {Branchini}, {Brau-Nogue}, {Brescia}, {Candini}, {Capobianco},
  {Carbone}, {Carlberg}, {Carretero}, {Casas}, {Castander}, {Castellano},
  {Cavuoti}, {Cimatti}, {Cledassou}, {Congedo}, {Conselice}, {Conversi},
  {Copin}, {Corcione}, {Costille}, {Courbin}, {Degaudenzi}, {Douspis},
  {Dubath}, {Duncan}, {Dusini}, {Farrens}, {Ferriol}, {Fosalba}, {Fourmanoit},
  {Frailis}, {Franceschi}, {Franzetti}, {Fumana}, {Gillis}, {Giocoli},
  {Grazian}, {Grupp}, {Haugan}, {Holmes}, {Hormuth}, {Hudelot}, {Kermiche},
  {Kiessling}, {Kilbinger}, {Kohley}, {Kubik}, {K{\"u}mmel}, {Kunz},
  {Kurki-Suonio}, {Lahav}, {Ligori}, {Lilje}, {Lloro}, {Mansutti}, {Marggraf},
  {Markovic}, {Marulli}, {Massey}, {Maurogordato}, {Melchior}, {Merlin},
  {Meylan}, {Mohr}, {Moresco}, {Morin}, {Moscardini}, {Munari}, {Nichol},
  {Padilla}, {Paltani}, {Peacock}, {Pedersen}, {Pettorino}, {Pires}, {Poncet},
  {Popa}, {Pozzetti}, {Raison}, {Rebolo}, {Rhodes}, {Rix}, {Roncarelli},
  {Rossetti}, {Saglia}, {Schneider}, {Schrabback}, {Secroun}, {Seidel},
  {Serrano}, {Sirignano}, {Sirri}, {Skottfelt}, {Stanco}, {Starck},
  {Tallada-Cresp{\'\i}}, {Tavagnacco}, {Taylor}, {Teplitz}, {Toledo-Moreo},
  {Torradeflot}, {Trifoglio}, {Valentijn}, {Valenziano}, {Verdoes Kleijn},
  {Wang}, {Welikala}, {Weller}, {Wetzstein}, {Zamorani}, {Zoubian}, {Andreon},
  {Baldi}, {Bardelli}, {Boucaud}, {Camera}, {Di Ferdinando}, {Fabbian},
  {Farinelli}, {Galeotta}, {Graci{\'a}-Carpio}, {Maino}, {Medinaceli}, {Mei},
  {Neissner}, {Polenta}, {Renzi}, {Romelli}, {Rosset}, {Sureau}, {Tenti},
  {Vassallo}, {Zucca}, {Baccigalupi}, {Balaguera-Antol{\'\i}nez}, {Battaglia},
  {Biviano}, {Borgani}, {Bozzo}, {Cabanac}, \&
  {Cappi}}]{Euclid-Collaboration:2022ab}
{Euclid Collaboration}, {Scaramella}, R., {Amiaux}, J., {et~al.} 2022,
  \bibinfo{title}{{Euclid preparation. I. The Euclid Wide Survey},} \aap, 662,
  A112, \dodoi{10.1051/0004-6361/202141938}

\bibitem[{A.~L. {Faisst} {et~al.}(2017){Faisst}, {Capak}, {Yan}, {Pavesi},
  {Riechers}, {Bari{\v{s}}i{\'c}}, {Cooke}, {Kartaltepe}, \&
  {Masters}}]{Faisst:2017aa}
{Faisst}, A.~L., {Capak}, P.~L., {Yan}, L., {et~al.} 2017, \bibinfo{title}{{Are
  High-redshift Galaxies Hot? Temperature of z > 5 Galaxies and Implications
  for Their Dust Properties},} \apj, 847, 21, \dodoi{10.3847/1538-4357/aa886c}

\bibitem[{A.~L. {Faisst} {et~al.}(2025){Faisst}, {Yang}, {Brinch}, {Casey},
  {Chartab}, {Dessauges-Zavadsky}, {Drakos}, {Gillman}, {Gozaliasl}, {Hayward},
  {Ilbert}, {Jablonka}, {Kaminsky}, {Kartaltepe}, {Koekemoer}, {Kokorev},
  {Lambrides}, {Liu}, {Maraston}, {Martin}, {Renzini}, {Robertson}, {Sanders},
  {Sattari}, {Scoville}, {Urry}, {Vijayan}, {Weaver}, {Akins}, {Allen},
  {Arango-Toro}, {Cooper}, {Franco}, {Gentile}, {Harish}, {Hirschmann},
  {Khostovan}, {Laigle}, {Larson}, {Lee}, {Liu}, {Long}, {Magdis}, {Massey},
  {McCracken}, {McKinney}, {Paquereau}, {Rhodes}, {Rich}, {Shuntov},
  {Silverman}, {Talia}, {Toft}, \& {Zavala}}]{Faisst:2025aa}
{Faisst}, A.~L., {Yang}, L., {Brinch}, M., {et~al.} 2025,
  \bibinfo{title}{{COSMOS-Web: The Role of Galaxy Interactions and Disk
  Instabilities in Producing Starbursts at z < 4},} \apj, 980, 204,
  \dodoi{10.3847/1538-4357/ada566}

\bibitem[{E.~E. {Falco} {et~al.}(1985){Falco}, {Gorenstein}, \&
  {Shapiro}}]{Falco:1985aa}
{Falco}, E.~E., {Gorenstein}, M.~V., \& {Shapiro}, I.~I. 1985,
  \bibinfo{title}{{On model-dependent bounds on H 0 from gravitational images :
  application to Q 0957+561 A, B.},} \apjl, 289, L1, \dodoi{10.1086/184422}

\bibitem[{N. {Falstad} {et~al.}(2021){Falstad}, {Aalto}, {K{\"o}nig}, {Onishi},
  {Muller}, {Gorski}, {Sato}, {Stanley}, {Combes}, {Gonz{\'a}lez-Alfonso},
  {Mangum}, {Evans}, {Barcos-Mu{\~n}oz}, {Privon}, {Linden},
  {D{\'\i}az-Santos}, {Mart{\'\i}n}, {Sakamoto}, {Harada}, {Fuller},
  {Gallagher}, {van der Werf}, {Viti}, {Greve}, {Garc{\'\i}a-Burillo},
  {Henkel}, {Imanishi}, {Izumi}, {Nishimura}, {Ricci}, \&
  {M{\"u}hle}}]{Falstad:2021aa}
{Falstad}, N., {Aalto}, S., {K{\"o}nig}, S., {et~al.} 2021,
  \bibinfo{title}{{CON-quest. Searching for the most obscured galaxy nuclei},}
  \aap, 649, A105, \dodoi{10.1051/0004-6361/202039291}

\bibitem[{J. {Fohlmeister} {et~al.}(2008){Fohlmeister}, {Kochanek}, {Falco},
  {Morgan}, \& {Wambsganss}}]{Fohlmeister:2008aa}
{Fohlmeister}, J., {Kochanek}, C.~S., {Falco}, E.~E., {Morgan}, C.~W., \&
  {Wambsganss}, J. 2008, \bibinfo{title}{{The Rewards of Patience: An 822 Day
  Time Delay in the Gravitational Lens SDSS J1004+4112},} \apj, 676, 761,
  \dodoi{10.1086/528789}

\bibitem[{J. {Fohlmeister} {et~al.}(2007){Fohlmeister}, {Kochanek}, {Falco},
  {Wambsganss}, {Morgan}, {Morgan}, {Ofek}, {Maoz}, {Keeton}, {Barentine},
  {Dalton}, {Dembicky}, {Ketzeback}, {McMillan}, \&
  {Peters}}]{Fohlmeister:2007aa}
{Fohlmeister}, J., {Kochanek}, C.~S., {Falco}, E.~E., {et~al.} 2007,
  \bibinfo{title}{{A Time Delay for the Cluster-Lensed Quasar SDSS
  J1004+4112},} \apj, 662, 62, \dodoi{10.1086/518018}

\bibitem[{N. {Foo} {et~al.}(2025){Foo}, {Harrington}, {Frye}, {Kamieneski},
  {Yun}, {Pascale}, {Yoon}, {Noble}, {Windhorst}, {Cohen}, {Lowenthal},
  {Kaasinen}, {Alcalde Pampliega}, {Liu}, {Cooper}, {Garcia Diaz}, {Diaz},
  {Diego}, {Garuda}, {Jim{\'e}nez-Andrade}, {Leimbach}, {Vishwas}, {Wang},
  {Zhou}, \& {Zitrin}}]{Foo:2025aa}
{Foo}, N., {Harrington}, K.~C., {Frye}, B., {et~al.} 2025,
  \bibinfo{title}{{PASSAGES: The Discovery of a Strongly Lensed Protocluster
  Core Candidate at Cosmic Noon},} arXiv e-prints, arXiv:2504.05617.
\newblock \doarXiv{2504.05617}

\bibitem[{O.~D. {Fox} {et~al.}(2021){Fox}, {Khandrika}, {Rubin}, {Casper},
  {Li}, {Szalai}, {Armus}, {Filippenko}, {Skrutskie}, {Strolger}, \& {Van
  Dyk}}]{Fox:2021ab}
{Fox}, O.~D., {Khandrika}, H., {Rubin}, D., {et~al.} 2021, \bibinfo{title}{{A
  Spitzer survey for dust-obscured supernovae},} \mnras, 506, 4199,
  \dodoi{10.1093/mnras/stab1740}

\bibitem[{W.~L. {Freedman} {et~al.}(2025){Freedman}, {Madore}, {Hoyt}, {Jang},
  {Lee}, \& {Owens}}]{Freedman:2025aa}
{Freedman}, W.~L., {Madore}, B.~F., {Hoyt}, T.~J., {et~al.} 2025,
  \bibinfo{title}{{Status Report on the Chicago-Carnegie Hubble Program (CCHP):
  Measurement of the Hubble Constant Using the Hubble and James Webb Space
  Telescopes},} \apj, 985, 203, \dodoi{10.3847/1538-4357/adce78}

\bibitem[{W.~L. {Freedman} {et~al.}(2019){Freedman}, {Madore}, {Hatt}, {Hoyt},
  {Jang}, {Beaton}, {Burns}, {Lee}, {Monson}, {Neeley}, {Phillips}, {Rich}, \&
  {Seibert}}]{Freedman:2019aa}
{Freedman}, W.~L., {Madore}, B.~F., {Hatt}, D., {et~al.} 2019,
  \bibinfo{title}{{The Carnegie-Chicago Hubble Program. VIII. An Independent
  Determination of the Hubble Constant Based on the Tip of the Red Giant
  Branch},} \apj, 882, 34, \dodoi{10.3847/1538-4357/ab2f73}

\bibitem[{B.~L. {Frye} {et~al.}(2019){Frye}, {Pascale}, {Qin}, {Zitrin},
  {Diego}, {Walth}, {Yan}, {Conselice}, {Alpaslan}, {Bauer}, {Busoni}, {Coe},
  {Cohen}, {Dole}, {Donahue}, {Georgiev}, {Jansen}, {Limousin}, {Livermore},
  {Norman}, {Rabien}, \& {Windhorst}}]{Frye:2019aa}
{Frye}, B.~L., {Pascale}, M., {Qin}, Y., {et~al.} 2019, \bibinfo{title}{{PLCK
  G165.7+67.0: Analysis of a Massive Lensing Cluster in a Hubble Space
  Telescope Census of Submillimeter Giant Arcs Selected Using
  Planck/Herschel},} \apj, 871, 51, \dodoi{10.3847/1538-4357/aaeff7}

\bibitem[{B.~L. {Frye} {et~al.}(2023){Frye}, {Pascale}, {Cohen}, {Summers},
  {Foo}, {Kamieneski}, {Carleton}, {Jansen}, {Pierel}, {Engesser}, {Chen},
  {Austin}, {Marshall}, {Trussler}, {Meena}, {Leimbach}, {Garuda}, {Honor},
  {Furtak}, {Strolger}, {Windhorst}, {Koekemoer}, {Zitrin}, {Diego}, {Kelly},
  {Coe}, {Conselice}, {Dai}, {D{\^a}Silva}, {Dole}, {Driver}, {Grogin},
  {Nonino}, {Pirzkal}, {Polletta}, {Robotham}, {Rutkowski}, {Ryan}, {Tompkins},
  {Willmer}, {Willner}, {Yan}, \& {Yun}}]{Frye:2023ab}
{Frye}, B.~L., {Pascale}, M., {Cohen}, S., {et~al.} 2023, \bibinfo{title}{{SN
  H0pe: three images of a SN detected near the central region of the galaxy
  cluster field PLCK G165.7+67.0},} Transient Name Server AstroNote, 96, 1

\bibitem[{B.~L. {Frye} {et~al.}(2024){Frye}, {Pascale}, {Pierel}, {Chen},
  {Foo}, {Leimbach}, {Garuda}, {Cohen}, {Kamieneski}, {Windhorst}, {Koekemoer},
  {Kelly}, {Summers}, {Engesser}, {Liu}, {Furtak}, {del Carmen Polletta},
  {Harrington}, {Willner}, {Diego}, {Jansen}, {Coe}, {Conselice}, {Dai},
  {Dole}, {D'Silva}, {Driver}, {Grogin}, {Marshall}, {Meena}, {Nonino},
  {Ortiz}, {Pirzkal}, {Robotham}, {Ryan}, {Strolger}, {Tompkins}, {Willmer},
  {Yan}, {Yun}, \& {Zitrin}}]{Frye:2024aa}
{Frye}, B.~L., {Pascale}, M., {Pierel}, J., {et~al.} 2024, \bibinfo{title}{{The
  JWST Discovery of the Triply Imaged Type Ia ``Supernova H0pe'' and
  Observations of the Galaxy Cluster PLCK G165.7+67.0},} \apj, 961, 171,
  \dodoi{10.3847/1538-4357/ad1034}

\bibitem[{S. {Fujimoto} {et~al.}(2017){Fujimoto}, {Ouchi}, {Shibuya}, \&
  {Nagai}}]{Fujimoto:2017aa}
{Fujimoto}, S., {Ouchi}, M., {Shibuya}, T., \& {Nagai}, H. 2017,
  \bibinfo{title}{{Demonstrating a New Census of Infrared Galaxies with ALMA
  (DANCING-ALMA). I. FIR Size and Luminosity Relation at z = 0-6 Revealed with
  1034 ALMA Sources},} \apj, 850, 83, \dodoi{10.3847/1538-4357/aa93e6}

\bibitem[{S. {Fujimoto} {et~al.}(2024){Fujimoto}, {Ouchi}, {Kohno},
  {Valentino}, {Gim{\'e}nez-Arteaga}, {Brammer}, {Furtak}, {Kohandel}, {Oguri},
  {Pallottini}, {Richard}, {Zitrin}, {Bauer}, {Boylan-Kolchin},
  {Dessauges-Zavadsky}, {Egami}, {Finkelstein}, {Ma}, {Smail}, {Watson},
  {Hutchison}, {Rigby}, {Welch}, {Ao}, {Bradley}, {Caminha}, {Caputi},
  {Espada}, {Endsley}, {Fudamoto}, {Gonz{\'a}lez-L{\'o}pez}, {Hatsukade},
  {Koekemoer}, {Kokorev}, {Laporte}, {Lee}, {Magdis}, {Ono}, {Rizzo},
  {Shibuya}, {Shimasaku}, {Sun}, {Toft}, {Umehata}, {Wang}, \&
  {Yajima}}]{Fujimoto:2024aa}
{Fujimoto}, S., {Ouchi}, M., {Kohno}, K., {et~al.} 2024,
  \bibinfo{title}{{Primordial Rotating Disk Composed of $\geq$15 Dense
  Star-Forming Clumps at Cosmic Dawn},} arXiv e-prints, arXiv:2402.18543,
  \dodoi{10.48550/arXiv.2402.18543}

\bibitem[{A. {Gal-Yam} {et~al.}(2002){Gal-Yam}, {Maoz}, \&
  {Sharon}}]{Gal-Yam:2002aa}
{Gal-Yam}, A., {Maoz}, D., \& {Sharon}, K. 2002, \bibinfo{title}{{Supernovae in
  deep Hubble Space Telescope galaxy cluster fields: cluster rates and field
  counts},} \mnras, 332, 37, \dodoi{10.1046/j.1365-8711.2002.05274.x}

\bibitem[{R. {Genzel} {et~al.}(2015){Genzel}, {Tacconi}, {Lutz}, {Saintonge},
  {Berta}, {Magnelli}, {Combes}, {Garc{\'\i}a-Burillo}, {Neri}, {Bolatto},
  {Contini}, {Lilly}, {Boissier}, {Boone}, {Bouch{\'e}}, {Bournaud}, {Burkert},
  {Carollo}, {Colina}, {Cooper}, {Cox}, {Feruglio}, {F{\"o}rster Schreiber},
  {Freundlich}, {Gracia-Carpio}, {Juneau}, {Kovac}, {Lippa}, {Naab}, {Salome},
  {Renzini}, {Sternberg}, {Walter}, {Weiner}, {Weiss}, \&
  {Wuyts}}]{Genzel:2015aa}
{Genzel}, R., {Tacconi}, L.~J., {Lutz}, D., {et~al.} 2015,
  \bibinfo{title}{{Combined CO and Dust Scaling Relations of Depletion Time and
  Molecular Gas Fractions with Cosmic Time, Specific Star-formation Rate, and
  Stellar Mass},} \apj, 800, 20, \dodoi{10.1088/0004-637X/800/1/20}

\bibitem[{C. {Gim{\'e}nez-Arteaga} {et~al.}(2023){Gim{\'e}nez-Arteaga},
  {Oesch}, {Brammer}, {Valentino}, {Mason}, {Weibel}, {Barrufet}, {Fujimoto},
  {Heintz}, {Nelson}, {Strait}, {Suess}, \& {Gibson}}]{Gimenez-Arteaga:2023aa}
{Gim{\'e}nez-Arteaga}, C., {Oesch}, P.~A., {Brammer}, G.~B., {et~al.} 2023,
  \bibinfo{title}{{Spatially Resolved Properties of Galaxies at 5 < z < 9 in
  the SMACS 0723 JWST ERO Field},} \apj, 948, 126,
  \dodoi{10.3847/1538-4357/acc5ea}

\bibitem[{M. {Giulietti} {et~al.}(2024){Giulietti}, {Gandolfi}, {Massardi},
  {Behiri}, \& {Lapi}}]{Giulietti:2024ab}
{Giulietti}, M., {Gandolfi}, G., {Massardi}, M., {Behiri}, M., \& {Lapi}, A.
  2024, \bibinfo{title}{{Observing Dusty Star-Forming Galaxies at the Cosmic
  Noon through Gravitational Lensing: Perspectives from New-Generation
  Telescopes},} Galaxies, 12, 9, \dodoi{10.3390/galaxies12020009}

\bibitem[{D.~A. {Goldstein} \& P.~E. {Nugent}(2017){Goldstein} \&
  {Nugent}}]{Goldstein:2017aa}
{Goldstein}, D.~A., \& {Nugent}, P.~E. 2017, \bibinfo{title}{{How to Find
  Gravitationally Lensed Type Ia Supernovae},} \apjl, 834, L5,
  \dodoi{10.3847/2041-8213/834/1/L5}

\bibitem[{D.~A. {Goldstein} {et~al.}(2019){Goldstein}, {Nugent}, \&
  {Goobar}}]{Goldstein:2019aa}
{Goldstein}, D.~A., {Nugent}, P.~E., \& {Goobar}, A. 2019,
  \bibinfo{title}{{Rates and Properties of Supernovae Strongly Gravitationally
  Lensed by Elliptical Galaxies in Time-domain Imaging Surveys},} \apjs, 243,
  6, \dodoi{10.3847/1538-4365/ab1fe0}

\bibitem[{A. {Goobar} {et~al.}(2017){Goobar}, {Amanullah}, {Kulkarni},
  {Nugent}, {Johansson}, {Steidel}, {Law}, {M{\"o}rtsell}, {Quimby},
  {Blagorodnova}, {Brand eker}, {Cao}, {Cooray}, {Ferretti}, {Fremling},
  {Hangard}, {Kasliwal}, {Kupfer}, {Lunnan}, {Masci}, {Miller}, {Nayyeri},
  {Neill}, {Ofek}, {Papadogiannakis}, {Petrushevska}, {Ravi}, {Sollerman},
  {Sullivan}, {Taddia}, {Walters}, {Wilson}, {Yan}, \& {Yaron}}]{Goobar:2017aa}
{Goobar}, A., {Amanullah}, R., {Kulkarni}, S.~R., {et~al.} 2017,
  \bibinfo{title}{{iPTF16geu: A multiply imaged, gravitationally lensed type Ia
  supernova},} Science, 356, 291, \dodoi{10.1126/science.aal2729}

\bibitem[{A. {Goobar} {et~al.}(2023){Goobar}, {Johansson}, {Schulze},
  {Arendse}, {Carracedo}, {Dhawan}, {M{\"o}rtsell}, {Fremling}, {Yan},
  {Perley}, {Sollerman}, {Joseph}, {Hinds}, {Meynardie}, {Andreoni}, {Bellm},
  {Bloom}, {Collett}, {Drake}, {Graham}, {Kasliwal}, {Kulkarni}, {Lemon},
  {Miller}, {Neill}, {Nordin}, {Pierel}, {Richard}, {Riddle}, {Rigault},
  {Rusholme}, {Sharma}, {Stein}, {Stewart}, {Townsend}, {Vinko}, {Wheeler}, \&
  {Wold}}]{Goobar:2023aa}
{Goobar}, A., {Johansson}, J., {Schulze}, S., {et~al.} 2023,
  \bibinfo{title}{{Uncovering a population of gravitational lens galaxies with
  magnified standard candle SN Zwicky},} Nature Astronomy, 7, 1098,
  \dodoi{10.1038/s41550-023-01981-3}

\bibitem[{A.~A. {Goobar} {et~al.}(2022){Goobar}, {Johansson}, {Dhawan},
  {Schulze}, {Arendse}, {Carracedo}, {Joseph}, {Nordin}, \&
  {Townsend}}]{Goobar:2022aa}
{Goobar}, A.~A., {Johansson}, J., {Dhawan}, S., {et~al.} 2022,
  \bibinfo{title}{{SN Zwicky (SN2022qmx): a Strongly Lensed Type Ia at z=0.35
  discovered by ZTF},} Transient Name Server AstroNote, 180, 1

\bibitem[{M.~V. {Gorenstein} {et~al.}(1988){Gorenstein}, {Falco}, \&
  {Shapiro}}]{Gorenstein:1988aa}
{Gorenstein}, M.~V., {Falco}, E.~E., \& {Shapiro}, I.~I. 1988,
  \bibinfo{title}{{Degeneracies in Parameter Estimates for Models of
  Gravitational Lens Systems},} \apj, 327, 693, \dodoi{10.1086/166226}

\bibitem[{T.~R. {Greve} {et~al.}(2005){Greve}, {Bertoldi}, {Smail}, {Neri},
  {Chapman}, {Blain}, {Ivison}, {Genzel}, {Omont}, {Cox}, {Tacconi}, \&
  {Kneib}}]{Greve:2005aa}
{Greve}, T.~R., {Bertoldi}, F., {Smail}, I., {et~al.} 2005, \bibinfo{title}{{An
  interferometric CO survey of luminous submillimetre galaxies},} \mnras, 359,
  1165, \dodoi{10.1111/j.1365-2966.2005.08979.x}

\bibitem[{C. {Grillo} {et~al.}(2024){Grillo}, {Pagano}, {Rosati}, \&
  {Suyu}}]{Grillo:2024aa}
{Grillo}, C., {Pagano}, L., {Rosati}, P., \& {Suyu}, S.~H. 2024,
  \bibinfo{title}{{Cosmography with supernova Refsdal through time-delay
  cluster lensing: Independent measurements of the Hubble constant and geometry
  of the Universe},} \aap, 684, L23, \dodoi{10.1051/0004-6361/202449278}

\bibitem[{C. {Grillo} {et~al.}(2018){Grillo}, {Rosati}, {Suyu}, {Balestra},
  {Caminha}, {Halkola}, {Kelly}, {Lombardi}, {Mercurio}, {Rodney}, \&
  {Treu}}]{Grillo:2018aa}
{Grillo}, C., {Rosati}, P., {Suyu}, S.~H., {et~al.} 2018,
  \bibinfo{title}{{Measuring the Value of the Hubble Constant
  {\textquotedblleft}{\`a} la Refsdal{\textquotedblright}},} \apj, 860, 94,
  \dodoi{10.3847/1538-4357/aac2c9}

\bibitem[{C.~A. {Hales} {et~al.}(2012){Hales}, {Murphy}, {Curran},
  {Middelberg}, {Gaensler}, \& {Norris}}]{Hales:2012aa}
{Hales}, C.~A., {Murphy}, T., {Curran}, J.~R., {et~al.} 2012,
  \bibinfo{title}{{BLOBCAT: software to catalogue flood-filled blobs in radio
  images of total intensity and linear polarization},} \mnras, 425, 979,
  \dodoi{10.1111/j.1365-2966.2012.21373.x}

\bibitem[{K.~C. {Harrington} {et~al.}(2016){Harrington}, {Yun}, {Cybulski},
  {Wilson}, {Aretxaga}, {Chavez}, {De la Luz}, {Erickson}, {Ferrusca},
  {Gallup}, {Hughes}, {Monta{\~n}a}, {Narayanan}, {S{\'a}nchez-Arg{\"u}elles},
  {Schloerb}, {Souccar}, {Terlevich}, {Terlevich}, {Zeballos}, \&
  {Zavala}}]{Harrington:2016aa}
{Harrington}, K.~C., {Yun}, M.~S., {Cybulski}, R., {et~al.} 2016,
  \bibinfo{title}{{Early science with the Large Millimeter Telescope:
  observations of extremely luminous high-z sources identified by Planck},}
  \mnras, 458, 4383, \dodoi{10.1093/mnras/stw614}

\bibitem[{K.~C. {Harrington} {et~al.}(2021){Harrington}, {Weiss}, {Yun},
  {Magnelli}, {Sharon}, {Leung}, {Vishwas}, {Wang}, {Frayer},
  {Jim{\'e}nez-Andrade}, {Liu}, {Garc{\'\i}a}, {Romano-D{\'\i}az}, {Frye},
  {Jarugula}, {B{\u{a}}descu}, {Berman}, {Dannerbauer},
  {D{\'\i}az-S{\'a}nchez}, {Grassitelli}, {Kamieneski}, {Kim}, {Kirkpatrick},
  {Lowenthal}, {Messias}, {Puschnig}, {Stacey}, {Torne}, \&
  {Bertoldi}}]{Harrington:2021aa}
{Harrington}, K.~C., {Weiss}, A., {Yun}, M.~S., {et~al.} 2021,
  \bibinfo{title}{{Turbulent Gas in Lensed Planck-selected Starbursts at z
  {\ensuremath{\sim}} 1-3.5},} \apj, 908, 95, \dodoi{10.3847/1538-4357/abcc01}

\bibitem[{T. {Harvey} {et~al.}(2025){Harvey}, {Conselice}, {Adams}, {Austin},
  {Li}, {Rusakov}, {Westcott}, {Goolsby}, {Lovell}, {Cochrane}, {Vijayan}, \&
  {Trussler}}]{Harvey:2025ac}
{Harvey}, T., {Conselice}, C.~J., {Adams}, N.~J., {et~al.} 2025,
  \bibinfo{title}{{Behind the Spotlight: A systematic assessment of outshining
  using NIRCam medium-bands in the JADES Origins Field},} arXiv e-prints,
  arXiv:2504.05244, \dodoi{10.48550/arXiv.2504.05244}

\bibitem[{Y.~D. {Hezaveh} {et~al.}(2012){Hezaveh}, {Marrone}, \&
  {Holder}}]{Hezaveh:2012aa}
{Hezaveh}, Y.~D., {Marrone}, D.~P., \& {Holder}, G.~P. 2012,
  \bibinfo{title}{{Size Bias and Differential Lensing of Strongly Lensed, Dusty
  Galaxies Identified in Wide-Field Surveys},} \apj, 761, 20,
  \dodoi{10.1088/0004-637X/761/1/20}

\bibitem[{S. {Hilbert} {et~al.}(2007){Hilbert}, {White}, {Hartlap}, \&
  {Schneider}}]{Hilbert:2007aa}
{Hilbert}, S., {White}, S.~D.~M., {Hartlap}, J., \& {Schneider}, P. 2007,
  \bibinfo{title}{{Strong lensing optical depths in a {$\Lambda$}CDM
  universe},} \mnras, 382, 121, \dodoi{10.1111/j.1365-2966.2007.12391.x}

\bibitem[{S. {Hilbert} {et~al.}(2008){Hilbert}, {White}, {Hartlap}, \&
  {Schneider}}]{Hilbert:2008aa}
{Hilbert}, S., {White}, S. D.~M., {Hartlap}, J., \& {Schneider}, P. 2008,
  \bibinfo{title}{{Strong-lensing optical depths in a {\ensuremath{\Lambda}}CDM
  universe - II. The influence of the stellar mass in galaxies},} \mnras, 386,
  1845, \dodoi{10.1111/j.1365-2966.2008.13190.x}

\bibitem[{J.~A. {Hodge} {et~al.}(2016){Hodge}, {Swinbank}, {Simpson}, {Smail},
  {Walter}, {Alexander}, {Bertoldi}, {Biggs}, {Brandt}, {Chapman}, {Chen},
  {Coppin}, {Cox}, {Dannerbauer}, {Edge}, {Greve}, {Ivison}, {Karim},
  {Knudsen}, {Menten}, {Rix}, {Schinnerer}, {Wardlow}, {Weiss}, \& {van der
  Werf}}]{Hodge:2016aa}
{Hodge}, J.~A., {Swinbank}, A.~M., {Simpson}, J.~M., {et~al.} 2016,
  \bibinfo{title}{{Kiloparsec-scale Dust Disks in High-redshift Luminous
  Submillimeter Galaxies},} \apj, 833, 103, \dodoi{10.3847/1538-4357/833/1/103}

\bibitem[{J.~A. {Hodge} {et~al.}(2019){Hodge}, {Smail}, {Walter}, {da Cunha},
  {Swinbank}, {Rybak}, {Venemans}, {Brandt}, {Calistro Rivera}, {Chapman},
  {Chen}, {Cox}, {Dannerbauer}, {Decarli}, {Greve}, {Knudsen}, {Menten},
  {Schinnerer}, {Simpson}, {van der Werf}, {Wardlow}, \&
  {Weiss}}]{Hodge:2019aa}
{Hodge}, J.~A., {Smail}, I., {Walter}, F., {et~al.} 2019, \bibinfo{title}{{ALMA
  Reveals Potential Evidence for Spiral Arms, Bars, and Rings in High-redshift
  Submillimeter Galaxies},} \apj, 876, 130, \dodoi{10.3847/1538-4357/ab1846}

\bibitem[{B.~W. {Holwerda} {et~al.}(2021){Holwerda}, {Knabel}, {Steele},
  {Strolger}, {Kielkopf}, {Jacques}, \& {Roemer}}]{Holwerda:2021aa}
{Holwerda}, B.~W., {Knabel}, S., {Steele}, R.~C., {et~al.} 2021,
  \bibinfo{title}{{The observable supernova rate in galaxy-galaxy lensing
  systems with the TESS satellite},} \mnras, 505, 1316,
  \dodoi{10.1093/mnras/stab1370}

\bibitem[{R. {Hounsell} {et~al.}(2018){Hounsell}, {Scolnic}, {Foley},
  {Kessler}, {Miranda}, {Avelino}, {Bohlin}, {Filippenko}, {Frieman}, {Jha},
  {Kelly}, {Kirshner}, {Mandel}, {Rest}, {Riess}, {Rodney}, \&
  {Strolger}}]{Hounsell:2018aa}
{Hounsell}, R., {Scolnic}, D., {Foley}, R.~J., {et~al.} 2018,
  \bibinfo{title}{{Simulations of the WFIRST Supernova Survey and Forecasts of
  Cosmological Constraints},} \apj, 867, 23, \dodoi{10.3847/1538-4357/aac08b}

\bibitem[{S. {Huber} {et~al.}(2019){Huber}, {Suyu}, {Noebauer}, {Bonvin},
  {Rothchild}, {Chan}, {Awan}, {Courbin}, {Kromer}, {Marshall}, {Oguri},
  {Ribeiro}, \& {LSST Dark Energy Science Collaboration}}]{Huber:2019aa}
{Huber}, S., {Suyu}, S.~H., {Noebauer}, U.~M., {et~al.} 2019,
  \bibinfo{title}{{Strongly lensed SNe Ia in the era of LSST: observing cadence
  for lens discoveries and time-delay measurements},} \aap, 631, A161,
  \dodoi{10.1051/0004-6361/201935370}

\bibitem[{S. {Huber} {et~al.}(2022){Huber}, {Suyu}, {Ghoshdastidar},
  {Taubenberger}, {Bonvin}, {Chan}, {Kromer}, {Noebauer}, {Sim}, \&
  {Leal-Taix{\'e}}}]{Huber:2022aa}
{Huber}, S., {Suyu}, S.~H., {Ghoshdastidar}, D., {et~al.} 2022,
  \bibinfo{title}{{HOLISMOKES. VII. Time-delay measurement of strongly lensed
  Type Ia supernovae using machine learning},} \aap, 658, A157,
  \dodoi{10.1051/0004-6361/202141956}

\bibitem[{C.-L. {Hung} {et~al.}(2016){Hung}, {Casey}, {Chiang}, {Capak},
  {Cowley}, {Darvish}, {Kacprzak}, {Kova{\v{c}}}, {Lilly}, {Nanayakkara},
  {Spitler}, {Tran}, \& {Yuan}}]{Hung:2016aa}
{Hung}, C.-L., {Casey}, C.~M., {Chiang}, Y.-K., {et~al.} 2016,
  \bibinfo{title}{{Large-scale Structure around a z=2.1 Cluster},} \apj, 826,
  130, \dodoi{10.3847/0004-637X/826/2/130}

\bibitem[{J.~D. {Hunter}(2007){Hunter}}]{Hunter:2007aa}
{Hunter}, J.~D. 2007, \bibinfo{title}{{Matplotlib: A 2D Graphics Environment},}
  Computing in Science and Engineering, 9, 90, \dodoi{10.1109/MCSE.2007.55}

\bibitem[{A. {Immeli} {et~al.}(2004){Immeli}, {Samland}, {Gerhard}, \&
  {Westera}}]{Immeli:2004aa}
{Immeli}, A., {Samland}, M., {Gerhard}, O., \& {Westera}, P. 2004,
  \bibinfo{title}{{Gas physics, disk fragmentation, and bulge formation in
  young galaxies},} \aap, 413, 547, \dodoi{10.1051/0004-6361:20034282}

\bibitem[{M.~J. {Irwin} {et~al.}(1989){Irwin}, {Webster}, {Hewett}, {Corrigan},
  \& {Jedrzejewski}}]{Irwin:1989aa}
{Irwin}, M.~J., {Webster}, R.~L., {Hewett}, P.~C., {Corrigan}, R.~T., \&
  {Jedrzejewski}, R.~I. 1989, \bibinfo{title}{{Photometric Variations in the
  Q2237+0305 System: First Detection of a Microlensing Event},} \aj, 98, 1989,
  \dodoi{10.1086/115272}

\bibitem[{T.~H. {Jarrett} {et~al.}(2013){Jarrett}, {Masci}, {Tsai}, {Petty},
  {Cluver}, {Assef}, {Benford}, {Blain}, {Bridge}, {Donoso}, {Eisenhardt},
  {Koribalski}, {Lake}, {Neill}, {Seibert}, {Sheth}, {Stanford}, \&
  {Wright}}]{Jarrett:2013aa}
{Jarrett}, T.~H., {Masci}, F., {Tsai}, C.~W., {et~al.} 2013,
  \bibinfo{title}{{Extending the Nearby Galaxy Heritage with WISE: First
  Results from the WISE Enhanced Resolution Galaxy Atlas},} \aj, 145, 6,
  \dodoi{10.1088/0004-6256/145/1/6}

\bibitem[{J.~E. {Jencson} {et~al.}(2019){Jencson}, {Kasliwal}, {Adams}, {Bond},
  {De}, {Johansson}, {Karambelkar}, {Lau}, {Tinyanont}, {Ryder}, {Cody},
  {Masci}, {Bally}, {Blagorodnova}, {Castell{\'o}n}, {Fremling}, {Gehrz},
  {Helou}, {Kilpatrick}, {Milne}, {Morrell}, {Perley}, {Phillips}, {Smith},
  {van Dyk}, \& {Williams}}]{Jencson:2019aa}
{Jencson}, J.~E., {Kasliwal}, M.~M., {Adams}, S.~M., {et~al.} 2019,
  \bibinfo{title}{{The SPIRITS Sample of Luminous Infrared Transients:
  Uncovering Hidden Supernovae and Dusty Stellar Outbursts in Nearby
  Galaxies},} \apj, 886, 40, \dodoi{10.3847/1538-4357/ab4a01}

\bibitem[{Z. {Ji} {et~al.}(2025){Ji}, {Alberts}, {Zhu}, {Vanzella},
  {Giavalisco}, {Hainline}, {Baker}, {Bunker}, {Helton}, {Lyu}, {Rinaldi},
  {Robertson}, {Simmonds}, {Tacchella}, {Williams}, {Willmer}, \&
  {Witstok}}]{Ji:2025aa}
{Ji}, Z., {Alberts}, S., {Zhu}, Y., {et~al.} 2025, \bibinfo{title}{{The
  Importance of Dust Distribution in Ionizing-photon Escape: NIRCam and MIRI
  Imaging of a Lyman Continuum-emitting Galaxy at z
  \raisebox{-0.5ex}\textasciitilde 3.8},} arXiv e-prints, arXiv:2504.01067.
\newblock \doarXiv{2504.01067}

\bibitem[{E.~F. {Jim{\'e}nez-Andrade} {et~al.}(2019){Jim{\'e}nez-Andrade},
  {Magnelli}, {Karim}, {Zamorani}, {Bondi}, {Schinnerer}, {Sargent},
  {Romano-D{\'\i}az}, {Novak}, {Lang}, {Bertoldi}, {Vardoulaki}, {Toft},
  {Smol{\v{c}}i{\'c}}, {Harrington}, {Leslie}, {Delhaize}, {Liu}, {Karoumpis},
  {Kartaltepe}, \& {Koekemoer}}]{Jimenez-Andrade:2019aa}
{Jim{\'e}nez-Andrade}, E.~F., {Magnelli}, B., {Karim}, A., {et~al.} 2019,
  \bibinfo{title}{{Radio continuum size evolution of star-forming galaxies over
  0.35 < z < 2.25},} \aap, 625, A114, \dodoi{10.1051/0004-6361/201935178}

\bibitem[{T.~L. {Johnson} \& K. {Sharon}(2016){Johnson} \&
  {Sharon}}]{Johnson:2016aa}
{Johnson}, T.~L., \& {Sharon}, K. 2016, \bibinfo{title}{{The Systematics of
  Strong Lens Modeling Quantified: The Effects of Constraint Selection and
  Redshift Information on Magnification, Mass, and Multiple Image
  Predictability},} \apj, 832, 82, \dodoi{10.3847/0004-637X/832/1/82}

\bibitem[{E. {Jullo} \& J.-P. {Kneib}(2009){Jullo} \& {Kneib}}]{Jullo:2009aa}
{Jullo}, E., \& {Kneib}, J.-P. 2009, \bibinfo{title}{{Multiscale cluster lens
  mass mapping - I. Strong lensing modelling},} \mnras, 395, 1319,
  \dodoi{10.1111/j.1365-2966.2009.14654.x}

\bibitem[{E. {Jullo} {et~al.}(2007){Jullo}, {Kneib}, {Limousin},
  {El{\'{\i}}asd{\'o}ttir}, {Marshall}, \& {Verdugo}}]{Jullo:2007aa}
{Jullo}, E., {Kneib}, J.-P., {Limousin}, M., {et~al.} 2007, \bibinfo{title}{{A
  Bayesian approach to strong lensing modelling of galaxy clusters},} New
  Journal of Physics, 9, 447, \dodoi{10.1088/1367-2630/9/12/447}

\bibitem[{B.~S. {Kalita} {et~al.}(2025){Kalita}, {Silverman}, {Daddi},
  {Mercier}, {Ho}, \& {Ding}}]{Kalita:2025ac}
{Kalita}, B.~S., {Silverman}, J.~D., {Daddi}, E., {et~al.} 2025,
  \bibinfo{title}{{Near-IR clumps and their properties in high-z galaxies with
  JWST/NIRCam},} \mnras, 537, 402, \dodoi{10.1093/mnras/staf031}

\bibitem[{P.~S. {Kamieneski}(2024){Kamieneski}}]{Kamieneski:2024ab}
{Kamieneski}, P.~S. 2024, \bibinfo{title}{{Where are the Eddington-limited
  starbursts? Gravitational lensing provides a way forward for sub-kiloparsec
  views of star formation},} IAU Symposium, 381, 147,
  \dodoi{10.1017/S174392132300368X}

\bibitem[{P.~S. {Kamieneski} {et~al.}(2024{\natexlab{a}}){Kamieneski}, {Yun},
  {Harrington}, {Lowenthal}, {Wang}, {Frye}, {Jim{\'e}nez-Andrade}, {Vishwas},
  {Cooper}, {Pascale}, {Foo}, {Berman}, {Englert}, \& {Garcia
  Diaz}}]{Kamieneski:2024aa}
{Kamieneski}, P.~S., {Yun}, M.~S., {Harrington}, K.~C., {et~al.}
  2024{\natexlab{a}}, \bibinfo{title}{{PASSAGES: The Wide-ranging, Extreme
  Intrinsic Properties of Planck-selected, Lensed Dusty Star-forming
  Galaxies},} \apj, 961, 2, \dodoi{10.3847/1538-4357/acf930}

\bibitem[{P.~S. {Kamieneski} {et~al.}(2024{\natexlab{b}}){Kamieneski}, {Frye},
  {Windhorst}, {Harrington}, {Yun}, {Noble}, {Pascale}, {Foo}, {Cohen},
  {Jansen}, {Carleton}, {Koekemoer}, {Willmer}, {Summers}, {Garuda},
  {Leimbach}, {Holwerda}, {Pierel}, {Jim{\'e}nez-Andrade}, {Willner}, {Alcalde
  Pampliega}, {Vishwas}, {Keel}, {Wang}, {Cheng}, {Coe}, {Conselice},
  {D'Silva}, {Driver}, {Grogin}, {Hinrichs}, {Lowenthal}, {Marshall}, {Nonino},
  {Ortiz}, {Pigarelli}, {Pirzkal}, {Polletta}, {Robotham}, {Ryan}, \&
  {Yan}}]{Kamieneski:2024ac}
{Kamieneski}, P.~S., {Frye}, B.~L., {Windhorst}, R.~A., {et~al.}
  2024{\natexlab{b}}, \bibinfo{title}{{Birds of a Feather: Resolving Stellar
  Mass Assembly with JWST/NIRCam in a Pair of Kindred z {\ensuremath{\sim}} 2
  Dusty Star-forming Galaxies Lensed by the PLCK G165.7+67.0 Cluster},} \apj,
  973, 25, \dodoi{10.3847/1538-4357/ad5d59}

\bibitem[{E. {Kankare} {et~al.}(2008){Kankare}, {Mattila}, {Ryder},
  {P{\'e}rez-Torres}, {Alberdi}, {Romero-Canizales}, {D{\'\i}az-Santos},
  {V{\"a}is{\"a}nen}, {Efstathiou}, {Alonso-Herrero}, {Colina}, \&
  {Kotilainen}}]{Kankare:2008aa}
{Kankare}, E., {Mattila}, S., {Ryder}, S., {et~al.} 2008,
  \bibinfo{title}{{Discovery of a Very Highly Extinguished Supernova in a
  Luminous Infrared Galaxy},} \apjl, 689, L97, \dodoi{10.1086/595820}

\bibitem[{E. {Kankare} {et~al.}(2012){Kankare}, {Mattila}, {Ryder},
  {V{\"a}is{\"a}nen}, {Alberdi}, {Alonso-Herrero}, {Colina}, {Efstathiou},
  {Kotilainen}, {Melinder}, {P{\'e}rez-Torres}, {Romero-Ca{\~n}izales}, \&
  {Takalo}}]{Kankare:2012aa}
{Kankare}, E., {Mattila}, S., {Ryder}, S., {et~al.} 2012,
  \bibinfo{title}{{Discovery of Two Supernovae in the Nuclear Regions of the
  Luminous Infrared Galaxy IC 883},} \apjl, 744, L19,
  \dodoi{10.1088/2041-8205/744/2/L19}

\bibitem[{E. {Kankare} {et~al.}(2021){Kankare}, {Efstathiou}, {Kotak}, {Kool},
  {Kangas}, {O'Neill}, {Mattila}, {V{\"a}is{\"a}nen}, {Ramphul}, {Mogotsi},
  {Ryder}, {Parker}, {Reynolds}, {Fraser}, {Pastorello}, {Cappellaro},
  {Mazzali}, {Ochner}, {Tomasella}, {Turatto}, {Kotilainen}, {Kuncarayakti},
  {P{\'e}rez-Torres}, {Randriamanakoto}, {Romero-Ca{\~n}izales}, {Berton},
  {Cartier}, {Chen}, {Galbany}, {Gromadzki}, {Inserra}, {Maguire}, {Moran},
  {M{\"u}ller-Bravo}, {Nicholl}, {Reguitti}, \& {Young}}]{Kankare:2021aa}
{Kankare}, E., {Efstathiou}, A., {Kotak}, R., {et~al.} 2021,
  \bibinfo{title}{{Core-collapse supernova subtypes in luminous infrared
  galaxies},} \aap, 649, A134, \dodoi{10.1051/0004-6361/202039240}

\bibitem[{G. {Kauffmann} {et~al.}(1999){Kauffmann}, {Colberg}, {Diaferio}, \&
  {White}}]{Kauffmann:1999ab}
{Kauffmann}, G., {Colberg}, J.~M., {Diaferio}, A., \& {White}, S. D.~M. 1999,
  \bibinfo{title}{{Clustering of galaxies in a hierarchical universe - II.
  Evolution to high redshift},} \mnras, 307, 529,
  \dodoi{10.1046/j.1365-8711.1999.02711.x}

\bibitem[{C.~R. {Keeton} {et~al.}(2000){Keeton}, {Falco}, {Impey}, {Kochanek},
  {Leh{\'a}r}, {McLeod}, {Rix}, {Mu{\~n}oz}, \& {Peng}}]{Keeton:2000aa}
{Keeton}, C.~R., {Falco}, E.~E., {Impey}, C.~D., {et~al.} 2000,
  \bibinfo{title}{{The Host Galaxy of the Lensed Quasar Q0957+561},} \apj, 542,
  74, \dodoi{10.1086/309517}

\bibitem[{P.~L. {Kelly} {et~al.}(2015){Kelly}, {Rodney}, {Treu}, {Foley},
  {Brammer}, {Schmidt}, {Zitrin}, {Sonnenfeld}, {Strolger}, {Graur},
  {Filippenko}, {Jha}, {Riess}, {Bradac}, {Weiner}, {Scolnic}, {Malkan}, {von
  der Linden}, {Trenti}, {Hjorth}, {Gavazzi}, {Fontana}, {Merten}, {McCully},
  {Jones}, {Postman}, {Dressler}, {Patel}, {Cenko}, {Graham}, \&
  {Tucker}}]{Kelly:2015aa}
{Kelly}, P.~L., {Rodney}, S.~A., {Treu}, T., {et~al.} 2015,
  \bibinfo{title}{{Multiple images of a highly magnified supernova formed by an
  early-type cluster galaxy lens},} Science, 347, 1123,
  \dodoi{10.1126/science.aaa3350}

\bibitem[{P.~L. {Kelly} {et~al.}(2016){Kelly}, {Rodney}, {Treu}, {Strolger},
  {Foley}, {Jha}, {Selsing}, {Brammer}, {Brada{\v c}}, {Cenko}, {Graur},
  {Filippenko}, {Hjorth}, {McCully}, {Molino}, {Nonino}, {Riess}, {Schmidt},
  {Tucker}, {von der Linden}, {Weiner}, \& {Zitrin}}]{Kelly:2016aa}
{Kelly}, P.~L., {Rodney}, S.~A., {Treu}, T., {et~al.} 2016,
  \bibinfo{title}{{Deja Vu All Over Again: The Reappearance of Supernova
  Refsdal},} \apjl, 819, L8, \dodoi{10.3847/2041-8205/819/1/L8}

\bibitem[{P.~L. {Kelly} {et~al.}(2023{\natexlab{a}}){Kelly}, {Rodney}, {Treu},
  {Oguri}, {Chen}, {Zitrin}, {Birrer}, {Bonvin}, {Dessart}, {Diego},
  {Filippenko}, {Foley}, {Gilman}, {Hjorth}, {Jauzac}, {Mandel}, {Millon},
  {Pierel}, {Sharon}, {Thorp}, {Williams}, {Broadhurst}, {Dressler}, {Graur},
  {Jha}, {McCully}, {Postman}, {Schmidt}, {Tucker}, \& {von der
  Linden}}]{Kelly:2023ab}
{Kelly}, P.~L., {Rodney}, S., {Treu}, T., {et~al.} 2023{\natexlab{a}},
  \bibinfo{title}{{Constraints on the Hubble constant from supernova Refsdal's
  reappearance},} Science, 380, abh1322, \dodoi{10.1126/science.abh1322}

\bibitem[{P.~L. {Kelly} {et~al.}(2023{\natexlab{b}}){Kelly}, {Rodney}, {Treu},
  {Birrer}, {Bonvin}, {Dessart}, {Foley}, {Filippenko}, {Gilman}, {Jha},
  {Hjorth}, {Mandel}, {Millon}, {Pierel}, {Thorp}, {Zitrin}, {Broadhurst},
  {Chen}, {Diego}, {Dressler}, {Graur}, {Jauzac}, {Malkan}, {McCully}, {Oguri},
  {Postman}, {Schmidt}, {Sharon}, {Tucker}, {von der Linden}, \&
  {Wambsganss}}]{Kelly:2023aa}
{Kelly}, P.~L., {Rodney}, S., {Treu}, T., {et~al.} 2023{\natexlab{b}},
  \bibinfo{title}{{The Magnificent Five Images of Supernova Refsdal: Time Delay
  and Magnification Measurements},} \apj, 948, 93,
  \dodoi{10.3847/1538-4357/ac4ccb}

\bibitem[{J. {Kennicutt}(1998){Kennicutt}}]{Kennicutt:1998ac}
{Kennicutt}, Robert~C., J. 1998, \bibinfo{title}{{The Global Schmidt Law in
  Star-forming Galaxies},} \apj, 498, 541, \dodoi{10.1086/305588}

\bibitem[{L.~J. {Kewley} {et~al.}(2002){Kewley}, {Geller}, {Jansen}, \&
  {Dopita}}]{Kewley:2002ab}
{Kewley}, L.~J., {Geller}, M.~J., {Jansen}, R.~A., \& {Dopita}, M.~A. 2002,
  \bibinfo{title}{{The H{\ensuremath{\alpha}} and Infrared Star Formation Rates
  for the Nearby Field Galaxy Survey},} \aj, 124, 3135, \dodoi{10.1086/344487}

\bibitem[{C.-G. {Kim} \& E.~C. {Ostriker}(2015){Kim} \&
  {Ostriker}}]{Kim:2015aa}
{Kim}, C.-G., \& {Ostriker}, E.~C. 2015, \bibinfo{title}{{Momentum Injection by
  Supernovae in the Interstellar Medium},} \apj, 802, 99,
  \dodoi{10.1088/0004-637X/802/2/99}

\bibitem[{J.-P. {Kneib} {et~al.}(1996){Kneib}, {Ellis}, {Smail}, {Couch}, \&
  {Sharples}}]{Kneib:1996aa}
{Kneib}, J.-P., {Ellis}, R.~S., {Smail}, I., {Couch}, W.~J., \& {Sharples},
  R.~M. 1996, \bibinfo{title}{{Hubble Space Telescope Observations of the
  Lensing Cluster Abell 2218},} \apj, 471, 643, \dodoi{10.1086/177995}

\bibitem[{J.~P. {Kneib} {et~al.}(1993){Kneib}, {Mellier}, {Fort}, \&
  {Mathez}}]{Kneib:1993aa}
{Kneib}, J.~P., {Mellier}, Y., {Fort}, B., \& {Mathez}, G. 1993,
  \bibinfo{title}{{The Distribution of Dark Matter in Distant Cluster Lenses -
  Modelling A:370},} \aap, 273, 367

\bibitem[{C. {Kochanek} {et~al.}(2006){Kochanek}, {Schneider}, , \&
  {Wambsganss}}]{Kochanek:2006aa}
{Kochanek}, C., {Schneider}, P., , \& {Wambsganss}, J. 2006, {Gravitational
  Lensing: Strong, Weak and Micro}, 1st edn., ed. G.~{Meylan}, P.~{Jetzer}, \&
  e.~a. {North}, P., Saas-Fee Advanced Courses (Springer).
\newblock \doarXiv{astro-ph/0407232}

\bibitem[{C.~S. {Kochanek}(2002){Kochanek}}]{Kochanek:2002aa}
{Kochanek}, C.~S. 2002, \bibinfo{title}{{What Do Gravitational Lens Time Delays
  Measure?},} \apj, 578, 25, \dodoi{10.1086/342476}

\bibitem[{C.~S. {Kochanek}(2020){Kochanek}}]{Kochanek:2020ab}
{Kochanek}, C.~S. 2020, \bibinfo{title}{{Overconstrained gravitational lens
  models and the Hubble constant},} \mnras, 493, 1725,
  \dodoi{10.1093/mnras/staa344}

\bibitem[{E.~C. {Kool} {et~al.}(2018){Kool}, {Ryder}, {Kankare}, {Mattila},
  {Reynolds}, {McDermid}, {P{\'e}rez-Torres}, {Herrero-Illana}, {Schirmer},
  {Efstathiou}, {Bauer}, {Kotilainen}, {V{\"a}is{\"a}nen}, {Baldwin},
  {Romero-Ca{\~n}izales}, \& {Alberdi}}]{Kool:2018aa}
{Kool}, E.~C., {Ryder}, S., {Kankare}, E., {et~al.} 2018,
  \bibinfo{title}{{First results from GeMS/GSAOI for project SUNBIRD:
  Supernovae UNmasked By Infra-Red Detection},} \mnras, 473, 5641,
  \dodoi{10.1093/mnras/stx2463}

\bibitem[{L.~V.~E. {Koopmans} {et~al.}(2003){Koopmans}, {Treu}, {Fassnacht},
  {Blandford}, \& {Surpi}}]{Koopmans:2003aa}
{Koopmans}, L.~V.~E., {Treu}, T., {Fassnacht}, C.~D., {Blandford}, R.~D., \&
  {Surpi}, G. 2003, \bibinfo{title}{{The Hubble Constant from the Gravitational
  Lens B1608+656},} \apj, 599, 70, \dodoi{10.1086/379226}

\bibitem[{P. {Kroupa}(2001){Kroupa}}]{Kroupa:2001aa}
{Kroupa}, P. 2001, \bibinfo{title}{{On the variation of the initial mass
  function},} \mnras, 322, 231, \dodoi{10.1046/j.1365-8711.2001.04022.x}

\bibitem[{A.~J. {Lee} {et~al.}(2021){Lee}, {Freedman}, {Madore}, {Owens},
  {Monson}, \& {Hoyt}}]{Lee:2021aa}
{Lee}, A.~J., {Freedman}, W.~L., {Madore}, B.~F., {et~al.} 2021,
  \bibinfo{title}{{The Astrophysical Distance Scale. III. Distance to the Local
  Group Galaxy WLM Using Multiwavelength Observations of the Tip of the Red
  Giant Branch, Cepheids, and JAGB Stars},} \apj, 907, 112,
  \dodoi{10.3847/1538-4357/abd253}

\bibitem[{E. {Leitet} {et~al.}(2013){Leitet}, {Bergvall}, {Hayes}, {Linn{\'e}},
  \& {Zackrisson}}]{Leitet:2013aa}
{Leitet}, E., {Bergvall}, N., {Hayes}, M., {Linn{\'e}}, S., \& {Zackrisson}, E.
  2013, \bibinfo{title}{{Escape of Lyman continuum radiation from local
  galaxies. Detection of leakage from the young starburst Tol 1247-232},} \aap,
  553, A106, \dodoi{10.1051/0004-6361/201118370}

\bibitem[{R.~C. {Levy} {et~al.}(2024){Levy}, {Bolatto}, {Mayya},
  {Cuevas-Otahola}, {Tarantino}, {Boyer}, {Boogaard}, {B{\"o}ker}, {Cronin},
  {Dale}, {Donaghue}, {Emig}, {Fisher}, {Glover}, {Herrera-Camus},
  {Jim{\'e}nez-Donaire}, {Klessen}, {Lenki{\'c}}, {Leroy}, {De Looze}, {Meier},
  {Mills}, {Ott}, {Rela{\~n}o}, {Veilleux}, {Villanueva}, {Walter}, \& {van der
  Werf}}]{Levy:2024aa}
{Levy}, R.~C., {Bolatto}, A.~D., {Mayya}, D., {et~al.} 2024,
  \bibinfo{title}{{JWST Observations of Starbursts: Massive Star Clusters in
  the Central Starburst of M82},} \apjl, 973, L55,
  \dodoi{10.3847/2041-8213/ad7af3}

\bibitem[{A.~J.~R. {Lewis} {et~al.}(2018){Lewis}, {Ivison}, {Best}, {Simpson},
  {Weiss}, {Oteo}, {Zhang}, {Arumugam}, {Bremer}, {Chapman}, {Clements},
  {Dannerbauer}, {Dunne}, {Eales}, {Maddox}, {Oliver}, {Omont}, {Riechers},
  {Serjeant}, {Valiante}, {Wardlow}, {van der Werf}, \& {De
  Zotti}}]{Lewis:2018aa}
{Lewis}, A.~J.~R., {Ivison}, R.~J., {Best}, P.~N., {et~al.} 2018,
  \bibinfo{title}{{Ultra-red Galaxies Signpost Candidate Protoclusters at High
  Redshift},} \apj, 862, 96, \dodoi{10.3847/1538-4357/aacc25}

\bibitem[{L. {Liang} {et~al.}(2021){Liang}, {Feldmann}, {Hayward}, {Narayanan},
  {{\c{C}}atmabacak}, {Kere{\v{s}}}, {Faucher-Gigu{\`e}re}, \&
  {Hopkins}}]{Liang:2021aa}
{Liang}, L., {Feldmann}, R., {Hayward}, C.~C., {et~al.} 2021,
  \bibinfo{title}{{The IRX-{\ensuremath{\beta}} relation of high-redshift
  galaxies},} \mnras, 502, 3210, \dodoi{10.1093/mnras/stab096}

\bibitem[{S.~T. {Linden} {et~al.}(2017){Linden}, {Evans}, {Rich}, {Larson},
  {Armus}, {D{\'\i}az-Santos}, {Privon}, {Howell}, {Inami}, {Kim}, {Chien},
  {Vavilkin}, {Mazzarella}, {Modica}, {Surace}, {Manning}, {Abdullah}, {Blake},
  {Yarber}, \& {Lambert}}]{Linden:2017aa}
{Linden}, S.~T., {Evans}, A.~S., {Rich}, J., {et~al.} 2017,
  \bibinfo{title}{{Massive Star Cluster Formation and Destruction in Luminous
  Infrared Galaxies in GOALS},} \apj, 843, 91, \dodoi{10.3847/1538-4357/aa7266}

\bibitem[{D. {Liu} {et~al.}(2019){Liu}, {Schinnerer}, {Groves}, {Magnelli},
  {Lang}, {Leslie}, {Jim{\'e}nez-Andrade}, {Riechers}, {Popping}, {Magdis},
  {Daddi}, {Sargent}, {Gao}, {Fudamoto}, {Oesch}, \& {Bertoldi}}]{Liu:2019ae}
{Liu}, D., {Schinnerer}, E., {Groves}, B., {et~al.} 2019,
  \bibinfo{title}{{Automated Mining of the ALMA Archive in the COSMOS Field
  (A$^{3}$COSMOS). II. Cold Molecular Gas Evolution out to Redshift 6},} \apj,
  887, 235, \dodoi{10.3847/1538-4357/ab578d}

\bibitem[{D. {Liu} {et~al.}(2024){Liu}, {F{\"o}rster Schreiber}, {Harrington},
  {Lee}, {Kamieneski}, {Davies}, {Lutz}, {Renzini}, {Wuyts}, {Tacconi},
  {Genzel}, {Burkert}, {Herrera-Camus}, {Alcalde Pampliega}, {Vishwas},
  {Kaasinen}, {Wang}, {Jim{\'e}nez-Andrade}, {Lowenthal}, {Foo}, {Frye},
  {Shangguan}, {Cao}, {Agapito}, {Berbel}, {Barfety}, {Baruffolo}, {Berman},
  {Black}, {Bonaglia}, {Briguglio}, {Carbonaro}, {Chapman}, {Chen}, {Cikota},
  {Concas}, {Cooper}, {Cresci}, {Dallilar}, {Deysenroth}, {Di Antonio}, {Di
  Cianno}, {Di Rico}, {Doelman}, {Dolci}, {Eisenhauer}, {Espejo}, {Esposito},
  {Fantinel}, {Ferruzzi}, {Feuchtgruber}, {Gao}, {Garcia Diaz}, {Gillessen},
  {Grani}, {Hartl}, {Henry}, {Huber}, {Jolly}, {Keller}, {Kenworthy},
  {Kravchenko}, {Lee}, {Lightfoot}, {Lunney}, {Macintosh}, {Mannucci}, {Ott},
  {Pascale}, {Pastras}, {Pearson}, {Puglisi}, {Pulsoni}, {Rabien}, {Rau},
  {Riccardi}, {Salasnich}, {Shimizu}, {Snik}, {Sturm}, {Taylor}, {Valentini},
  {Waring}, {Wiezorrek}, {Xompero}, \& {Yun}}]{Liu:2024ae}
{Liu}, D., {F{\"o}rster Schreiber}, N.~M., {Harrington}, K.~C., {et~al.} 2024,
  \bibinfo{title}{{Detailed study of a rare hyperluminous rotating disk in an
  Einstein ring 10 billion years ago},} Nature Astronomy, 8, 1181,
  \dodoi{10.1038/s41550-024-02296-7}

\bibitem[{Z. {Liu} {et~al.}(2024){Liu}, {Silverman}, {Daddi}, {Puglisi},
  {Renzini}, {Kalita}, {Kartaltepe}, {Kashino}, {Rodighiero}, {Rujopakarn},
  {Suzuki}, {Tanaka}, {Valentino}, {Andika}, {Casey}, {Faisst}, {Franco},
  {Gozaliasl}, {Gillman}, {Hayward}, {Koekemoer}, {Kokorev}, {Lambrides},
  {Lee}, {Magdis}, {Harish}, {McCracken}, {Rhodes}, {Shuntov}, \&
  {Ding}}]{Liu:2024af}
{Liu}, Z., {Silverman}, J.~D., {Daddi}, E., {et~al.} 2024,
  \bibinfo{title}{{JWST and ALMA Discern the Assembly of Structural and
  Obscured Components in a High-redshift Starburst Galaxy},} \apj, 968, 15,
  \dodoi{10.3847/1538-4357/ad4096}

\bibitem[{A.~S. {Long} {et~al.}(2020){Long}, {Cooray}, {Ma}, {Casey},
  {Wardlow}, {Nayyeri}, {Ivison}, {Farrah}, \& {Dannerbauer}}]{Long:2020aa}
{Long}, A.~S., {Cooray}, A., {Ma}, J., {et~al.} 2020,
  \bibinfo{title}{{Emergence of an Ultrared, Ultramassive Galaxy Cluster Core
  at z = 4},} \apj, 898, 133, \dodoi{10.3847/1538-4357/ab9d1f}

\bibitem[{J.~M. {Lotz} {et~al.}(2017){Lotz}, {Koekemoer}, {Coe}, {Grogin},
  {Capak}, {Mack}, {Anderson}, {Avila}, {Barker}, {Borncamp}, {Brammer},
  {Durbin}, {Gunning}, {Hilbert}, {Jenkner}, {Khandrika}, {Levay}, {Lucas},
  {MacKenty}, {Ogaz}, {Porterfield}, {Reid}, {Robberto}, {Royle}, {Smith},
  {Storrie-Lombardi}, {Sunnquist}, {Surace}, {Taylor}, {Williams}, {Bullock},
  {Dickinson}, {Finkelstein}, {Natarajan}, {Richard}, {Robertson}, {Tumlinson},
  {Zitrin}, {Flanagan}, {Sembach}, {Soifer}, \& {Mountain}}]{Lotz:2017aa}
{Lotz}, J.~M., {Koekemoer}, A., {Coe}, D., {et~al.} 2017, \bibinfo{title}{{The
  Frontier Fields: Survey Design and Initial Results},} \apj, 837, 97,
  \dodoi{10.3847/1538-4357/837/1/97}

\bibitem[{T. {Louis} {et~al.}(2025){Louis}, {La Posta}, {Atkins}, {Jense},
  {Abril-Cabezas}, {Addison}, {Ade}, {Aiola}, {Alford}, {Alonso}, {Amiri},
  {An}, {Austermann}, {Barbavara}, {Battaglia}, {Battistelli}, {Beall}, {Bean},
  {Beheshti}, {Beringue}, {Bhandarkar}, {Biermann}, {Bolliet}, {Bond},
  {Calabrese}, {Capalbo}, {Carrero}, {Chen}, {Chesmore}, {Cho}, {Choi},
  {Clark}, {Cothard}, {Coughlin}, {Coulton}, {Crichton}, {Crowley}, {Darwish},
  {Devlin}, {Dicker}, {Duell}, {Duff}, {Duivenvoorden}, {Dunkley}, {Dunner},
  {Embil Villagra}, {Fankhanel}, {Farren}, {Ferraro}, {Foster}, {Freundt},
  {Fuzia}, {Gallardo}, {Garrido}, {Gerbino}, {Giardiello}, {Gill}, {Givans},
  {Gluscevic}, {Goldstein}, {Golec}, {Gong}, {Guan}, {Halpern}, {Harrison},
  {Hasselfield}, {Healy}, {Henderson}, {Hensley}, {Herv{\'\i}as-Caimapo},
  {Hill}, {Hilton}, {Hilton}, {Hincks}, {Hlo{\v{z}}ek}, {Ho}, {Hood},
  {Hornecker}, {Huber}, {Hubmayr}, {Huffenberger}, {Hughes}, {Ikape}, {Irwin},
  {Isopi}, {Joshi}, {Keller}, {Kim}, {Knowles}, {Koopman}, {Kosowsky},
  {Kramer}, {Kusiak}, {Lague}, {Lakey}, {Lee}, {Li}, {Li}, {Limon}, {Lokken},
  {Lungu}, {MacCrann}, {MacInnis}, {Madhavacheril}, {Maldonado}, {Maldonado},
  {Mallaby-Kay}, {Marques}, {van Marrewijk}, {McCarthy}, {McMahon}, {Mehta},
  {Menanteau}, {Moodley}, {Morris}, {Mroczkowski}, {Naess}, {Namikawa}, {Nati},
  {Nerval}, {Newburgh}, {Nicola}, {Niemack}, {Nolta}, {Orlowski-Scherer},
  {Pagano}, {Page}, {Pandey}, {Partridge}, {Perez Sarmiento}, {Prince},
  {Puddu}, {Qu}, {Ragavan}, {Ried Guachalla}, {Rogers}, {Rojas}, {Sakuma},
  {Schaan}, {Schmitt}, {Sehgal}, {Shaikh}, {Sherwin}, {Sierra}, {Sievers},
  {Sif{\'o}n}, {Simon}, {Sonka}, {Spergel}, {Staggs}, {Storer}, {Surrao},
  {Switzer}, {Tampier}, {Thornton}, {Trac}, {Tucker}, {Ullom}, {Vale}, {Van
  Engelen}, {Van Lanen}, {Vargas}, {Vavagiakis}, {Wagoner}, {Wang}, {Wenzl},
  {Wollack}, \& {Zheng}}]{Louis:2025aa}
{Louis}, T., {La Posta}, A., {Atkins}, Z., {et~al.} 2025, \bibinfo{title}{{The
  Atacama Cosmology Telescope: DR6 Power Spectra, Likelihoods and $Λ$CDM
  Parameters},} arXiv e-prints, arXiv:2503.14452,
  \dodoi{10.48550/arXiv.2503.14452}

\bibitem[{ {LSST Science Collaboration} {et~al.}(2009){LSST Science
  Collaboration}, {Abell}, {Allison}, {Anderson}, {Andrew}, {Angel}, {Armus},
  {Arnett}, {Asztalos}, {Axelrod}, {Bailey}, {Ballantyne}, {Bankert},
  {Barkhouse}, {Barr}, {Barrientos}, {Barth}, {Bartlett}, {Becker}, {Becla},
  {Beers}, {Bernstein}, {Biswas}, {Blanton}, {Bloom}, {Bochanski}, {Boeshaar},
  {Borne}, {Bradac}, {Brandt}, {Bridge}, {Brown}, {Brunner}, {Bullock},
  {Burgasser}, {Burge}, {Burke}, {Cargile}, {Chandrasekharan}, {Chartas},
  {Chesley}, {Chu}, {Cinabro}, {Claire}, {Claver}, {Clowe}, {Connolly}, {Cook},
  {Cooke}, {Cooray}, {Covey}, {Culliton}, {de Jong}, {de Vries}, {Debattista},
  {Delgado}, {Dell'Antonio}, {Dhital}, {Di Stefano}, {Dickinson}, {Dilday},
  {Djorgovski}, {Dobler}, {Donalek}, {Dubois-Felsmann}, {Durech},
  {Eliasdottir}, {Eracleous}, {Eyer}, {Falco}, {Fan}, {Fassnacht}, {Ferguson},
  {Fernandez}, {Fields}, {Finkbeiner}, {Figueroa}, {Fox}, {Francke}, {Frank},
  {Frieman}, {Fromenteau}, {Furqan}, {Galaz}, {Gal-Yam}, {Garnavich},
  {Gawiser}, {Geary}, {Gee}, {Gibson}, {Gilmore}, {Grace}, {Green}, {Gressler},
  {Grillmair}, {Habib}, {Haggerty}, {Hamuy}, {Harris}, {Hawley}, {Heavens},
  {Hebb}, {Henry}, {Hileman}, {Hilton}, {Hoadley}, {Holberg}, {Holman},
  {Howell}, {Infante}, {Ivezic}, {Jacoby}, {Jain}, {R}, {Jedicke}, {Jee},
  {Garrett Jernigan}, {Jha}, {Johnston}, {Jones}, {Juric}, {Kaasalainen},
  {Styliani}, {Kafka}, {Kahn}, {Kaib}, {Kalirai}, {Kantor}, {Kasliwal},
  {Keeton}, {Kessler}, {Knezevic}, {Kowalski}, {Krabbendam}, {Krughoff},
  {Kulkarni}, {Kuhlman}, {Lacy}, {Lepine}, {Liang}, {Lien}, {Lira}, {Long},
  {Lorenz}, {Lotz}, {Lupton}, {Lutz}, {Macri}, {Mahabal}, {Mandelbaum},
  {Marshall}, {May}, {McGehee}, {Meadows}, {Meert}, {Milani}, {Miller},
  {Miller}, {Mills}, {Minniti}, {Monet}, {Mukadam}, {Nakar}, {Neill}, {Newman},
  {Nikolaev}, {Nordby}, {O'Connor}, {Oguri}, {Oliver}, {Olivier}, {Olsen},
  {Olsen}, {Olszewski}, {Oluseyi}, {Padilla}, {Parker}, {Pepper}, {Peterson},
  {Petry}, {Pinto}, {Pizagno}, {Popescu}, {Prsa}, {Radcka}, {Raddick},
  {Rasmussen}, {Rau}, {Rho}, {Rhoads}, {Richards}, {Ridgway}, {Robertson},
  {Roskar}, {Saha}, {Sarajedini}, {Scannapieco}, {Schalk}, {Schindler}, \&
  {Schmidt}}]{LSST-Science-Collaboration:2009aa}
{LSST Science Collaboration}, {Abell}, P.~A., {Allison}, J., {et~al.} 2009,
  \bibinfo{title}{{LSST Science Book, Version 2.0},} arXiv e-prints,
  arXiv:0912.0201, \dodoi{10.48550/arXiv.0912.0201}

\bibitem[{W.~E. {Lucas} {et~al.}(2020){Lucas}, {Bonnell}, \&
  {Dale}}]{Lucas:2020aa}
{Lucas}, W.~E., {Bonnell}, I.~A., \& {Dale}, J.~E. 2020,
  \bibinfo{title}{{Supernova feedback and the energy deposition in molecular
  clouds},} \mnras, 493, 4700, \dodoi{10.1093/mnras/staa451}

\bibitem[{P. {Madau} {et~al.}(1998){Madau}, {della Valle}, \&
  {Panagia}}]{Madau:1998aa}
{Madau}, P., {della Valle}, M., \& {Panagia}, N. 1998, \bibinfo{title}{{On the
  evolution of the cosmic supernova rates},} \mnras, 297, L17,
  \dodoi{10.1046/j.1365-8711.1998.01697.x}

\bibitem[{P. {Madau} \& M. {Dickinson}(2014){Madau} \&
  {Dickinson}}]{Madau:2014aa}
{Madau}, P., \& {Dickinson}, M. 2014, \bibinfo{title}{{Cosmic Star-Formation
  History},} \araa, 52, 415, \dodoi{10.1146/annurev-astro-081811-125615}

\bibitem[{M.~R. {Magee} {et~al.}(2023){Magee}, {Sainz de Murieta}, {Collett},
  \& {Enzi}}]{Magee:2023aa}
{Magee}, M.~R., {Sainz de Murieta}, A., {Collett}, T.~E., \& {Enzi}, W. 2023,
  \bibinfo{title}{{Identifying gravitationally lensed supernovae within the
  Zwicky Transient Facility public survey},} arXiv e-prints, arXiv:2303.15439,
  \dodoi{10.48550/arXiv.2303.15439}

\bibitem[{F. {Mannucci} {et~al.}(2006){Mannucci}, {Della Valle}, \&
  {Panagia}}]{Mannucci:2006aa}
{Mannucci}, F., {Della Valle}, M., \& {Panagia}, N. 2006, \bibinfo{title}{{Two
  populations of progenitors for Type Ia supernovae?},} \mnras, 370, 773,
  \dodoi{10.1111/j.1365-2966.2006.10501.x}

\bibitem[{F. {Mannucci} {et~al.}(2005){Mannucci}, {Della Valle}, {Panagia},
  {Cappellaro}, {Cresci}, {Maiolino}, {Petrosian}, \&
  {Turatto}}]{Mannucci:2005aa}
{Mannucci}, F., {Della Valle}, M., {Panagia}, N., {et~al.} 2005,
  \bibinfo{title}{{The supernova rate per unit mass},} \aap, 433, 807,
  \dodoi{10.1051/0004-6361:20041411}

\bibitem[{F. {Mannucci} {et~al.}(2003){Mannucci}, {Maiolino}, {Cresci}, {Della
  Valle}, {Vanzi}, {Ghinassi}, {Ivanov}, {Nagar}, \&
  {Alonso-Herrero}}]{Mannucci:2003aa}
{Mannucci}, F., {Maiolino}, R., {Cresci}, G., {et~al.} 2003,
  \bibinfo{title}{{The infrared supernova rate in starburst galaxies},} \aap,
  401, 519, \dodoi{10.1051/0004-6361:20030198}

\bibitem[{I. {M{\"a}ntynen} {et~al.}(2025){M{\"a}ntynen}, {Kankare}, {Mattila},
  {Efstathiou}, {Ryder}, {Reynolds}, {Vassallo}, \&
  {V{\"a}is{\"a}nen}}]{Mantynen:2025aa}
{M{\"a}ntynen}, I., {Kankare}, E., {Mattila}, S., {et~al.} 2025,
  \bibinfo{title}{{The undetectable fraction of core-collapse supernovae in
  luminous infrared galaxies},} arXiv e-prints, arXiv:2506.18518,
  \dodoi{10.48550/arXiv.2506.18518}

\bibitem[{D. {Maoz} {et~al.}(2014){Maoz}, {Mannucci}, \&
  {Nelemans}}]{Maoz:2014aa}
{Maoz}, D., {Mannucci}, F., \& {Nelemans}, G. 2014,
  \bibinfo{title}{{Observational Clues to the Progenitors of Type Ia
  Supernovae},} \araa, 52, 107, \dodoi{10.1146/annurev-astro-082812-141031}

\bibitem[{I. {Mart{\'\i}-Vidal} {et~al.}(2007){Mart{\'\i}-Vidal}, {Marcaide},
  {Alberdi}, {Guirado}, {Lara}, {P{\'e}rez-Torres}, {Ros}, {Argo}, {Beswick},
  {Muxlow}, {Pedlar}, {Shapiro}, {Stockdale}, {Sramek}, {Weiler}, \&
  {Vinko}}]{Marti-Vidal:2007aa}
{Mart{\'\i}-Vidal}, I., {Marcaide}, J.~M., {Alberdi}, A., {et~al.} 2007,
  \bibinfo{title}{{8.4 GHz VLBI observations of SN{\,}2004et in NGC{\,}6946},}
  \aap, 470, 1071, \dodoi{10.1051/0004-6361:20077522}

\bibitem[{S. {Mattila} \& W.~P.~S. {Meikle}(2001){Mattila} \&
  {Meikle}}]{Mattila:2001aa}
{Mattila}, S., \& {Meikle}, W.~P.~S. 2001, \bibinfo{title}{{Supernovae in the
  nuclear regions of starburst galaxies},} \mnras, 324, 325,
  \dodoi{10.1046/j.1365-8711.2001.04255.x}

\bibitem[{S. {Mattila} {et~al.}(2007){Mattila}, {V{\"a}is{\"a}nen}, {Farrah},
  {Efstathiou}, {Meikle}, {Dahlen}, {Fransson}, {Lira}, {Lundqvist},
  {{\"O}stlin}, {Ryder}, \& {Sollerman}}]{Mattila:2007aa}
{Mattila}, S., {V{\"a}is{\"a}nen}, P., {Farrah}, D., {et~al.} 2007,
  \bibinfo{title}{{Adaptive Optics Discovery of Supernova 2004ip in the Nuclear
  Regions of the Luminous Infrared Galaxy IRAS 18293-3413},} \apjl, 659, L9,
  \dodoi{10.1086/516821}

\bibitem[{W. {McKinney}(2010){McKinney}}]{McKinney:2010aa}
{McKinney}, W. 2010, \bibinfo{title}{{D}ata {S}tructures for {S}tatistical
  {C}omputing in {P}ython,} in {P}roceedings of the 9th {P}ython in {S}cience
  {C}onference, ed. {S}t\'efan van~der {W}alt \& {J}arrod {M}illman, 56 -- 61,
  \dodoi{10.25080/Majora-92bf1922-00a}

\bibitem[{J.~P. {McMullin} {et~al.}(2007){McMullin}, {Waters}, {Schiebel},
  {Young}, \& {Golap}}]{McMullin:2007aa}
{McMullin}, J.~P., {Waters}, B., {Schiebel}, D., {Young}, W., \& {Golap}, K.
  2007, \bibinfo{title}{{CASA Architecture and Applications},} in Astronomical
  Society of the Pacific Conference Series, Vol. 376, Astronomical Data
  Analysis Software and Systems XVI, ed. R.~A. {Shaw}, F.~{Hill}, \& D.~J.
  {Bell}, 127

\bibitem[{K.~B.~W. {McQuinn} {et~al.}(2012){McQuinn}, {Skillman}, {Dalcanton},
  {Cannon}, {Dolphin}, {Holtzman}, {Weisz}, \& {Williams}}]{McQuinn:2012aa}
{McQuinn}, K. B.~W., {Skillman}, E.~D., {Dalcanton}, J.~J., {et~al.} 2012,
  \bibinfo{title}{{The Nature of Starbursts. III. The Spatial Distribution of
  Star Formation},} \apj, 759, 77, \dodoi{10.1088/0004-637X/759/1/77}

\bibitem[{J. {Melinder} {et~al.}(2012){Melinder}, {Dahlen}, {Menc{\'\i}a
  Trinchant}, {{\"O}stlin}, {Mattila}, {Sollerman}, {Fransson}, {Hayes},
  {Kankare}, \& {Nasoudi-Shoar}}]{Melinder:2012aa}
{Melinder}, J., {Dahlen}, T., {Menc{\'\i}a Trinchant}, L., {et~al.} 2012,
  \bibinfo{title}{{The rate of supernovae at redshift 0.1-1.0. The Stockholm
  VIMOS Supernova Survey III},} \aap, 545, A96,
  \dodoi{10.1051/0004-6361/201219364}

\bibitem[{A. {Melo} {et~al.}(2025){Melo}, {Ca{\~n}ameras}, {Schuldt}, {Suyu},
  {Andika}, {Bag}, \& {Taubenberger}}]{Melo:2025aa}
{Melo}, A., {Ca{\~n}ameras}, R., {Schuldt}, S., {et~al.} 2025,
  \bibinfo{title}{{HOLISMOKES: XV. Search for strong gravitational lenses
  combining ground-based and space-based imaging},} \aap, 698, A264,
  \dodoi{10.1051/0004-6361/202453195}

\bibitem[{S. {Mengel} {et~al.}(2002){Mengel}, {Lehnert}, {Thatte}, \&
  {Genzel}}]{Mengel:2002aa}
{Mengel}, S., {Lehnert}, M.~D., {Thatte}, N., \& {Genzel}, R. 2002,
  \bibinfo{title}{{Dynamical masses of young star clusters in NGC 4038/4039},}
  \aap, 383, 137, \dodoi{10.1051/0004-6361:20011704}

\bibitem[{G.~R. {Meurer} {et~al.}(1999){Meurer}, {Heckman}, \&
  {Calzetti}}]{Meurer:1999aa}
{Meurer}, G.~R., {Heckman}, T.~M., \& {Calzetti}, D. 1999,
  \bibinfo{title}{{Dust Absorption and the Ultraviolet Luminosity Density at z
  \raisebox{-0.5ex}\textasciitilde 3 as Calibrated by Local Starburst
  Galaxies},} \apj, 521, 64, \dodoi{10.1086/307523}

\bibitem[{O. {Miettinen} {et~al.}(2017){Miettinen}, {Delvecchio}, {Smol{\v
  c}i{\'c}}, {Aravena}, {Brisbin}, {Karim}, {Magnelli}, {Novak}, {Schinnerer},
  {Albrecht}, {Aussel}, {Bertoldi}, {Capak}, {Casey}, {Hayward}, {Ilbert},
  {Intema}, {Jiang}, {Le F{\`e}vre}, {McCracken}, {Mu{\~n}oz Arancibia},
  {Navarrete}, {Padilla}, {Riechers}, {Salvato}, {Scott}, {Sheth}, \&
  {Tasca}}]{Miettinen:2017aa}
{Miettinen}, O., {Delvecchio}, I., {Smol{\v c}i{\'c}}, V., {et~al.} 2017,
  \bibinfo{title}{{An ALMA survey of submillimetre galaxies in the COSMOS
  field: Physical properties derived from energy balance spectral energy
  distribution modelling},} \aap, 606, A17, \dodoi{10.1051/0004-6361/201730762}

\bibitem[{G. {Miley} \& C. {De Breuck}(2008){Miley} \& {De
  Breuck}}]{Miley:2008aa}
{Miley}, G., \& {De Breuck}, C. 2008, \bibinfo{title}{{Distant radio galaxies
  and their environments},} \aapr, 15, 67, \dodoi{10.1007/s00159-007-0008-z}

\bibitem[{T.~B. {Miller} {et~al.}(2015){Miller}, {Hayward}, {Chapman}, \&
  {Behroozi}}]{Miller:2015aa}
{Miller}, T.~B., {Hayward}, C.~C., {Chapman}, S.~C., \& {Behroozi}, P.~S. 2015,
  \bibinfo{title}{{The bias of the submillimetre galaxy population: SMGs are
  poor tracers of the most-massive structures in the z {\ensuremath{\sim}} 2
  Universe},} \mnras, 452, 878, \dodoi{10.1093/mnras/stv1267}

\bibitem[{T.~B. {Miller} {et~al.}(2018){Miller}, {Chapman}, {Aravena}, {Ashby},
  {Hayward}, {Vieira}, {Wei{\ss}}, {Babul}, {B{\'e}thermin}, {Bradford},
  {Brodwin}, {Carlstrom}, {Chen}, {Cunningham}, {De Breuck}, {Gonzalez},
  {Greve}, {Harnett}, {Hezaveh}, {Lacaille}, {Litke}, {Ma}, {Malkan},
  {Marrone}, {Morningstar}, {Murphy}, {Narayanan}, {Pass}, {Perry}, {Phadke},
  {Rennehan}, {Rotermund}, {Simpson}, {Spilker}, {Sreevani}, {Stark},
  {Strandet}, \& {Strom}}]{Miller:2018ab}
{Miller}, T.~B., {Chapman}, S.~C., {Aravena}, M., {et~al.} 2018,
  \bibinfo{title}{{A massive core for a cluster of galaxies at a redshift of
  4.3},} \nat, 556, 469, \dodoi{10.1038/s41586-018-0025-2}

\bibitem[{M. {Miluzio} {et~al.}(2013){Miluzio}, {Cappellaro}, {Botticella},
  {Cresci}, {Greggio}, {Mannucci}, {Benetti}, {Bufano}, {Elias-Rosa},
  {Pastorello}, {Turatto}, \& {Zampieri}}]{Miluzio:2013aa}
{Miluzio}, M., {Cappellaro}, E., {Botticella}, M.~T., {et~al.} 2013,
  \bibinfo{title}{{HAWK-I infrared supernova search in starburst galaxies},}
  \aap, 554, A127, \dodoi{10.1051/0004-6361/201321192}

\bibitem[{I. {Mitsuhashi} {et~al.}(2024){Mitsuhashi}, {Tadaki}, {Ikeda},
  {Herrera-Camus}, {Aravena}, {De Looze}, {F{\"o}rster Schreiber},
  {Gonz{\'a}lez-L{\'o}pez}, {Spilker}, {Assef}, {Bouwens}, {Barcos-Munoz},
  {Birkin}, {Bowler}, {Calistro Rivera}, {Davies}, {Da Cunha},
  {D{\'\i}az-Santos}, {Ferrara}, {Fisher}, {Lee}, {Li}, {Lutz}, {Rela{\~n}o},
  {Naab}, {Palla}, {Posses}, {Solimano}, {Tacconi}, {{\"U}bler}, {van der
  Giessen}, \& {Veilleux}}]{Mitsuhashi:2024aa}
{Mitsuhashi}, I., {Tadaki}, K.-i., {Ikeda}, R., {et~al.} 2024,
  \bibinfo{title}{{The ALMA-CRISTAL survey: Widespread dust-obscured star
  formation in typical star-forming galaxies at z = 4{\textendash}6},} \aap,
  690, A197, \dodoi{10.1051/0004-6361/202348782}

\bibitem[{S.~D. {Mork} {et~al.}(2025){Mork}, {Gladders}, {Khullar}, {Sharon},
  {Chicoine}, {Cloonan}, {Dahle}, {Garza}, {Glusman}, {Gozman}, {Horwath},
  {Levine}, {Liang}, {Mahronic}, {Manwadkar}, {Martinez}, {Masegian}, {Matthews
  Acu{\~n}a}, {Merz}, {Pan}, {Sanchez}, {Sierra}, {Kavin Stein}, {Sukay},
  {Tamargo-Arizmendi}, {Tavangar}, {Tu}, {Wagner}, {Zaborowski}, {Zhang}, \&
  {Cool-Lamps Collaboration}}]{Mork:2025aa}
{Mork}, S.~D., {Gladders}, M.~D., {Khullar}, G., {et~al.} 2025,
  \bibinfo{title}{{COOL-LAMPS. VII. Quantifying Strong-lens Scaling Relations
  with 177 Cluster-scale Strong Gravitational Lenses in DECaLS},} \apj, 979,
  184, \dodoi{10.3847/1538-4357/ada24c}

\bibitem[{E. {M{\"o}rtsell} {et~al.}(2022{\natexlab{a}}){M{\"o}rtsell},
  {Goobar}, {Johansson}, \& {Dhawan}}]{Mortsell:2022aa}
{M{\"o}rtsell}, E., {Goobar}, A., {Johansson}, J., \& {Dhawan}, S.
  2022{\natexlab{a}}, \bibinfo{title}{{Sensitivity of the Hubble Constant
  Determination to Cepheid Calibration},} \apj, 933, 212,
  \dodoi{10.3847/1538-4357/ac756e}

\bibitem[{E. {M{\"o}rtsell} {et~al.}(2022{\natexlab{b}}){M{\"o}rtsell},
  {Goobar}, {Johansson}, \& {Dhawan}}]{Mortsell:2022ab}
{M{\"o}rtsell}, E., {Goobar}, A., {Johansson}, J., \& {Dhawan}, S.
  2022{\natexlab{b}}, \bibinfo{title}{{The Hubble Tension Revisited: Additional
  Local Distance Ladder Uncertainties},} \apj, 935, 58,
  \dodoi{10.3847/1538-4357/ac7c19}

\bibitem[{J.~A. {Mu{\~n}oz} {et~al.}(2022){Mu{\~n}oz}, {Kochanek},
  {Fohlmeister}, {Wambsganss}, {Falco}, \& {For{\'e}s-Toribio}}]{Munoz:2022aa}
{Mu{\~n}oz}, J.~A., {Kochanek}, C.~S., {Fohlmeister}, J., {et~al.} 2022,
  \bibinfo{title}{{The Longest Delay: A 14.5 yr Campaign to Determine the Third
  Time Delay in the Lensing Cluster SDSS J1004+4112},} \apj, 937, 34,
  \dodoi{10.3847/1538-4357/ac8877}

\bibitem[{K. {Napier} {et~al.}(2023){Napier}, {Sharon}, {Dahle}, {Bayliss},
  {Gladders}, {Mahler}, {Rigby}, \& {Florian}}]{Napier:2023aa}
{Napier}, K., {Sharon}, K., {Dahle}, H., {et~al.} 2023, \bibinfo{title}{{Hubble
  Constant Measurement from Three Large-separation Quasars Strongly Lensed by
  Galaxy Clusters},} \apj, 959, 134, \dodoi{10.3847/1538-4357/ad045a}

\bibitem[{R. {Narayan} \& M. {Bartelmann}(1996){Narayan} \&
  {Bartelmann}}]{Narayan:1996aa}
{Narayan}, R., \& {Bartelmann}, M. 1996, \bibinfo{title}{{Lectures on
  Gravitational Lensing},} ArXiv Astrophysics e-prints

\bibitem[{D. {Narayanan} {et~al.}(2018){Narayanan}, {Dav{\'e}}, {Johnson},
  {Thompson}, {Conroy}, \& {Geach}}]{Narayanan:2018ab}
{Narayanan}, D., {Dav{\'e}}, R., {Johnson}, B.~D., {et~al.} 2018,
  \bibinfo{title}{{The IRX-{\ensuremath{\beta}} dust attenuation relation in
  cosmological galaxy formation simulations},} \mnras, 474, 1718,
  \dodoi{10.1093/mnras/stx2860}

\bibitem[{M. {Negrello} {et~al.}(2010){Negrello}, {Hopwood}, {De Zotti},
  {Cooray}, {Verma}, {Bock}, {Frayer}, {Gurwell}, {Omont}, {Neri},
  {Dannerbauer}, {Leeuw}, {Barton}, {Cooke}, {Kim}, {da Cunha}, {Rodighiero},
  {Cox}, {Bonfield}, {Jarvis}, {Serjeant}, {Ivison}, {Dye}, {Aretxaga},
  {Hughes}, {Ibar}, {Bertoldi}, {Valtchanov}, {Eales}, {Dunne}, {Driver},
  {Auld}, {Buttiglione}, {Cava}, {Grady}, {Clements}, {Dariush}, {Fritz},
  {Hill}, {Hornbeck}, {Kelvin}, {Lagache}, {Lopez-Caniego}, {Gonzalez-Nuevo},
  {Maddox}, {Pascale}, {Pohlen}, {Rigby}, {Robotham}, {Simpson}, {Smith},
  {Temi}, {Thompson}, {Woodgate}, {York}, {Aguirre}, {Beelen}, {Blain},
  {Baker}, {Birkinshaw}, {Blundell}, {Bradford}, {Burgarella}, {Danese},
  {Dunlop}, {Fleuren}, {Glenn}, {Harris}, {Kamenetzky}, {Lupu}, {Maddalena},
  {Madore}, {Maloney}, {Matsuhara}, {Micha{\l}owski}, {Murphy}, {Naylor},
  {Nguyen}, {Popescu}, {Rawlings}, {Rigopoulou}, {Scott}, {Scott}, {Seibert},
  {Smail}, {Tuffs}, {Vieira}, {van der Werf}, \&
  {Zmuidzinas}}]{Negrello:2010aa}
{Negrello}, M., {Hopwood}, R., {De Zotti}, G., {et~al.} 2010,
  \bibinfo{title}{{The Detection of a Population of Submillimeter-Bright,
  Strongly Lensed Galaxies},} Science, 330, 800,
  \dodoi{10.1126/science.1193420}

\bibitem[{M. {Negrello} {et~al.}(2014){Negrello}, {Hopwood}, {Dye}, {da Cunha},
  {Serjeant}, {Fritz}, {Rowlands}, {Fleuren}, {Bussmann}, {Cooray},
  {Dannerbauer}, {Gonzalez-Nuevo}, {Lapi}, {Omont}, {Amber}, {Auld}, {Baes},
  {Buttiglione}, {Cava}, {Danese}, {Dariush}, {De Zotti}, {Dunne}, {Eales},
  {Ibar}, {Ivison}, {Kim}, {Leeuw}, {Maddox}, {Micha{\l}owski}, {Massardi},
  {Pascale}, {Pohlen}, {Rigby}, {Smith}, {Sutherland}, {Temi}, \&
  {Wardlow}}]{Negrello:2014aa}
{Negrello}, M., {Hopwood}, R., {Dye}, S., {et~al.} 2014,
  \bibinfo{title}{{Herschel *-ATLAS: deep HST/WFC3 imaging of strongly lensed
  submillimetre galaxies},} \mnras, 440, 1999, \dodoi{10.1093/mnras/stu413}

\bibitem[{M. {Negrello} {et~al.}(2017){Negrello}, {Amber}, {Amvrosiadis},
  {Cai}, {Lapi}, {Gonzalez-Nuevo}, {De Zotti}, {Furlanetto}, {Maddox}, {Allen},
  {Bakx}, {Bussmann}, {Cooray}, {Covone}, {Danese}, {Dannerbauer}, {Fu},
  {Greenslade}, {Gurwell}, {Hopwood}, {Koopmans}, {Napolitano}, {Nayyeri},
  {Omont}, {Petrillo}, {Riechers}, {Serjeant}, {Tortora}, {Valiante}, {Verdoes
  Kleijn}, {Vernardos}, {Wardlow}, {Baes}, {Baker}, {Bourne}, {Clements},
  {Crawford}, {Dye}, {Dunne}, {Eales}, {Ivison}, {Marchetti}, {Micha{\l}owski},
  {Smith}, {Vaccari}, \& {van der Werf}}]{Negrello:2017aa}
{Negrello}, M., {Amber}, S., {Amvrosiadis}, A., {et~al.} 2017,
  \bibinfo{title}{{The Herschel-ATLAS: a sample of 500 {$\mu$}m-selected lensed
  galaxies over 600 deg$^{2}$},} \mnras, 465, 3558,
  \dodoi{10.1093/mnras/stw2911}

\bibitem[{A.~B. {Newman} {et~al.}(2018){Newman}, {Belli}, {Ellis}, \&
  {Patel}}]{Newman:2018aa}
{Newman}, A.~B., {Belli}, S., {Ellis}, R.~S., \& {Patel}, S.~G. 2018,
  \bibinfo{title}{{Resolving Quiescent Galaxies at z {\ensuremath{\gtrsim}} 2.
  I. Search for Gravitationally Lensed Sources and Characterization of Their
  Structure, Stellar Populations, and Line Emission},} \apj, 862, 125,
  \dodoi{10.3847/1538-4357/aacd4d}

\bibitem[{M. {Oguri}(2007){Oguri}}]{Oguri:2007aa}
{Oguri}, M. 2007, \bibinfo{title}{{Gravitational Lens Time Delays: A
  Statistical Assessment of Lens Model Dependences and Implications for the
  Global Hubble Constant},} \apj, 660, 1, \dodoi{10.1086/513093}

\bibitem[{M. {Oguri} \& P.~J. {Marshall}(2010){Oguri} \&
  {Marshall}}]{Oguri:2010ac}
{Oguri}, M., \& {Marshall}, P.~J. 2010, \bibinfo{title}{{Gravitationally lensed
  quasars and supernovae in future wide-field optical imaging surveys},}
  \mnras, 405, 2579, \dodoi{10.1111/j.1365-2966.2010.16639.x}

\bibitem[{V. {Olivares} {et~al.}(2016){Olivares}, {Treister}, {Privon},
  {Alaghband-Zadeh}, {Casey}, {Schawinski}, {Kurczynski}, {Gawiser}, {Nagar},
  {Chapman}, {Bauer}, \& {Sanders}}]{Olivares:2016aa}
{Olivares}, V., {Treister}, E., {Privon}, G.~C., {et~al.} 2016,
  \bibinfo{title}{{Spatially Resolved Spectroscopy of Submillimeter Galaxies at
  z ≃ 2},} \apj, 827, 57, \dodoi{10.3847/0004-637X/827/1/57}

\bibitem[{D.~E. {Osterbrock} \& G.~J. {Ferland}(2006){Osterbrock} \&
  {Ferland}}]{Osterbrock:2006aa}
{Osterbrock}, D.~E., \& {Ferland}, G.~J. 2006, {Astrophysics of gaseous nebulae
  and active galactic nuclei} (University Science Books)

\bibitem[{I. {Oteo} {et~al.}(2018){Oteo}, {Ivison}, {Dunne}, {Manilla-Robles},
  {Maddox}, {Lewis}, {de Zotti}, {Bremer}, {Clements}, {Cooray}, {Dannerbauer},
  {Eales}, {Greenslade}, {Omont}, {Perez{\textendash}Fourn{\'o}n}, {Riechers},
  {Scott}, {van der Werf}, {Weiss}, \& {Zhang}}]{Oteo:2018aa}
{Oteo}, I., {Ivison}, R.~J., {Dunne}, L., {et~al.} 2018, \bibinfo{title}{{An
  Extreme Protocluster of Luminous Dusty Starbursts in the Early Universe},}
  \apj, 856, 72, \dodoi{10.3847/1538-4357/aaa1f1}

\bibitem[{R.~A. {Overzier}(2016){Overzier}}]{Overzier:2016aa}
{Overzier}, R.~A. 2016, \bibinfo{title}{{The realm of the galaxy protoclusters.
  A review},} \aapr, 24, 14, \dodoi{10.1007/s00159-016-0100-3}

\bibitem[{P.~A. {Palicio} {et~al.}(2024){Palicio}, {Matteucci}, {Della Valle},
  \& {Spitoni}}]{Palicio:2024aa}
{Palicio}, P.~A., {Matteucci}, F., {Della Valle}, M., \& {Spitoni}, E. 2024,
  \bibinfo{title}{{Cosmic Type Ia supernova rate and constraints on supernova
  Ia progenitors},} \aap, 689, A203, \dodoi{10.1051/0004-6361/202449740}

\bibitem[{M. {Pascale} {et~al.}(2023){Pascale}, {Dai}, {McKee}, \&
  {Tsang}}]{Pascale:2023ab}
{Pascale}, M., {Dai}, L., {McKee}, C.~F., \& {Tsang}, B. T.~H. 2023,
  \bibinfo{title}{{Nitrogen-enriched, Highly Pressurized Nebular Clouds
  Surrounding a Super Star Cluster at Cosmic Noon},} \apj, 957, 77,
  \dodoi{10.3847/1538-4357/acf75c}

\bibitem[{M. {Pascale} {et~al.}(2022){Pascale}, {Frye}, {Dai}, {Foo}, {Qin},
  {Leimbach}, {Bauer}, {Merlin}, {Coe}, {Diego}, {Yan}, {Zitrin}, {Cohen},
  {Conselice}, {Dole}, {Harrington}, {Jansen}, {Kamieneski}, {Windhorst}, \&
  {Yun}}]{Pascale:2022aa}
{Pascale}, M., {Frye}, B.~L., {Dai}, L., {et~al.} 2022,
  \bibinfo{title}{{Possible Ongoing Merger Discovered by Photometry and
  Spectroscopy in the Field of the Galaxy Cluster PLCK G165.7+67.0},} \apj,
  932, 85, \dodoi{10.3847/1538-4357/ac6ce9}

\bibitem[{M. {Pascale} {et~al.}(2025){Pascale}, {Frye}, {Pierel}, {Chen},
  {Kelly}, {Cohen}, {Windhorst}, {Riess}, {Kamieneski}, {Diego}, {Meena},
  {Cha}, {Oguri}, {Zitrin}, {Jee}, {Foo}, {Leimbach}, {Koekemoer}, {Conselice},
  {Dai}, {Goobar}, {Siebert}, {Strolger}, \& {Willner}}]{Pascale:2025aa}
{Pascale}, M., {Frye}, B.~L., {Pierel}, J. D.~R., {et~al.} 2025,
  \bibinfo{title}{{SN H0pe: The First Measurement of H$_{0}$ from a Multiply
  Imaged Type Ia Supernova, Discovered by JWST},} \apj, 979, 13,
  \dodoi{10.3847/1538-4357/ad9928}

\bibitem[{R. {Pavesi} {et~al.}(2018){Pavesi}, {Riechers}, {Sharon},
  {Smol{\v{c}}i{\'c}}, {Faisst}, {Schinnerer}, {Carilli}, {Capak}, {Scoville},
  \& {Stacey}}]{Pavesi:2018aa}
{Pavesi}, R., {Riechers}, D.~A., {Sharon}, C.~E., {et~al.} 2018,
  \bibinfo{title}{{Hidden in Plain Sight: A Massive, Dusty Starburst in a
  Galaxy Protocluster at z = 5.7 in the COSMOS Field},} \apj, 861, 43,
  \dodoi{10.3847/1538-4357/aac6b6}

\bibitem[{M. {P{\'e}rez-Torres} {et~al.}(2021){P{\'e}rez-Torres}, {Mattila},
  {Alonso-Herrero}, {Aalto}, \& {Efstathiou}}]{Perez-Torres:2021aa}
{P{\'e}rez-Torres}, M., {Mattila}, S., {Alonso-Herrero}, A., {Aalto}, S., \&
  {Efstathiou}, A. 2021, \bibinfo{title}{{Star formation and nuclear activity
  in luminous infrared galaxies: an infrared through radio review},} \aapr, 29,
  2, \dodoi{10.1007/s00159-020-00128-x}

\bibitem[{M.~A. {P{\'e}rez-Torres} {et~al.}(2009){P{\'e}rez-Torres},
  {Romero-Ca{\~n}izales}, {Alberdi}, \& {Polatidis}}]{Perez-Torres:2009aa}
{P{\'e}rez-Torres}, M.~A., {Romero-Ca{\~n}izales}, C., {Alberdi}, A., \&
  {Polatidis}, A. 2009, \bibinfo{title}{{An extremely prolific supernova
  factory in the buried nucleus of the starburst galaxy IC 694},} \aap, 507,
  L17, \dodoi{10.1051/0004-6361/200912964}

\bibitem[{F. {Perrotta} {et~al.}(2002){Perrotta}, {Baccigalupi}, {Bartelmann},
  {De Zotti}, \& {Granato}}]{Perrotta:2002aa}
{Perrotta}, F., {Baccigalupi}, C., {Bartelmann}, M., {De Zotti}, G., \&
  {Granato}, G.~L. 2002, \bibinfo{title}{{Gravitational lensing of extended
  high-redshift sources by dark matter haloes},} \mnras, 329, 445,
  \dodoi{10.1046/j.1365-8711.2002.05009.x}

\bibitem[{D.~W. {Pesce} {et~al.}(2020){Pesce}, {Braatz}, {Reid}, {Riess},
  {Scolnic}, {Condon}, {Gao}, {Henkel}, {Impellizzeri}, {Kuo}, \&
  {Lo}}]{Pesce:2020aa}
{Pesce}, D.~W., {Braatz}, J.~A., {Reid}, M.~J., {et~al.} 2020,
  \bibinfo{title}{{The Megamaser Cosmology Project. XIII. Combined Hubble
  Constant Constraints},} \apjl, 891, L1, \dodoi{10.3847/2041-8213/ab75f0}

\bibitem[{T. {Petrushevska}(2020){Petrushevska}}]{Petrushevska:2020aa}
{Petrushevska}, T. 2020, \bibinfo{title}{{Strongly Lensed Supernovae in
  Well-Studied Galaxy Clusters with the Vera C. Rubin Observatory},} Symmetry,
  12, 1966, \dodoi{10.3390/sym12121966}

\bibitem[{T. {Petrushevska} {et~al.}(2018{\natexlab{a}}){Petrushevska},
  {Okamura}, {Kawamata}, {Hangard}, {Mahler}, \&
  {Goobar}}]{Petrushevska:2018ab}
{Petrushevska}, T., {Okamura}, T., {Kawamata}, R., {et~al.} 2018{\natexlab{a}},
  \bibinfo{title}{{Prospects for Strongly Lensed Supernovae Behind Hubble
  Frontier Fields Galaxy Clusters with the James Webb Space Telescope},}
  Astronomy Reports, 62, 917, \dodoi{10.1134/S1063772918120272}

\bibitem[{T. {Petrushevska} {et~al.}(2016){Petrushevska}, {Amanullah},
  {Goobar}, {Fabbro}, {Johansson}, {Kjellsson}, {Lidman}, {Paech}, {Richard},
  {Dahle}, {Ferretti}, {Kneib}, {Limousin}, {Nordin}, \&
  {Stanishev}}]{Petrushevska:2016aa}
{Petrushevska}, T., {Amanullah}, R., {Goobar}, A., {et~al.} 2016,
  \bibinfo{title}{{High-redshift supernova rates measured with the
  gravitational telescope A 1689},} \aap, 594, A54,
  \dodoi{10.1051/0004-6361/201628925}

\bibitem[{T. {Petrushevska} {et~al.}(2018{\natexlab{b}}){Petrushevska},
  {Goobar}, {Lagattuta}, {Amanullah}, {Hangard}, {Fabbro}, {Lidman}, {Paech},
  {Richard}, \& {Kneib}}]{Petrushevska:2018aa}
{Petrushevska}, T., {Goobar}, A., {Lagattuta}, D.~J., {et~al.}
  2018{\natexlab{b}}, \bibinfo{title}{{Searching for supernovae in the
  multiply-imaged galaxies behind the gravitational telescope A370},} \aap,
  614, A103, \dodoi{10.1051/0004-6361/201731552}

\bibitem[{J.~D.~R. {Pierel} \& S. {Rodney}(2019){Pierel} \&
  {Rodney}}]{Pierel:2019aa}
{Pierel}, J.~D.~R., \& {Rodney}, S. 2019, \bibinfo{title}{{Turning
  Gravitationally Lensed Supernovae into Cosmological Probes},} \apj, 876, 107,
  \dodoi{10.3847/1538-4357/ab164a}

\bibitem[{J.~D.~R. {Pierel} {et~al.}(2021){Pierel}, {Rodney}, {Vernardos},
  {Oguri}, {Kessler}, \& {Anguita}}]{Pierel:2021ab}
{Pierel}, J.~D.~R., {Rodney}, S., {Vernardos}, G., {et~al.} 2021,
  \bibinfo{title}{{Projected Cosmological Constraints from Strongly Lensed
  Supernovae with the Roman Space Telescope},} \apj, 908, 190,
  \dodoi{10.3847/1538-4357/abd8d3}

\bibitem[{J.~D.~R. {Pierel} {et~al.}(2023){Pierel}, {Arendse}, {Ertl}, {Huang},
  {Moustakas}, {Schuldt}, {Shajib}, {Shu}, {Birrer}, {Bronikowski}, {Hjorth},
  {Suyu}, {Agarwal}, {Agnello}, {Bolton}, {Chakrabarti}, {Cold}, {Courbin},
  {Della Costa}, {Dhawan}, {Engesser}, {Fox}, {Gall}, {Gomez}, {Goobar}, {Jha},
  {Jimenez}, {Johansson}, {Larison}, {Li}, {Marques-Chaves}, {Mao}, {Mazzali},
  {Perez-Fournon}, {Petrushevska}, {Poidevin}, {Rest}, {Sheu}, {Shirley},
  {Silver}, {Storfer}, {Strolger}, {Treu}, {Wojtak}, \&
  {Zenati}}]{Pierel:2023aa}
{Pierel}, J.~D.~R., {Arendse}, N., {Ertl}, S., {et~al.} 2023,
  \bibinfo{title}{{LensWatch. I. Resolved HST Observations and Constraints on
  the Strongly Lensed Type Ia Supernova 2022qmx (``SN Zwicky'')},} \apj, 948,
  115, \dodoi{10.3847/1538-4357/acc7a6}

\bibitem[{J.~D.~R. {Pierel} {et~al.}(2024{\natexlab{a}}){Pierel}, {Frye},
  {Pascale}, {Caminha}, {Chen}, {Dhawan}, {Gilman}, {Grayling}, {Huber},
  {Kelly}, {Thorp}, {Arendse}, {Birrer}, {Bronikowski}, {Ca{\~n}ameras}, {Coe},
  {Cohen}, {Conselice}, {Driver}, {D{\'S}ilva}, {Engesser}, {Foo}, {Gall},
  {Garuda}, {Grillo}, {Grogin}, {Henderson}, {Hjorth}, {Jansen}, {Johansson},
  {Kamieneski}, {Koekemoer}, {Larison}, {Marshall}, {Moustakas}, {Nonino},
  {Ortiz}, {Petrushevska}, {Pirzkal}, {Robotham}, {Ryan}, {Schuldt},
  {Strolger}, {Summers}, {Suyu}, {Treu}, {Willmer}, {Windhorst}, {Yan},
  {Zitrin}, {Acebron}, {Chakrabarti}, {Coulter}, {Fox}, {Huang}, {Jha}, {Li},
  {Mazzali}, {Meena}, {P{\'e}rez-Fournon}, {Poidevin}, {Rest}, \&
  {Riess}}]{Pierel:2024aa}
{Pierel}, J.~D.~R., {Frye}, B.~L., {Pascale}, M., {et~al.} 2024{\natexlab{a}},
  \bibinfo{title}{{JWST Photometric Time-delay and Magnification Measurements
  for the Triply Imaged Type Ia ``SN H0pe'' at z = 1.78},} \apj, 967, 50,
  \dodoi{10.3847/1538-4357/ad3c43}

\bibitem[{J.~D.~R. {Pierel} {et~al.}(2024{\natexlab{b}}){Pierel}, {Newman},
  {Dhawan}, {Gu}, {Joshi}, {Li}, {Schuldt}, {Strolger}, {Suyu}, {Caminha},
  {Cohen}, {Diego}, {D{\'S}ilva}, {Ertl}, {Frye}, {Granata}, {Grillo},
  {Koekemoer}, {Li}, {Robotham}, {Summers}, {Treu}, {Windhorst}, {Zitrin},
  {Agarwal}, {Agrawal}, {Arendse}, {Belli}, {Burns}, {Ca{\~n}ameras},
  {Chakrabarti}, {Chen}, {Collett}, {Coulter}, {Ellis}, {Engesser}, {Foo},
  {Fox}, {Gall}, {Garuda}, {Gezari}, {Gomez}, {Glazebrook}, {Hjorth}, {Huang},
  {Jha}, {Kamieneski}, {Kelly}, {Larison}, {Moustakas}, {Pascale},
  {P{\'e}rez-Fournon}, {Petrushevska}, {Poidevin}, {Rest}, {Shahbandeh},
  {Shajib}, {Siebert}, {Storfer}, {Talbot}, {Wang}, {Wevers}, \&
  {Zenati}}]{Pierel:2024ab}
{Pierel}, J.~D.~R., {Newman}, A.~B., {Dhawan}, S., {et~al.} 2024{\natexlab{b}},
  \bibinfo{title}{{Lensed Type Ia Supernova ``Encore'' at z = 2: The First
  Instance of Two Multiply Imaged Supernovae in the Same Host Galaxy},} \apjl,
  967, L37, \dodoi{10.3847/2041-8213/ad4648}

\bibitem[{ {Planck Collaboration} {et~al.}(2015){Planck Collaboration},
  {Aghanim}, {Altieri}, {Arnaud}, {Ashdown}, {Aumont}, {Baccigalupi}, {Banday},
  {Barreiro}, {Bartolo}, {Battaner}, {Beelen}, {Benabed}, {Benoit-L{\'e}vy},
  {Bernard}, {Bersanelli}, {Bethermin}, {Bielewicz}, {Bonavera}, {Bond},
  {Borrill}, {Bouchet}, {Boulanger}, {Burigana}, {Calabrese}, {Canameras},
  {Cardoso}, {Catalano}, {Chamballu}, {Chary}, {Chiang}, {Christensen},
  {Clements}, {Colombi}, {Couchot}, {Crill}, {Curto}, {Danese}, {Dassas},
  {Davies}, {Davis}, {de Bernardis}, {de Rosa}, {de Zotti}, {Delabrouille},
  {Diego}, {Dole}, {Donzelli}, {Dor{\'e}}, {Douspis}, {Ducout}, {Dupac},
  {Efstathiou}, {Elsner}, {En{\ss}lin}, {Falgarone}, {Flores-Cacho}, {Forni},
  {Frailis}, {Fraisse}, {Franceschi}, {Frejsel}, {Frye}, {Galeotta}, {Galli},
  {Ganga}, {Giard}, {Gjerl{\o}w}, {Gonz{\'a}lez-Nuevo}, {G{\'o}rski},
  {Gregorio}, {Gruppuso}, {Gu{\'e}ry}, {Hansen}, {Hanson}, {Harrison}, {Helou},
  {Hern{\'a}ndez-Monteagudo}, {Hildebrandt}, {Hivon}, {Hobson}, {Holmes},
  {Hovest}, {Huffenberger}, {Hurier}, {Jaffe}, {Jaffe}, {Keih{\"a}nen},
  {Keskitalo}, {Kisner}, {Kneissl}, {Knoche}, {Kunz}, {Kurki-Suonio},
  {Lagache}, {Lamarre}, {Lasenby}, {Lattanzi}, {Lawrence}, {Le Floc'h},
  {Leonardi}, {Levrier}, {Liguori}, {Lilje}, {Linden-V{\o}rnle},
  {L{\'o}pez-Caniego}, {Lubin}, {Mac{\'\i}as-P{\'e}rez}, {MacKenzie}, {Maffei},
  {Mandolesi}, {Maris}, {Martin}, {Martinache}, {Mart{\'\i}nez-Gonz{\'a}lez},
  {Masi}, {Matarrese}, {Mazzotta}, {Melchiorri}, {Mennella}, {Migliaccio},
  {Moneti}, {Montier}, {Morgante}, {Mortlock}, {Munshi}, {Murphy}, {Natoli},
  {Negrello}, {Nesvadba}, {Novikov}, {Novikov}, {Omont}, {Pagano}, {Pajot},
  {Pasian}, {Perdereau}, {Perotto}, {Perrotta}, {Pettorino}, {Piacentini},
  {Piat}, {Plaszczynski}, {Pointecouteau}, {Polenta}, {Popa}, {Pratt},
  {Prunet}, {Puget}, {Rachen}, {Reach}, {Reinecke}, {Remazeilles}, {Renault},
  {Ristorcelli}, {Rocha}, {Roudier}, {Rusholme}, {Sandri}, {Santos}, {Savini},
  {Scott}, {Spencer}, {Stolyarov}, {Sunyaev}, {Sutton}, {Sygnet}, {Tauber},
  {Terenzi}, {Toffolatti}, {Tomasi}, {Tristram}, {Tucci}, {Umana},
  {Valenziano}, {Valiviita}, {Valtchanov}, {Van Tent}, {Vieira}, {Vielva},
  {Wade}, {Wandelt}, {Wehus}, {Welikala}, {Zacchei}, \&
  {Zonca}}]{Planck-Collaboration:2015aa}
{Planck Collaboration}, {Aghanim}, N., {Altieri}, B., {et~al.} 2015,
  \bibinfo{title}{{Planck intermediate results. XXVII. High-redshift infrared
  galaxy overdensity candidates and lensed sources discovered by Planck and
  confirmed by Herschel-SPIRE},} \aap, 582, A30,
  \dodoi{10.1051/0004-6361/201424790}

\bibitem[{ {Planck Collaboration} {et~al.}(2020){Planck Collaboration},
  {Aghanim}, {Akrami}, {Ashdown}, {Aumont}, {Baccigalupi}, {Ballardini},
  {Banday}, {Barreiro}, {Bartolo}, {Basak}, {Battye}, {Benabed}, {Bernard},
  {Bersanelli}, {Bielewicz}, {Bock}, {Bond}, {Borrill}, {Bouchet}, {Boulanger},
  {Bucher}, {Burigana}, {Butler}, {Calabrese}, {Cardoso}, {Carron},
  {Challinor}, {Chiang}, {Chluba}, {Colombo}, {Combet}, {Contreras}, {Crill},
  {Cuttaia}, {de Bernardis}, {de Zotti}, {Delabrouille}, {Delouis}, {Di
  Valentino}, {Diego}, {Dor{\'e}}, {Douspis}, {Ducout}, {Dupac}, {Dusini},
  {Efstathiou}, {Elsner}, {En{\ss}lin}, {Eriksen}, {Fantaye}, {Farhang},
  {Fergusson}, {Fernandez-Cobos}, {Finelli}, {Forastieri}, {Frailis},
  {Fraisse}, {Franceschi}, {Frolov}, {Galeotta}, {Galli}, {Ganga},
  {G{\'e}nova-Santos}, {Gerbino}, {Ghosh}, {Gonz{\'a}lez-Nuevo}, {G{\'o}rski},
  {Gratton}, {Gruppuso}, {Gudmundsson}, {Hamann}, {Handley}, {Hansen},
  {Herranz}, {Hildebrandt}, {Hivon}, {Huang}, {Jaffe}, {Jones}, {Karakci},
  {Keih{\"a}nen}, {Keskitalo}, {Kiiveri}, {Kim}, {Kisner}, {Knox},
  {Krachmalnicoff}, {Kunz}, {Kurki-Suonio}, {Lagache}, {Lamarre}, {Lasenby},
  {Lattanzi}, {Lawrence}, {Le Jeune}, {Lemos}, {Lesgourgues}, {Levrier},
  {Lewis}, {Liguori}, {Lilje}, {Lilley}, {Lindholm}, {L{\'o}pez-Caniego},
  {Lubin}, {Ma}, {Mac{\'\i}as-P{\'e}rez}, {Maggio}, {Maino}, {Mandolesi},
  {Mangilli}, {Marcos-Caballero}, {Maris}, {Martin}, {Martinelli},
  {Mart{\'\i}nez-Gonz{\'a}lez}, {Matarrese}, {Mauri}, {McEwen}, {Meinhold},
  {Melchiorri}, {Mennella}, {Migliaccio}, {Millea}, {Mitra},
  {Miville-Desch{\^e}nes}, {Molinari}, {Montier}, {Morgante}, {Moss}, {Natoli},
  {N{\o}rgaard-Nielsen}, {Pagano}, {Paoletti}, {Partridge}, {Patanchon},
  {Peiris}, {Perrotta}, {Pettorino}, {Piacentini}, {Polastri}, {Polenta},
  {Puget}, {Rachen}, {Reinecke}, {Remazeilles}, {Renzi}, {Rocha}, {Rosset},
  {Roudier}, {Rubi{\~n}o-Mart{\'\i}n}, {Ruiz-Granados}, {Salvati}, {Sandri},
  {Savelainen}, {Scott}, {Shellard}, {Sirignano}, {Sirri}, {Spencer},
  {Sunyaev}, {Suur-Uski}, {Tauber}, {Tavagnacco}, {Tenti}, {Toffolatti},
  {Tomasi}, {Trombetti}, {Valenziano}, {Valiviita}, {Van Tent}, {Vibert},
  {Vielva}, {Villa}, {Vittorio}, {Wandelt}, {Wehus}, {White}, {White},
  {Zacchei}, \& {Zonca}}]{Planck-Collaboration:2020aa}
{Planck Collaboration}, {Aghanim}, N., {Akrami}, Y., {et~al.} 2020,
  \bibinfo{title}{{Planck 2018 results. VI. Cosmological parameters},} \aap,
  641, A6, \dodoi{10.1051/0004-6361/201833910}

\bibitem[{M. {Polletta} {et~al.}(2023){Polletta}, {Nonino}, {Frye}, {Gargiulo},
  {Bisogni}, {Garuda}, {Thompson}, {Lehnert}, {Pascale}, {Willner},
  {Kamieneski}, {Leimbach}, {Cheng}, {Coe}, {Cohen}, {Conselice}, {Dai},
  {Diego}, {Dole}, {Driver}, {D'Silva}, {Fontana}, {Foo}, {Furtak}, {Grogin},
  {Harrington}, {Hathi}, {Jansen}, {Kelly}, {Koekemoer}, {Mancini}, {Marshall},
  {Pierel}, {Pirzkal}, {Robotham}, {Rutkowski}, {Ryan}, {Snigula}, {Summers},
  {Tompkins}, {Willmer}, {Windhorst}, {Yan}, {Yun}, \&
  {Zitrin}}]{Polletta:2023aa}
{Polletta}, M., {Nonino}, M., {Frye}, B., {et~al.} 2023,
  \bibinfo{title}{{Spectroscopy of the supernova H0pe host galaxy at redshift
  1.78},} \aap, 675, L4, \dodoi{10.1051/0004-6361/202346964}

\bibitem[{M. {Polletta} {et~al.}(2024){Polletta}, {Frye}, {Garuda}, {Willner},
  {Berta}, {Kneissl}, {Dole}, {Jansen}, {Lehnert}, {Cohen}, {Summers},
  {Windhorst}, {D'Silva}, {Koekemoer}, {Coe}, {Conselice}, {Driver}, {Grogin},
  {Marshall}, {Nonino}, {Ortiz}, {Pirzkal}, {Robotham}, {Ryan}, {Willmer},
  {Yan}, {Arumugam}, {Cheng}, {Gim}, {Hathi}, {Holwerda}, {Kamieneski}, {Keel},
  {Li}, {Pascale}, {Rottgering}, {Smith}, \& {Yun}}]{Polletta:2024ab}
{Polletta}, M., {Frye}, B.~L., {Garuda}, N., {et~al.} 2024,
  \bibinfo{title}{{JWST's PEARLS: Resolved study of the stellar and dust
  components in starburst galaxies at cosmic noon},} \aap, 690, A285,
  \dodoi{10.1051/0004-6361/202450671}

\bibitem[{K.~M. {Pontoppidan} {et~al.}(2016){Pontoppidan}, {Pickering},
  {Laidler}, {Gilbert}, {Sontag}, {Slocum}, {Sienkiewicz}, {Hanley}, {Earl},
  {Pueyo}, {Ravindranath}, {Karakla}, {Robberto}, {Noriega-Crespo}, \&
  {Barker}}]{Pontoppidan:2016aa}
{Pontoppidan}, K.~M., {Pickering}, T.~E., {Laidler}, V.~G., {et~al.} 2016,
  \bibinfo{title}{{Pandeia: a multi-mission exposure time calculator for JWST
  and WFIRST},} in Society of Photo-Optical Instrumentation Engineers (SPIE)
  Conference Series, Vol. 9910, Observatory Operations: Strategies, Processes,
  and Systems VI, ed. A.~B. {Peck}, R.~L. {Seaman}, \& C.~R. {Benn}, 991016,
  \dodoi{10.1117/12.2231768}

\bibitem[{S.~H. {Price} {et~al.}(2014){Price}, {Kriek}, {Brammer}, {Conroy},
  {F{\"o}rster Schreiber}, {Franx}, {Fumagalli}, {Lundgren}, {Momcheva},
  {Nelson}, {Skelton}, {van Dokkum}, {Whitaker}, \& {Wuyts}}]{Price:2014aa}
{Price}, S.~H., {Kriek}, M., {Brammer}, G.~B., {et~al.} 2014,
  \bibinfo{title}{{Direct Measurements of Dust Attenuation in z \~{} 1.5
  Star-forming Galaxies from 3D-HST: Implications for Dust Geometry and Star
  Formation Rates},} \apj, 788, 86, \dodoi{10.1088/0004-637X/788/1/86}

\bibitem[{A. {Puglisi} {et~al.}(2016){Puglisi}, {Rodighiero}, {Franceschini},
  {Talia}, {Cimatti}, {Baronchelli}, {Daddi}, {Renzini}, {Schawinski},
  {Mancini}, {Silverman}, {Gruppioni}, {Lutz}, {Berta}, \&
  {Oliver}}]{Puglisi:2016aa}
{Puglisi}, A., {Rodighiero}, G., {Franceschini}, A., {et~al.} 2016,
  \bibinfo{title}{{Dust attenuation in z \raisebox{-0.5ex}\textasciitilde 1
  galaxies from Herschel and 3D-HST H{\ensuremath{\alpha}} measurements},}
  \aap, 586, A83, \dodoi{10.1051/0004-6361/201526782}

\bibitem[{A. {Puglisi} {et~al.}(2017){Puglisi}, {Daddi}, {Renzini},
  {Rodighiero}, {Silverman}, {Kashino}, {Rodr{\'\i}guez-Mu{\~n}oz}, {Mancini},
  {Mainieri}, {Man}, {Franceschini}, {Valentino}, {Calabr{\`o}}, {Jin},
  {Darvish}, {Maier}, {Kartaltepe}, \& {Sanders}}]{Puglisi:2017aa}
{Puglisi}, A., {Daddi}, E., {Renzini}, A., {et~al.} 2017, \bibinfo{title}{{The
  Bright and Dark Sides of High-redshift Starburst Galaxies from Herschel and
  Subaru Observations},} \apjl, 838, L18, \dodoi{10.3847/2041-8213/aa66c9}

\bibitem[{R.~M. {Quimby} {et~al.}(2013){Quimby}, {Werner}, {Oguri}, {More},
  {More}, {Tanaka}, {Nomoto}, {Moriya}, {Folatelli}, {Maeda}, \&
  {Bersten}}]{Quimby:2013aa}
{Quimby}, R.~M., {Werner}, M.~C., {Oguri}, M., {et~al.} 2013,
  \bibinfo{title}{{Extraordinary Magnification of the Ordinary Type Ia
  Supernova PS1-10afx},} \apjl, 768, L20, \dodoi{10.1088/2041-8205/768/1/L20}

\bibitem[{R.~M. {Quimby} {et~al.}(2014){Quimby}, {Oguri}, {More}, {More},
  {Moriya}, {Werner}, {Tanaka}, {Folatelli}, {Bersten}, {Maeda}, \&
  {Nomoto}}]{Quimby:2014aa}
{Quimby}, R.~M., {Oguri}, M., {More}, A., {et~al.} 2014,
  \bibinfo{title}{{Detection of the Gravitational Lens Magnifying a Type Ia
  Supernova},} Science, 344, 396, \dodoi{10.1126/science.1250903}

\bibitem[{N.~A. {Reddy} {et~al.}(2015){Reddy}, {Kriek}, {Shapley}, {Freeman},
  {Siana}, {Coil}, {Mobasher}, {Price}, {Sanders}, \& {Shivaei}}]{Reddy:2015aa}
{Reddy}, N.~A., {Kriek}, M., {Shapley}, A.~E., {et~al.} 2015,
  \bibinfo{title}{{The MOSDEF Survey: Measurements of Balmer Decrements and the
  Dust Attenuation Curve at Redshifts z \~{} 1.4-2.6},} \apj, 806, 259,
  \dodoi{10.1088/0004-637X/806/2/259}

\bibitem[{S. {Refsdal}(1964){Refsdal}}]{Refsdal:1964aa}
{Refsdal}, S. 1964, \bibinfo{title}{{On the possibility of determining Hubble's
  parameter and the masses of galaxies from the gravitational lens effect},}
  \mnras, 128, 307, \dodoi{10.1093/mnras/128.4.307}

\bibitem[{M.~J. {Reid} {et~al.}(2009){Reid}, {Braatz}, {Condon}, {Greenhill},
  {Henkel}, \& {Lo}}]{Reid:2009aa}
{Reid}, M.~J., {Braatz}, J.~A., {Condon}, J.~J., {et~al.} 2009,
  \bibinfo{title}{{The Megamaser Cosmology Project. I. Very Long Baseline
  Interferometric Observations of UGC 3789},} \apj, 695, 287,
  \dodoi{10.1088/0004-637X/695/1/287}

\bibitem[{M.~J. {Reid} {et~al.}(2013){Reid}, {Braatz}, {Condon}, {Lo}, {Kuo},
  {Impellizzeri}, \& {Henkel}}]{Reid:2013aa}
{Reid}, M.~J., {Braatz}, J.~A., {Condon}, J.~J., {et~al.} 2013,
  \bibinfo{title}{{The Megamaser Cosmology Project. IV. A Direct Measurement of
  the Hubble Constant from UGC 3789},} \apj, 767, 154,
  \dodoi{10.1088/0004-637X/767/2/154}

\bibitem[{J.~D. {Remolina Gonz{\'a}lez} {et~al.}(2020){Remolina Gonz{\'a}lez},
  {Sharon}, {Reed}, {Li}, {Mahler}, {Bleem}, {Gladders}, {Niemiec}, {Acebron},
  \& {Child}}]{Remolina-Gonzalez:2020aa}
{Remolina Gonz{\'a}lez}, J.~D., {Sharon}, K., {Reed}, B., {et~al.} 2020,
  \bibinfo{title}{{Efficient Mass Estimate at the Core of Strong Lensing Galaxy
  Clusters Using the Einstein Radius},} \apj, 902, 44,
  \dodoi{10.3847/1538-4357/abb2a1}

\bibitem[{D.~A. {Riechers} {et~al.}(2010){Riechers}, {Capak}, {Carilli}, {Cox},
  {Neri}, {Scoville}, {Schinnerer}, {Bertoldi}, \& {Yan}}]{Riechers:2010aa}
{Riechers}, D.~A., {Capak}, P.~L., {Carilli}, C.~L., {et~al.} 2010,
  \bibinfo{title}{{A Massive Molecular Gas Reservoir in the z = 5.3
  Submillimeter Galaxy AzTEC-3},} \apjl, 720, L131,
  \dodoi{10.1088/2041-8205/720/2/L131}

\bibitem[{A.~G. {Riess} {et~al.}(2022){Riess}, {Yuan}, {Macri}, {Scolnic},
  {Brout}, {Casertano}, {Jones}, {Murakami}, {Anand}, {Breuval}, {Brink},
  {Filippenko}, {Hoffmann}, {Jha}, {D'arcy Kenworthy}, {Mackenty}, {Stahl}, \&
  {Zheng}}]{Riess:2022aa}
{Riess}, A.~G., {Yuan}, W., {Macri}, L.~M., {et~al.} 2022, \bibinfo{title}{{A
  Comprehensive Measurement of the Local Value of the Hubble Constant with 1 km
  s$^{-1}$ Mpc$^{-1}$ Uncertainty from the Hubble Space Telescope and the SH0ES
  Team},} \apjl, 934, L7, \dodoi{10.3847/2041-8213/ac5c5b}

\bibitem[{A. {Robertson} {et~al.}(2020){Robertson}, {Smith}, {Massey}, {Eke},
  {Jauzac}, {Bianconi}, \& {Ryczanowski}}]{Robertson:2020ac}
{Robertson}, A., {Smith}, G.~P., {Massey}, R., {et~al.} 2020,
  \bibinfo{title}{{What does strong gravitational lensing? The mass and
  redshift distribution of high-magnification lenses},} \mnras, 495, 3727,
  \dodoi{10.1093/mnras/staa1429}

\bibitem[{C. {Robertson} {et~al.}(2024){Robertson}, {Holwerda}, {Young},
  {Keel}, {Berkheimer}, {Cook}, {Conselice}, {Frye}, {Grogin}, {Koekemoer},
  {Nasr}, {Patel}, {Roemer}, {Smith}, \& {Windhorst}}]{Robertson:2024aa}
{Robertson}, C., {Holwerda}, B.~W., {Young}, J., {et~al.} 2024,
  \bibinfo{title}{{Ground- and Space-based Dust Observations of VV 191
  Overlapping Galaxy Pair},} \aj, 167, 263, \dodoi{10.3847/1538-3881/ad39c4}

\bibitem[{T. {Robitaille}(2019){Robitaille}}]{Robitaille:2019aa}
{Robitaille}, T. 2019, {APLpy v2.0: The Astronomical Plotting Library in
  Python}, 2.0, Zenodo Zenodo, \dodoi{10.5281/zenodo.2567476}

\bibitem[{T. {Robitaille} \& E. {Bressert}(2012){Robitaille} \&
  {Bressert}}]{Robitaille:2012aa}
{Robitaille}, T., \& {Bressert}, E. 2012, {APLpy: Astronomical Plotting Library
  in Python},, Astrophysics Source Code Library \doeprint{1208.017}

\bibitem[{G. {Rodighiero} {et~al.}(2011){Rodighiero}, {Daddi}, {Baronchelli},
  {Cimatti}, {Renzini}, {Aussel}, {Popesso}, {Lutz}, {Andreani}, {Berta},
  {Cava}, {Elbaz}, {Feltre}, {Fontana}, {F{\"o}rster Schreiber},
  {Franceschini}, {Genzel}, {Grazian}, {Gruppioni}, {Ilbert}, {Le Floch},
  {Magdis}, {Magliocchetti}, {Magnelli}, {Maiolino}, {McCracken}, {Nordon},
  {Poglitsch}, {Santini}, {Pozzi}, {Riguccini}, {Tacconi}, {Wuyts}, \&
  {Zamorani}}]{Rodighiero:2011aa}
{Rodighiero}, G., {Daddi}, E., {Baronchelli}, I., {et~al.} 2011,
  \bibinfo{title}{{The Lesser Role of Starbursts in Star Formation at z = 2},}
  \apjl, 739, L40, \dodoi{10.1088/2041-8205/739/2/L40}

\bibitem[{S.~A. {Rodney} {et~al.}(2021){Rodney}, {Brammer}, {Pierel},
  {Richard}, {Toft}, {O'Connor}, {Akhshik}, \& {Whitaker}}]{Rodney:2021ab}
{Rodney}, S.~A., {Brammer}, G.~B., {Pierel}, J. D.~R., {et~al.} 2021,
  \bibinfo{title}{{A gravitationally lensed supernova with an observable
  two-decade time delay},} Nature Astronomy, \dodoi{10.1038/s41550-021-01450-9}

\bibitem[{S.~A. {Rodney} {et~al.}(2016){Rodney}, {Strolger}, {Kelly},
  {Brada{\v{c}}}, {Brammer}, {Filippenko}, {Foley}, {Graur}, {Hjorth}, {Jha},
  {McCully}, {Molino}, {Riess}, {Schmidt}, {Selsing}, {Sharon}, {Treu},
  {Weiner}, \& {Zitrin}}]{Rodney:2016aa}
{Rodney}, S.~A., {Strolger}, L.~G., {Kelly}, P.~L., {et~al.} 2016,
  \bibinfo{title}{{SN Refsdal: Photometry and Time Delay Measurements of the
  First Einstein Cross Supernova},} \apj, 820, 50,
  \dodoi{10.3847/0004-637X/820/1/50}

\bibitem[{N. {Roy} {et~al.}(2024){Roy}, {Heckman}, {Henry}, {Chisholm},
  {Flury}, {Leitherer}, {Hayes}, {Jaskot}, {Ji}, {Schaerer}, {Wang},
  {Borthakur}, {Xu}, \& {{\"O}stlin}}]{Roy:2024ae}
{Roy}, N., {Heckman}, T., {Henry}, A., {et~al.} 2024, \bibinfo{title}{{Lyman
  Continuum leakage from massive leaky starbursts: A different class of
  emitters?},} arXiv e-prints, arXiv:2410.13254,
  \dodoi{10.48550/arXiv.2410.13254}

\bibitem[{A.~J. {Ruiter} \& I.~R. {Seitenzahl}(2025){Ruiter} \&
  {Seitenzahl}}]{Ruiter:2025aa}
{Ruiter}, A.~J., \& {Seitenzahl}, I.~R. 2025, \bibinfo{title}{{Type Ia
  supernova progenitors: a contemporary view of a long-standing puzzle},}
  \aapr, 33, 1, \dodoi{10.1007/s00159-024-00158-9}

\bibitem[{W. {Rujopakarn} {et~al.}(2011){Rujopakarn}, {Rieke}, {Eisenstein}, \&
  {Juneau}}]{Rujopakarn:2011aa}
{Rujopakarn}, W., {Rieke}, G.~H., {Eisenstein}, D.~J., \& {Juneau}, S. 2011,
  \bibinfo{title}{{Morphology and Size Differences Between Local and
  High-redshift Luminous Infrared Galaxies},} \apj, 726, 93,
  \dodoi{10.1088/0004-637X/726/2/93}

\bibitem[{D. {Ryczanowski} {et~al.}(2023){Ryczanowski}, {Smith}, {Bianconi},
  {McGee}, {Robertson}, {Massey}, \& {Jauzac}}]{Ryczanowski:2023aa}
{Ryczanowski}, D., {Smith}, G.~P., {Bianconi}, M., {et~al.} 2023,
  \bibinfo{title}{{Enabling discovery of gravitationally lensed explosive
  transients: a new method to build an all-sky watch list of groups and
  clusters of galaxies},} \mnras, 520, 2547, \dodoi{10.1093/mnras/stad231}

\bibitem[{M. {Safarzadeh} {et~al.}(2017){Safarzadeh}, {Hayward}, \&
  {Ferguson}}]{Safarzadeh:2017ab}
{Safarzadeh}, M., {Hayward}, C.~C., \& {Ferguson}, H.~C. 2017,
  \bibinfo{title}{{The IRX-{\ensuremath{\beta}} Relation: Insights from
  Simulations},} \apj, 840, 15, \dodoi{10.3847/1538-4357/aa6c5b}

\bibitem[{A. {Sagu{\'e}s Carracedo} {et~al.}(2024){Sagu{\'e}s Carracedo},
  {Goobar}, {M{\"o}rtsell}, {Arendse}, {Johansson}, {Townsend}, {Dhawan},
  {Nordin}, {Sollerman}, \& {Schulze}}]{Sagues-Carracedo:2024aa}
{Sagu{\'e}s Carracedo}, A., {Goobar}, A., {M{\"o}rtsell}, E., {et~al.} 2024,
  \bibinfo{title}{{Detectability and Characterisation of Strongly Lensed
  Supernova Lightcurves in the Zwicky Transient Facility},} arXiv e-prints,
  arXiv:2406.00052, \dodoi{10.48550/arXiv.2406.00052}

\bibitem[{A. {Sainz de Murieta} {et~al.}(2024){Sainz de Murieta}, {Collett},
  {Magee}, {Pierel}, {Enzi}, {Lokken}, {Gagliano}, \&
  {Ryczanowski}}]{Sainz-de-Murieta:2024aa}
{Sainz de Murieta}, A., {Collett}, T.~E., {Magee}, M.~R., {et~al.} 2024,
  \bibinfo{title}{{Find the haystacks, then look for needles: the rate of
  strongly lensed supernovae in galaxy-galaxy strong gravitational lenses},}
  \mnras, 535, 2523, \dodoi{10.1093/mnras/stae2486}

\bibitem[{A. {Sainz de Murieta} {et~al.}(2023){Sainz de Murieta}, {Collett},
  {Magee}, {Weisenbach}, {Krawczyk}, \& {Enzi}}]{Sainz-de-Murieta:2023aa}
{Sainz de Murieta}, A., {Collett}, T.~E., {Magee}, M.~R., {et~al.} 2023,
  \bibinfo{title}{{Lensed Type Ia supernovae in light of SN Zwicky and
  iPTF16geu},} \mnras, 526, 4296, \dodoi{10.1093/mnras/stad3031}

\bibitem[{K. {Sakamoto} {et~al.}(2013){Sakamoto}, {Aalto}, {Costagliola},
  {Mart{\'\i}n}, {Ohyama}, {Wiedner}, \& {Wilner}}]{Sakamoto:2013aa}
{Sakamoto}, K., {Aalto}, S., {Costagliola}, F., {et~al.} 2013,
  \bibinfo{title}{{Submillimeter Interferometry of the Luminous Infrared Galaxy
  NGC 4418: A Hidden Hot Nucleus with an Inflow and an Outflow},} \apj, 764,
  42, \dodoi{10.1088/0004-637X/764/1/42}

\bibitem[{E.~E. {Salpeter}(1955){Salpeter}}]{Salpeter:1955aa}
{Salpeter}, E.~E. 1955, \bibinfo{title}{{The Luminosity Function and Stellar
  Evolution.},} \apj, 121, 161, \dodoi{10.1086/145971}

\bibitem[{E. {Scannapieco} \& L. {Bildsten}(2005){Scannapieco} \&
  {Bildsten}}]{Scannapieco:2005aa}
{Scannapieco}, E., \& {Bildsten}, L. 2005, \bibinfo{title}{{The Type Ia
  Supernova Rate},} \apjl, 629, L85, \dodoi{10.1086/452632}

\bibitem[{P.~L. {Schechter} {et~al.}(2003){Schechter}, {Udalski},
  {Szyma{\'n}ski}, {Kubiak}, {Pietrzy{\'n}ski}, {Soszy{\'n}ski}, {Wo{\'z}niak},
  {{\.Z}ebru{\'n}}, {Szewczyk}, \& {Wyrzykowski}}]{Schechter:2003aa}
{Schechter}, P.~L., {Udalski}, A., {Szyma{\'n}ski}, M., {et~al.} 2003,
  \bibinfo{title}{{Microlensing of Relativistic Knots in the Quasar HE
  1104-1805 AB},} \apj, 584, 657, \dodoi{10.1086/345716}

\bibitem[{M. {Schmidt}(1959){Schmidt}}]{Schmidt:1959aa}
{Schmidt}, M. 1959, \bibinfo{title}{{The Rate of Star Formation.},} \apj, 129,
  243, \dodoi{10.1086/146614}

\bibitem[{P. {Schneider}(1985){Schneider}}]{Schneider:1985aa}
{Schneider}, P. 1985, \bibinfo{title}{{A new formulation of gravitational lens
  theory, time-delay, and Fermat's principle},} \aap, 143, 413

\bibitem[{P. {Schneider} {et~al.}(1992){Schneider}, {Ehlers}, \&
  {Falco}}]{Schneider:1992aa}
{Schneider}, P., {Ehlers}, J., \& {Falco}, E.~E. 1992, {Gravitational Lenses}
  (Springer-Verlag), 112, \dodoi{10.1007/978-3-662-03758-4}

\bibitem[{P. {Schneider} \& D. {Sluse}(2013){Schneider} \&
  {Sluse}}]{Schneider:2013aa}
{Schneider}, P., \& {Sluse}, D. 2013, \bibinfo{title}{{Mass-sheet degeneracy,
  power-law models and external convergence: Impact on the determination of the
  Hubble constant from gravitational lensing},} \aap, 559, A37,
  \dodoi{10.1051/0004-6361/201321882}

\bibitem[{C. {Sedgwick} {et~al.}(2025){Sedgwick}, {Serjeant}, \&
  {Weiner}}]{Sedgwick:2025aa}
{Sedgwick}, C., {Serjeant}, S., \& {Weiner}, C. 2025, \bibinfo{title}{{The
  detection of strongly lensed submillimetre galaxies},} \mnras, 539, 1077,
  \dodoi{10.1093/mnras/staf414}

\bibitem[{I.~I. {Shapiro}(1964){Shapiro}}]{Shapiro:1964aa}
{Shapiro}, I.~I. 1964, \bibinfo{title}{{Fourth Test of General Relativity},}
  \prl, 13, 789, \dodoi{10.1103/PhysRevLett.13.789}

\bibitem[{H. {Shim} {et~al.}(2022){Shim}, {Lee}, {Kim}, {Scott}, {Serjeant},
  {Ao}, {Barrufet}, {Chapman}, {Clements}, {Conselice}, {Goto}, {Greve},
  {Hwang}, {Im}, {Jeong}, {Kim}, {Kim}, {Kim}, {Kong}, {Koprowski}, {Malkan},
  {Micha{\l}owski}, {Pearson}, {Seo}, {Takagi}, {Toba}, {White}, \&
  {Woo}}]{Shim:2022aa}
{Shim}, H., {Lee}, D., {Kim}, Y., {et~al.} 2022,
  \bibinfo{title}{{Multiwavelength properties of 850-{\ensuremath{\mu}}m
  selected sources from the North Ecliptic Pole SCUBA-2 survey},} \mnras, 514,
  2915, \dodoi{10.1093/mnras/stac1105}

\bibitem[{Y. {Shu} {et~al.}(2018){Shu}, {Bolton}, {Mao}, {Kang}, {Li}, \&
  {Soraisam}}]{Shu:2018ab}
{Shu}, Y., {Bolton}, A.~S., {Mao}, S., {et~al.} 2018,
  \bibinfo{title}{{Prediction of Supernova Rates in Known Galaxy-Galaxy
  Strong-lens Systems},} \apj, 864, 91, \dodoi{10.3847/1538-4357/aad5ea}

\bibitem[{Y. {Shu} {et~al.}(2021){Shu}, {Bolton}, {Mao}, {Kang}, {Li}, \&
  {Soraisam}}]{Shu:2021ab}
{Shu}, Y., {Bolton}, A.~S., {Mao}, S., {et~al.} 2021, \bibinfo{title}{{Erratum:
  ``Prediction of Supernova Rates in Known Galaxy-Galaxy Strong-lens Systems''
  (2018, ApJ, 864, 91)},} \apj, 919, 67, \dodoi{10.3847/1538-4357/ac24a4}

\bibitem[{S.~J. {Smartt}(2009){Smartt}}]{Smartt:2009aa}
{Smartt}, S.~J. 2009, \bibinfo{title}{{Progenitors of Core-Collapse
  Supernovae},} \araa, 47, 63, \dodoi{10.1146/annurev-astro-082708-101737}

\bibitem[{B.~M. {Smith} {et~al.}(2018){Smith}, {Windhorst}, {Jansen}, {Cohen},
  {Jiang}, {Dijkstra}, {Koekemoer}, {Bielby}, {Inoue}, {MacKenty}, {O'Connell},
  \& {Silk}}]{Smith:2018ac}
{Smith}, B.~M., {Windhorst}, R.~A., {Jansen}, R.~A., {et~al.} 2018,
  \bibinfo{title}{{Hubble Space Telescope Wide Field Camera 3 Observations of
  Escaping Lyman Continuum Radiation from Galaxies and Weak AGN at Redshifts z
  {\ensuremath{\sim}} 2.3-4.1},} \apj, 853, 191,
  \dodoi{10.3847/1538-4357/aaa3dc}

\bibitem[{B.~M. {Smith} {et~al.}(2020){Smith}, {Windhorst}, {Cohen},
  {Koekemoer}, {Jansen}, {White}, {Borthakur}, {Hathi}, {Jiang}, {Rutkowski},
  {Ryan}, {Inoue}, {O'Connell}, {MacKenty}, {Conselice}, \&
  {Silk}}]{Smith:2020ab}
{Smith}, B.~M., {Windhorst}, R.~A., {Cohen}, S.~H., {et~al.} 2020,
  \bibinfo{title}{{The Lyman Continuum Escape Fraction of Galaxies and AGN in
  the GOODS Fields},} \apj, 897, 41, \dodoi{10.3847/1538-4357/ab8811}

\bibitem[{B.~M. {Smith} {et~al.}(2024){Smith}, {Windhorst}, {Teplitz}, {Hayes},
  {Rafelski}, {Dickinson}, {Mehta}, {Hathi}, {MacKenty}, {Yung}, {Koekemoer},
  {Soto}, {Conselice}, {Lucas}, {Wang}, {Kim}, {Alavi}, {Grogin}, {Sunnquist},
  {Prichard}, {Jansen}, \& {Uvcandels Team}}]{Smith:2024aa}
{Smith}, B.~M., {Windhorst}, R.~A., {Teplitz}, H., {et~al.} 2024,
  \bibinfo{title}{{Lyman Continuum Emission from Active Galactic Nuclei at 2.3
  {\ensuremath{\lesssim}} z {\ensuremath{\lesssim}} 3.7 in the UVCANDELS
  Fields},} \apj, 964, 73, \dodoi{10.3847/1538-4357/ad1ef0}

\bibitem[{L.~J. {Smith} \& J.~S. {Gallagher}(2001){Smith} \&
  {Gallagher}}]{Smith:2001aa}
{Smith}, L.~J., \& {Gallagher}, J.~S. 2001, \bibinfo{title}{{M82-F: a doomed
  super star cluster?},} \mnras, 326, 1027,
  \dodoi{10.1046/j.1365-8711.2001.04627.x}

\bibitem[{M. {Smith} {et~al.}(2012){Smith}, {Nichol}, {Dilday}, {Marriner},
  {Kessler}, {Bassett}, {Cinabro}, {Frieman}, {Garnavich}, {Jha}, {Lampeitl},
  {Sako}, {Schneider}, \& {Sollerman}}]{Smith:2012aa}
{Smith}, M., {Nichol}, R.~C., {Dilday}, B., {et~al.} 2012, \bibinfo{title}{{The
  SDSS-II Supernova Survey: Parameterizing the Type Ia Supernova Rate as a
  Function of Host Galaxy Properties},} \apj, 755, 61,
  \dodoi{10.1088/0004-637X/755/1/61}

\bibitem[{V. {Smol{\v{c}}i{\'c}} {et~al.}(2017){Smol{\v{c}}i{\'c}},
  {Miettinen}, {Tomi{\v{c}}i{\'c}}, {Zamorani}, {Finoguenov}, {Lemaux},
  {Aravena}, {Capak}, {Chiang}, {Civano}, {Delvecchio}, {Ilbert}, {Jurlin},
  {Karim}, {Laigle}, {Le F{\`e}vre}, {Marchesi}, {McCracken}, {Riechers},
  {Salvato}, {Schinnerer}, {Tasca}, \& {Toft}}]{Smolcic:2017ab}
{Smol{\v{c}}i{\'c}}, V., {Miettinen}, O., {Tomi{\v{c}}i{\'c}}, N., {et~al.}
  2017, \bibinfo{title}{{(Sub)millimetre interferometric imaging of a sample of
  COSMOS/AzTEC submillimetre galaxies. III. Environments},} \aap, 597, A4,
  \dodoi{10.1051/0004-6361/201526989}

\bibitem[{J.~S. {Speagle} {et~al.}(2014){Speagle}, {Steinhardt}, {Capak}, \&
  {Silverman}}]{Speagle:2014aa}
{Speagle}, J.~S., {Steinhardt}, C.~L., {Capak}, P.~L., \& {Silverman}, J.~D.
  2014, \bibinfo{title}{{A Highly Consistent Framework for the Evolution of the
  Star-Forming ``Main Sequence'' from z \raisebox{-0.5ex}\textasciitilde 0-6},}
  \apjs, 214, 15, \dodoi{10.1088/0067-0049/214/2/15}

\bibitem[{V. {Stanishev} {et~al.}(2009){Stanishev}, {Goobar}, {Paech},
  {Amanullah}, {Dahl{\'e}n}, {J{\"o}nsson}, {Kneib}, {Lidman}, {Limousin},
  {M{\"o}rtsell}, {Nobili}, {Richard}, {Riehm}, \& {von
  Strauss}}]{Stanishev:2009aa}
{Stanishev}, V., {Goobar}, A., {Paech}, K., {et~al.} 2009,
  \bibinfo{title}{{Near-IR search for lensed supernovae behind galaxy clusters.
  I. Observations and transient detection efficiency},} \aap, 507, 61,
  \dodoi{10.1051/0004-6361/200911982}

\bibitem[{L.-G. {Strolger} {et~al.}(2015){Strolger}, {Dahlen}, {Rodney},
  {Graur}, {Riess}, {McCully}, {Ravindranath}, {Mobasher}, \&
  {Shahady}}]{Strolger:2015aa}
{Strolger}, L.-G., {Dahlen}, T., {Rodney}, S.~A., {et~al.} 2015,
  \bibinfo{title}{{The Rate of Core Collapse Supernovae to Redshift 2.5 from
  the CANDELS and CLASH Supernova Surveys},} \apj, 813, 93,
  \dodoi{10.1088/0004-637X/813/2/93}

\bibitem[{M. {Sullivan} {et~al.}(2006){Sullivan}, {Le Borgne}, {Pritchet},
  {Hodsman}, {Neill}, {Howell}, {Carlberg}, {Astier}, {Aubourg}, {Balam},
  {Basa}, {Conley}, {Fabbro}, {Fouchez}, {Guy}, {Hook}, {Pain},
  {Palanque-Delabrouille}, {Perrett}, {Regnault}, {Rich}, {Taillet}, {Baumont},
  {Bronder}, {Ellis}, {Filiol}, {Lusset}, {Perlmutter}, {Ripoche}, \&
  {Tao}}]{Sullivan:2006aa}
{Sullivan}, M., {Le Borgne}, D., {Pritchet}, C.~J., {et~al.} 2006,
  \bibinfo{title}{{Rates and Properties of Type Ia Supernovae as a Function of
  Mass and Star Formation in Their Host Galaxies},} \apj, 648, 868,
  \dodoi{10.1086/506137}

\bibitem[{S.~H. {Suyu} {et~al.}(2024){Suyu}, {Goobar}, {Collett}, {More}, \&
  {Vernardos}}]{Suyu:2024aa}
{Suyu}, S.~H., {Goobar}, A., {Collett}, T., {More}, A., \& {Vernardos}, G.
  2024, \bibinfo{title}{{Strong Gravitational Lensing and Microlensing of
  Supernovae},} \ssr, 220, 13, \dodoi{10.1007/s11214-024-01044-7}

\bibitem[{S.~H. {Suyu} {et~al.}(2020){Suyu}, {Huber}, {Ca{\~n}ameras},
  {Kromer}, {Schuldt}, {Taubenberger}, {Y{\i}ld{\i}r{\i}m}, {Bonvin}, {Chan},
  {Courbin}, {N{\"o}bauer}, {Sim}, \& {Sluse}}]{Suyu:2020aa}
{Suyu}, S.~H., {Huber}, S., {Ca{\~n}ameras}, R., {et~al.} 2020,
  \bibinfo{title}{{HOLISMOKES. I. Highly Optimised Lensing Investigations of
  Supernovae, Microlensing Objects, and Kinematics of Ellipticals and
  Spirals},} \aap, 644, A162, \dodoi{10.1051/0004-6361/202037757}

\bibitem[{A.~M. {Swinbank} {et~al.}(2004){Swinbank}, {Smail}, {Chapman},
  {Blain}, {Ivison}, \& {Keel}}]{Swinbank:2004aa}
{Swinbank}, A.~M., {Smail}, I., {Chapman}, S.~C., {et~al.} 2004,
  \bibinfo{title}{{The Rest-Frame Optical Spectra of SCUBA Galaxies},} \apj,
  617, 64, \dodoi{10.1086/425171}

\bibitem[{L.~J. {Tacconi} {et~al.}(2018){Tacconi}, {Genzel}, {Saintonge},
  {Combes}, {Garc{\'\i}a-Burillo}, {Neri}, {Bolatto}, {Contini}, {F{\"o}rster
  Schreiber}, {Lilly}, {Lutz}, {Wuyts}, {Accurso}, {Boissier}, {Boone},
  {Bouch{\'e}}, {Bournaud}, {Burkert}, {Carollo}, {Cooper}, {Cox}, {Feruglio},
  {Freundlich}, {Herrera-Camus}, {Juneau}, {Lippa}, {Naab}, {Renzini},
  {Salome}, {Sternberg}, {Tadaki}, {{\"U}bler}, {Walter}, {Weiner}, \&
  {Weiss}}]{Tacconi:2018aa}
{Tacconi}, L.~J., {Genzel}, R., {Saintonge}, A., {et~al.} 2018,
  \bibinfo{title}{{PHIBSS: Unified Scaling Relations of Gas Depletion Time and
  Molecular Gas Fractions},} \apj, 853, 179, \dodoi{10.3847/1538-4357/aaa4b4}

\bibitem[{K. {Tadaki} {et~al.}(2018){Tadaki}, {Iono}, {Yun}, {Aretxaga},
  {Hatsukade}, {Hughes}, {Ikarashi}, {Izumi}, {Kawabe}, {Kohno}, {Lee},
  {Matsuda}, {Nakanishi}, {Saito}, {Tamura}, {Ueda}, {Umehata}, {Wilson},
  {Michiyama}, {Ando}, \& {Kamieneski}}]{Tadaki:2018aa}
{Tadaki}, K., {Iono}, D., {Yun}, M.~S., {et~al.} 2018, \bibinfo{title}{{The
  gravitationally unstable gas disk of a starburst galaxy 12 billion years
  ago},} \nat, 560, 613, \dodoi{10.1038/s41586-018-0443-1}

\bibitem[{T. {Takata} {et~al.}(2006){Takata}, {Sekiguchi}, {Smail}, {Chapman},
  {Geach}, {Swinbank}, {Blain}, \& {Ivison}}]{Takata:2006aa}
{Takata}, T., {Sekiguchi}, K., {Smail}, I., {et~al.} 2006,
  \bibinfo{title}{{Rest-Frame Optical Spectroscopic Classifications for
  Submillimeter Galaxies},} \apj, 651, 713, \dodoi{10.1086/507985}

\bibitem[{ {The pandas Development Team}(2024){The pandas Development
  Team}}]{The-pandas-development-Team:2024aa}
{The pandas Development Team}. 2024, {pandas-dev/pandas: Pandas}, v2.2.3
  Zenodo, \dodoi{10.5281/zenodo.3509134}

\bibitem[{N. {Timmons} {et~al.}(2015){Timmons}, {Cooray}, {Nayyeri}, {Casey},
  {Calanog}, {Ma}, {Messias}, {Baes}, {Bussmann}, {Dunne}, {Dye}, {Eales},
  {Fu}, {Ivison}, {Maddox}, {Micha{\l}owski}, {Oteo}, {Riechers}, {Valiante},
  \& {Wardlow}}]{Timmons:2015aa}
{Timmons}, N., {Cooray}, A., {Nayyeri}, H., {et~al.} 2015,
  \bibinfo{title}{{Extinction and Nebular Line Properties of a
  Herschel-selected Lensed Dusty Starburst at z = 1.027},} \apj, 805, 140,
  \dodoi{10.1088/0004-637X/805/2/140}

\bibitem[{T. {Treu}(2010){Treu}}]{Treu:2010aa}
{Treu}, T. 2010, \bibinfo{title}{{Strong Lensing by Galaxies},} \araa, 48, 87,
  \dodoi{10.1146/annurev-astro-081309-130924}

\bibitem[{T. {Treu} \& P.~J. {Marshall}(2016){Treu} \&
  {Marshall}}]{Treu:2016aa}
{Treu}, T., \& {Marshall}, P.~J. 2016, \bibinfo{title}{{Time delay
  cosmography},} \aapr, 24, 11, \dodoi{10.1007/s00159-016-0096-8}

\bibitem[{T. {Treu} {et~al.}(2022){Treu}, {Suyu}, \& {Marshall}}]{Treu:2022ac}
{Treu}, T., {Suyu}, S.~H., \& {Marshall}, P.~J. 2022, \bibinfo{title}{{Strong
  lensing time-delay cosmography in the 2020s},} \aapr, 30, 8,
  \dodoi{10.1007/s00159-022-00145-y}

\bibitem[{T. {Trombetti} {et~al.}(2021){Trombetti}, {Burigana}, {Bonato},
  {Herranz}, {De Zotti}, {Negrello}, {Galluzzi}, \&
  {Massardi}}]{Trombetti:2021ab}
{Trombetti}, T., {Burigana}, C., {Bonato}, M., {et~al.} 2021,
  \bibinfo{title}{{Search for candidate strongly lensed dusty galaxies in the
  Planck satellite catalogues},} \aap, 653, A151,
  \dodoi{10.1051/0004-6361/202140830}

\bibitem[{T. {Tsujimoto} {et~al.}(1997){Tsujimoto}, {Yoshii}, {Nomoto},
  {Matteucci}, {Thielemann}, \& {Hashimoto}}]{Tsujimoto:1997aa}
{Tsujimoto}, T., {Yoshii}, Y., {Nomoto}, K., {et~al.} 1997, \bibinfo{title}{{A
  New Approach to Determine the Initial Mass Function in the Solar
  Neighborhood},} \apj, 483, 228, \dodoi{10.1086/304215}

\bibitem[{E.~L. {Turner} {et~al.}(1984){Turner}, {Ostriker}, \&
  {Gott}}]{Turner:1984aa}
{Turner}, E.~L., {Ostriker}, J.~P., \& {Gott}, III, J.~R. 1984,
  \bibinfo{title}{{The statistics of gravitational lenses - The distributions
  of image angular separations and lens redshifts},} \apj, 284, 1,
  \dodoi{10.1086/162379}

\bibitem[{H. {Umehata} {et~al.}(2015){Umehata}, {Tamura}, {Kohno}, {Ivison},
  {Alexander}, {Geach}, {Hatsukade}, {Hughes}, {Ikarashi}, {Kato}, {Izumi},
  {Kawabe}, {Kubo}, {Lee}, {Lehmer}, {Makiya}, {Matsuda}, {Nakanishi}, {Saito},
  {Smail}, {Yamada}, {Yamaguchi}, \& {Yun}}]{Umehata:2015aa}
{Umehata}, H., {Tamura}, Y., {Kohno}, K., {et~al.} 2015, \bibinfo{title}{{ALMA
  Deep Field in SSA22: A Concentration of Dusty Starbursts in a z = 3.09
  Protocluster Core},} \apjl, 815, L8, \dodoi{10.1088/2041-8205/815/1/L8}

\bibitem[{S.~A. {Urquhart} {et~al.}(2022){Urquhart}, {Bendo}, {Serjeant},
  {Bakx}, {Hagimoto}, {Cox}, {Neri}, {Lehnert}, {Sedgwick}, {Weiner},
  {Dannerbauer}, {Amvrosiadis}, {Andreani}, {Baker}, {Beelen}, {Berta},
  {Borsato}, {Buat}, {Butler}, {Cooray}, {De Zotti}, {Dunne}, {Dye}, {Eales},
  {Enia}, {Fan}, {Gavazzi}, {Gonz{\'a}lez-Nuevo}, {Harris}, {Herrera},
  {Hughes}, {Ismail}, {Ivison}, {Jin}, {Jones}, {Kohno}, {Krips}, {Lagache},
  {Marchetti}, {Massardi}, {Messias}, {Negrello}, {Omont}, {Perez-Fournon},
  {Riechers}, {Scott}, {Smith}, {Stanley}, {Tamura}, {Temi}, {Vlahakis},
  {Wei{\ss}}, {van der Werf}, {Verma}, {Yang}, \& {Young}}]{Urquhart:2022ab}
{Urquhart}, S.~A., {Bendo}, G.~J., {Serjeant}, S., {et~al.} 2022,
  \bibinfo{title}{{The bright extragalactic ALMA redshift survey (BEARS) I:
  redshifts of bright gravitationally lensed galaxies from the Herschel
  ATLAS},} \mnras, 511, 3017, \dodoi{10.1093/mnras/stac150}

\bibitem[{P. {V{\"a}is{\"a}nen} {et~al.}(2008){V{\"a}is{\"a}nen}, {Mattila},
  {Kniazev}, {Adamo}, {Efstathiou}, {Farrah}, {Johansson}, {{\"O}stlin},
  {Buckley}, {Burgh}, {Crause}, {Hashimoto}, {Lira}, {Loaring}, {Nordsieck},
  {Romero-Colmenero}, {Ryder}, {Still}, \& {Zijlstra}}]{Vaisanen:2008aa}
{V{\"a}is{\"a}nen}, P., {Mattila}, S., {Kniazev}, A., {et~al.} 2008,
  \bibinfo{title}{{Adaptive optics imaging and optical spectroscopy of a
  multiple merger in a luminous infrared galaxy},} \mnras, 384, 886,
  \dodoi{10.1111/j.1365-2966.2007.12703.x}

\bibitem[{E. {Varenius} {et~al.}(2019){Varenius}, {Conway}, {Batejat},
  {Mart{\'\i}-Vidal}, {P{\'e}rez-Torres}, {Aalto}, {Alberdi}, {Lonsdale}, \&
  {Diamond}}]{Varenius:2019aa}
{Varenius}, E., {Conway}, J.~E., {Batejat}, F., {et~al.} 2019,
  \bibinfo{title}{{The population of SNe/SNRs in the starburst galaxy Arp 220.
  A self-consistent analysis of 20 years of VLBI monitoring},} \aap, 623, A173,
  \dodoi{10.1051/0004-6361/201730631}

\bibitem[{B.~P. {Venemans} {et~al.}(2007){Venemans}, {R{\"o}ttgering}, {Miley},
  {van Breugel}, {de Breuck}, {Kurk}, {Pentericci}, {Stanford}, {Overzier},
  {Croft}, \& {Ford}}]{Venemans:2007aa}
{Venemans}, B.~P., {R{\"o}ttgering}, H.~J.~A., {Miley}, G.~K., {et~al.} 2007,
  \bibinfo{title}{{Protoclusters associated with z > 2 radio galaxies . I.
  Characteristics of high redshift protoclusters},} \aap, 461, 823,
  \dodoi{10.1051/0004-6361:20053941}

\bibitem[{L. {Verde} {et~al.}(2024){Verde}, {Sch{\"o}neberg}, \&
  {Gil-Mar{\'\i}n}}]{Verde:2024aa}
{Verde}, L., {Sch{\"o}neberg}, N., \& {Gil-Mar{\'\i}n}, H. 2024,
  \bibinfo{title}{{A Tale of Many H $_{0}$},} \araa, 62, 287,
  \dodoi{10.1146/annurev-astro-052622-033813}

\bibitem[{G. {Vernardos} {et~al.}(2024){Vernardos}, {Sluse}, {Pooley},
  {Schmidt}, {Millon}, {Weisenbach}, {Motta}, {Anguita}, {Saha}, {O'Dowd},
  {Peel}, \& {Schechter}}]{Vernardos:2024aa}
{Vernardos}, G., {Sluse}, D., {Pooley}, D., {et~al.} 2024,
  \bibinfo{title}{{Microlensing of Strongly Lensed Quasars},} \ssr, 220, 14,
  \dodoi{10.1007/s11214-024-01043-8}

\bibitem[{J. {Wambsganss} \& B. {Paczynski}(1994){Wambsganss} \&
  {Paczynski}}]{Wambsganss:1994aa}
{Wambsganss}, J., \& {Paczynski}, B. 1994, \bibinfo{title}{{Parameter
  Degeneracy in Models of the Quadruple Lens System Q2237+0305},} \aj, 108,
  1156, \dodoi{10.1086/117144}

\bibitem[{Q.~D. {Wang} {et~al.}(2024){Wang}, {Diaz}, {Kamieneski},
  {Harrington}, {Yun}, {Foo}, {Frye}, {Jimenez-Andrade}, {Liu}, {Lowenthal},
  {Pampliega}, {Pascale}, {Vishwas}, \& {Gurwell}}]{Wang:2024ab}
{Wang}, Q.~D., {Diaz}, C.~G., {Kamieneski}, P.~S., {et~al.} 2024,
  \bibinfo{title}{{X-ray detection of the most extreme star-forming galaxies at
  the cosmic noon via strong lensing},} \mnras, 527, 10584,
  \dodoi{10.1093/mnras/stad3827}

\bibitem[{T.-M. {Wang} {et~al.}(2022){Wang}, {Magnelli}, {Schinnerer}, {Liu},
  {Modak}, {Jim{\'e}nez-Andrade}, {Karoumpis}, {Kokorev}, \&
  {Bertoldi}}]{Wang:2022ac}
{Wang}, T.-M., {Magnelli}, B., {Schinnerer}, E., {et~al.} 2022,
  \bibinfo{title}{{A$^{3}$COSMOS: A census on the molecular gas mass and extent
  of main-sequence galaxies across cosmic time},} \aap, 660, A142,
  \dodoi{10.1051/0004-6361/202142299}

\bibitem[{X. {Wang} {et~al.}(2025){Wang}, {Teplitz}, {Smith}, {Windhorst},
  {Rafelski}, {Mehta}, {Alavi}, {Ji}, {Brammer}, {Colbert}, {Grogin}, {Hathi},
  {Koekemoer}, {Prichard}, {Scarlata}, {Sunnquist}, {Arrabal Haro},
  {Conselice}, {Gawiser}, {Guo}, {Hayes}, {Jansen}, {Lucas}, {O'Connell},
  {Robertson}, {Rutkowski}, {Siana}, {Vanzella}, {Ashcraft}, {Bagley},
  {Baronchelli}, {Barro}, {Blanche}, {Broussard}, {Carleton}, {Chartab},
  {Cheng}, {Codoreanu}, {Cohen}, {Dai}, {Darvish}, {Dav{\'e}}, {Degroot}, {de
  Mello}, {Dickinson}, {Emami}, {Ferguson}, {Ferreira}, {Finkelstein},
  {Finkelstein}, {Gardner}, {Gburek}, {Giavalisco}, {Grazian}, {Gronwall},
  {Hemmati}, {Howell}, {Iyer}, {Kaviraj}, {Kurczynski}, {Lazar}, {MacKenty},
  {Mantha}, {Martin}, {Martin}, {McCabe}, {Mobasher}, {Nedkova}, {Olsen},
  {Otteson}, {Ravindranath}, {Redshaw}, {Sattari}, {Soto}, {Yung}, {Zabelle},
  \& {UVCANDELS Team}}]{Wang:2025an}
{Wang}, X., {Teplitz}, H.~I., {Smith}, B.~M., {et~al.} 2025,
  \bibinfo{title}{{The Lyman Continuum Escape Fraction of Star-forming Galaxies
  at 2.4 {\ensuremath{\lesssim}} z {\ensuremath{\lesssim}} 3.0 from
  UVCANDELS},} \apj, 980, 74, \dodoi{10.3847/1538-4357/ada4ab}

\bibitem[{J.~L. {Wardlow} {et~al.}(2013){Wardlow}, {Cooray}, {De Bernardis},
  {Amblard}, {Arumugam}, {Aussel}, {Baker}, {B{\'e}thermin}, {Blundell},
  {Bock}, {Boselli}, {Bridge}, {Buat}, {Burgarella}, {Bussmann},
  {Cabrera-Lavers}, {Calanog}, {Carpenter}, {Casey}, {Castro-Rodr{\'{\i}}guez},
  {Cava}, {Chanial}, {Chapin}, {Chapman}, {Clements}, {Conley}, {Cox},
  {Dowell}, {Dye}, {Eales}, {Farrah}, {Ferrero}, {Franceschini}, {Frayer},
  {Frazer}, {Fu}, {Gavazzi}, {Glenn}, {Gonz{\'a}lez Solares}, {Griffin},
  {Gurwell}, {Harris}, {Hatziminaoglou}, {Hopwood}, {Hyde}, {Ibar}, {Ivison},
  {Kim}, {Lagache}, {Levenson}, {Marchetti}, {Marsden}, {Martinez-Navajas},
  {Negrello}, {Neri}, {Nguyen}, {O'Halloran}, {Oliver}, {Omont}, {Page},
  {Panuzzo}, {Papageorgiou}, {Pearson}, {P{\'e}rez-Fournon}, {Pohlen},
  {Riechers}, {Rigopoulou}, {Roseboom}, {Rowan-Robinson}, {Schulz}, {Scott},
  {Scoville}, {Seymour}, {Shupe}, {Smith}, {Streblyanska}, {Strom},
  {Symeonidis}, {Trichas}, {Vaccari}, {Vieira}, {Viero}, {Wang}, {Xu}, {Yan},
  \& {Zemcov}}]{Wardlow:2013aa}
{Wardlow}, J.~L., {Cooray}, A., {De Bernardis}, F., {et~al.} 2013,
  \bibinfo{title}{{HerMES: Candidate Gravitationally Lensed Galaxies and
  Lensing Statistics at Submillimeter Wavelengths},} \apj, 762, 59,
  \dodoi{10.1088/0004-637X/762/1/59}

\bibitem[{E.~J. {Watkins} {et~al.}(2023){Watkins}, {Barnes}, {Henny}, {Kim},
  {Kreckel}, {Meidt}, {Klessen}, {Glover}, {Williams}, {Keller}, {Leroy},
  {Rosolowsky}, {Lee}, {Anand}, {Belfiore}, {Bigiel}, {Blanc}, {Boquien},
  {Cao}, {Chandar}, {Chen}, {Chevance}, {Congiu}, {Dale}, {Deger}, {Egorov},
  {Emsellem}, {Faesi}, {Grasha}, {Groves}, {Hassani}, {Henshaw}, {Herrera},
  {Hughes}, {Jeffreson}, {Jim{\'e}nez-Donaire}, {Koch}, {Kruijssen}, {Larson},
  {Liu}, {Lopez}, {Pessa}, {Pety}, {Querejeta}, {Saito}, {Sandstrom},
  {Scheuermann}, {Schinnerer}, {Sormani}, {Stuber}, {Thilker}, {Usero}, \&
  {Whitmore}}]{Watkins:2023aa}
{Watkins}, E.~J., {Barnes}, A.~T., {Henny}, K., {et~al.} 2023,
  \bibinfo{title}{{PHANGS-JWST First Results: A Statistical View on Bubble
  Evolution in NGC 628},} \apjl, 944, L24, \dodoi{10.3847/2041-8213/aca6e4}

\bibitem[{A. {Weiss} {et~al.}(2013){Weiss}, {De Breuck}, {Marrone}, {Vieira},
  {Aguirre}, {Aird}, {Aravena}, {Ashby}, {Bayliss}, {Benson}, {B{\'e}thermin},
  {Biggs}, {Bleem}, {Bock}, {Bothwell}, {Bradford}, {Brodwin}, {Carlstrom},
  {Chang}, {Chapman}, {Crawford}, {Crites}, {de Haan}, {Dobbs}, {Downes},
  {Fassnacht}, {George}, {Gladders}, {Gonzalez}, {Greve}, {Halverson},
  {Hezaveh}, {High}, {Holder}, {Holzapfel}, {Hoover}, {Hrubes}, {Husband},
  {Keisler}, {Lee}, {Leitch}, {Lueker}, {Luong-Van}, {Malkan}, {McIntyre},
  {McMahon}, {Mehl}, {Menten}, {Meyer}, {Murphy}, {Padin}, {Plagge},
  {Reichardt}, {Rest}, {Rosenman}, {Ruel}, {Ruhl}, {Schaffer}, {Shirokoff},
  {Spilker}, {Stalder}, {Staniszewski}, {Stark}, {Story}, {Vanderlinde},
  {Welikala}, \& {Williamson}}]{Weiss:2013aa}
{Weiss}, A., {De Breuck}, C., {Marrone}, D.~P., {et~al.} 2013,
  \bibinfo{title}{{ALMA Redshifts of Millimeter-selected Galaxies from the SPT
  Survey: The Redshift Distribution of Dusty Star-forming Galaxies},} \apj,
  767, 88, \dodoi{10.1088/0004-637X/767/1/88}

\bibitem[{K.~E. {Whitaker} {et~al.}(2017){Whitaker}, {Pope}, {Cybulski},
  {Casey}, {Popping}, \& {Yun}}]{Whitaker:2017aa}
{Whitaker}, K.~E., {Pope}, A., {Cybulski}, R., {et~al.} 2017,
  \bibinfo{title}{{The Constant Average Relationship between Dust-obscured Star
  Formation and Stellar Mass from z = 0 to z = 2.5},} \apj, 850, 208,
  \dodoi{10.3847/1538-4357/aa94ce}

\bibitem[{K.~E. {Whitaker} {et~al.}(2012){Whitaker}, {van Dokkum}, {Brammer},
  \& {Franx}}]{Whitaker:2012aa}
{Whitaker}, K.~E., {van Dokkum}, P.~G., {Brammer}, G., \& {Franx}, M. 2012,
  \bibinfo{title}{{The Star Formation Mass Sequence Out to z = 2.5},} \apjl,
  754, L29, \dodoi{10.1088/2041-8205/754/2/L29}

\bibitem[{K.~E. {Whitaker} {et~al.}(2014){Whitaker}, {Franx}, {Leja}, {van
  Dokkum}, {Henry}, {Skelton}, {Fumagalli}, {Momcheva}, {Brammer}, {Labb{\'e}},
  {Nelson}, \& {Rigby}}]{Whitaker:2014ab}
{Whitaker}, K.~E., {Franx}, M., {Leja}, J., {et~al.} 2014,
  \bibinfo{title}{{Constraining the Low-mass Slope of the Star Formation
  Sequence at 0.5 < z < 2.5},} \apj, 795, 104,
  \dodoi{10.1088/0004-637X/795/2/104}

\bibitem[{B.~C. {Whitmore} {et~al.}(2010){Whitmore}, {Chandar}, {Schweizer},
  {Rothberg}, {Leitherer}, {Rieke}, {Rieke}, {Blair}, {Mengel}, \&
  {Alonso-Herrero}}]{Whitmore:2010aa}
{Whitmore}, B.~C., {Chandar}, R., {Schweizer}, F., {et~al.} 2010,
  \bibinfo{title}{{The Antennae Galaxies (NGC 4038/4039) Revisited: Advanced
  Camera for Surveys and NICMOS Observations of a Prototypical Merger},} \aj,
  140, 75, \dodoi{10.1088/0004-6256/140/1/75}

\bibitem[{V. {Wild} {et~al.}(2011){Wild}, {Charlot}, {Brinchmann}, {Heckman},
  {Vince}, {Pacifici}, \& {Chevallard}}]{Wild:2011aa}
{Wild}, V., {Charlot}, S., {Brinchmann}, J., {et~al.} 2011,
  \bibinfo{title}{{Empirical determination of the shape of dust attenuation
  curves in star-forming galaxies},} \mnras, 417, 1760,
  \dodoi{10.1111/j.1365-2966.2011.19367.x}

\bibitem[{R.~A. {Windhorst} {et~al.}(2023){Windhorst}, {Cohen}, {Jansen},
  {Summers}, {Tompkins}, {Conselice}, {Driver}, {Yan}, {Coe}, {Frye}, {Grogin},
  {Koekemoer}, {Marshall}, {O'Brien}, {Pirzkal}, {Robotham}, {Ryan}, {Willmer},
  {Carleton}, {Diego}, {Keel}, {Porto}, {Redshaw}, {Scheller}, {Wilkins},
  {Willner}, {Zitrin}, {Adams}, {Austin}, {Arendt}, {Beacom}, {Bhatawdekar},
  {Bradley}, {Broadhurst}, {Cheng}, {Civano}, {Dai}, {Dole}, {D'Silva},
  {Duncan}, {Fazio}, {Ferrami}, {Ferreira}, {Finkelstein}, {Furtak}, {Gim},
  {Griffiths}, {Hammel}, {Harrington}, {Hathi}, {Holwerda}, {Honor}, {Huang},
  {Hyun}, {Im}, {Joshi}, {Kamieneski}, {Kelly}, {Larson}, {Li}, {Lim}, {Ma},
  {Maksym}, {Manzoni}, {Meena}, {Milam}, {Nonino}, {Pascale}, {Petric},
  {Pierel}, {del Carmen Polletta}, {R{\"o}ttgering}, {Rutkowski}, {Smail},
  {Straughn}, {Strolger}, {Swirbul}, {Trussler}, {Wang}, {Welch}, {B. Wyithe},
  {Yun}, {Zackrisson}, {Zhang}, \& {Zhao}}]{Windhorst:2023aa}
{Windhorst}, R.~A., {Cohen}, S.~H., {Jansen}, R.~A., {et~al.} 2023,
  \bibinfo{title}{{JWST PEARLS. Prime Extragalactic Areas for Reionization and
  Lensing Science: Project Overview and First Results},} \aj, 165, 13,
  \dodoi{10.3847/1538-3881/aca163}

\bibitem[{H.~J. {Witt} {et~al.}(2000){Witt}, {Mao}, \& {Keeton}}]{Witt:2000ab}
{Witt}, H.~J., {Mao}, S., \& {Keeton}, C.~R. 2000, \bibinfo{title}{{Analytic
  Time Delays and H$_{0}$ Estimates for Gravitational Lenses},} \apj, 544, 98,
  \dodoi{10.1086/317201}

\bibitem[{H.~J. {Witt} {et~al.}(1995){Witt}, {Mao}, \&
  {Schechter}}]{Witt:1995ab}
{Witt}, H.~J., {Mao}, S., \& {Schechter}, P.~L. 1995, \bibinfo{title}{{On the
  Universality of Microlensing in Quadruple Gravitational Lenses},} \apj, 443,
  18, \dodoi{10.1086/175499}

\bibitem[{R. {Wojtak} {et~al.}(2019){Wojtak}, {Hjorth}, \&
  {Gall}}]{Wojtak:2019aa}
{Wojtak}, R., {Hjorth}, J., \& {Gall}, C. 2019, \bibinfo{title}{{Magnified or
  multiply imaged? - Search strategies for gravitationally lensed supernovae in
  wide-field surveys},} \mnras, 487, 3342, \dodoi{10.1093/mnras/stz1516}

\bibitem[{K.~C. {Wong} {et~al.}(2020){Wong}, {Suyu}, {Chen}, {Rusu}, {Millon},
  {Sluse}, {Bonvin}, {Fassnacht}, {Taubenberger}, {Auger}, {Birrer}, {Chan},
  {Courbin}, {Hilbert}, {Tihhonova}, {Treu}, {Agnello}, {Ding}, {Jee},
  {Komatsu}, {Shajib}, {Sonnenfeld}, {Blandford}, {Koopmans}, {Marshall}, \&
  {Meylan}}]{Wong:2020aa}
{Wong}, K.~C., {Suyu}, S.~H., {Chen}, G. C.~F., {et~al.} 2020,
  \bibinfo{title}{{H0LiCOW - XIII. A 2.4 per cent measurement of H$_{0}$ from
  lensed quasars: 5.3{\ensuremath{\sigma}} tension between early- and
  late-Universe probes},} \mnras, 498, 1420, \dodoi{10.1093/mnras/stz3094}

\bibitem[{E.~L. {Wright}(2006){Wright}}]{Wright:2006aa}
{Wright}, E.~L. 2006, \bibinfo{title}{{A Cosmology Calculator for the World
  Wide Web},} \pasp, 118, 1711, \dodoi{10.1086/510102}

\bibitem[{H. {Yan} {et~al.}(2018){Yan}, {Ma}, {Beacom}, \&
  {Runge}}]{Yan:2018aa}
{Yan}, H., {Ma}, Z., {Beacom}, J.~F., \& {Runge}, J. 2018,
  \bibinfo{title}{{Revealing Dusty Supernovae in High-redshift (Ultra)Luminous
  Infrared Galaxies through Near-infrared Integrated Light Variability},} \apj,
  867, 21, \dodoi{10.3847/1538-4357/aadf38}

\bibitem[{D.~R. {Young} {et~al.}(2008){Young}, {Smartt}, {Mattila}, {Tanvir},
  {Bersier}, {Chambers}, {Kaiser}, \& {Tonry}}]{Young:2008aa}
{Young}, D.~R., {Smartt}, S.~J., {Mattila}, S., {et~al.} 2008,
  \bibinfo{title}{{Core-collapse supernovae in low-metallicity environments and
  future all-sky transient surveys},} \aap, 489, 359,
  \dodoi{10.1051/0004-6361:20078662}

\bibitem[{F. {Zwicky}(1938){Zwicky}}]{Zwicky:1938aa}
{Zwicky}, F. 1938, \bibinfo{title}{{On the Frequency of Supernovae.},} \apj,
  88, 529, \dodoi{10.1086/144007}

\end{thebibliography}



\end{document}